\documentclass[12pt]{article}
\usepackage{latexsym}
\evensidemargin \oddsidemargin
\usepackage{graphicx}
\usepackage{longtable}
\usepackage{amsmath}
\usepackage{amssymb}
\usepackage{cite}
\usepackage{slashed}
\usepackage{tikz}
\usetikzlibrary{arrows,shapes,calc}

\usepackage{ulem}   
\usepackage{multirow}
\usepackage{longtable}
\usepackage{color}
\definecolor{orange}{RGB}{255,127,0}
\definecolor{brown}{RGB}{102,51,0}
\definecolor{myred}{RGB}{192,0,0}
\definecolor{Darkgreen}{RGB}{30,120,30}
\definecolor{Darkblue}{RGB}{0,0,200}
\newcommand\lsim{\mathrel{\rlap{\lower4pt\hbox{\hskip1pt$\sim$}}
    \raise1pt\hbox{$<$}}}
\newcommand\gsim{\mathrel{\rlap{\lower4pt\hbox{\hskip1pt$\sim$}}
    \raise1pt\hbox{$>$}}}
\newcommand{\ba}{\begin{array}}
\newcommand{\ea}{\end{array}}
\newcommand{\nn}{\nonumber}

\newcommand{\be}{\begin{equation}}
\newcommand{\ee}{\end{equation}}
\newcommand{\bear}{\begin{eqnarray}}
\newcommand{\eear}{\end{eqnarray}}

\newcommand{\ket}{\,\rangle}
\newcommand{\bra}{\langle \,}
\newcommand{\eqn}[1]{(\ref{#1})}
\newcommand{\cO}{{\cal O}}
\newcommand{\bel}[1]{\be\label{#1}}

\newcommand{\mL}{\mathcal{L}}

\newcommand{\mB}{\mathcal{B}}

\newcommand{\mF}{\mathcal{F}}
\newcommand{\mG}{\mathcal{G}}
\newcommand{\mH}{\mathcal{H}}

\newcommand{\mO}{\mathcal{O}}
\newcommand{\mP}{\mathcal{P}}

\newcommand{\mT}{\mathcal{T}}

\newcommand{\mX}{\mathcal{X}}
\newcommand{\mY}{\mathcal{Y}}
\newcommand{\wl}{\widetilde{\lambda}}

\newcommand{\Frac}[2]{\frac{\displaystyle #1}{\displaystyle #2}}

\def\bat{\begin{array}{cc}}

\begin{document}
\begin{titlepage}
\def\thefootnote{\fnsymbol{footnote}}
\begin{flushright}
{\small
IFIC/18-07; \quad FTUV/18-1026\\ FERMILAB-PUB-18-550-T \\ VBSCAN-PUB-06-18 \\
April 2019
}
\end{flushright}

\begin{center}
{\huge\sc {\bf
Colorful Imprints of Heavy States\\[10pt] in the Electroweak Effective Theory
}}
\end{center}

\vspace{0.5cm}
\begin{center}
{\sc
Claudius~Krause,$^{1,2}$\footnote{email: ckrause@fnal.gov}  \
Antonio~Pich,$^{1}$\footnote{email: pich@ific.uv.es}  \
Ignasi~Rosell,$^{3}$\footnote{email: rosell@uchceu.es} \
\\[5pt]
Joaqu\'{\i}n~Santos,$^{1}$\footnote{email: Joaquin.Santos@ific.uv.es}
and  \ Juan~Jos\'e~Sanz-Cillero$^{4}$\footnote{email: jjsanzcillero@ucm.es}
}

\vspace*{.7cm}

{\sl
$^1$ Departament de F\'\i sica Te\`orica, IFIC, Universitat de Val\`encia -- CSIC,\\
Apt. Correus 22085, E-46071 Val\`encia, Spain,

\vspace*{0.1cm}

$^{2}$
Theoretical Physics Department, Fermi National Accelerator Laboratory, Batavia, IL, 60510, USA

\vspace*{0.1cm}
$^3$
Departamento de Matem\'aticas, F\'\i sica y Ciencias Tecnol\'ogicas,\\
Universidad Cardenal Herrera-CEU, CEU Universities,
E-46115 Alfara del Patriarca, Val\`encia, Spain

\vspace*{0.1cm}

$^4$
Departamento de F\'\i sica Te\'orica and UPARCOS,  Universidad Complutense de Madrid, E-28040 Madrid, Spain
}
\end{center}

\vspace*{0.4cm}

\begin{abstract}
\vspace{0.1cm}\noindent
We analyze heavy states from generic ultraviolet completions of the Standard Model in a model-independent way and investigate their implications on the low-energy couplings of the electroweak effective theory. We build a general effective Lagrangian, implementing the electroweak symmetry breaking $SU(2)_L\otimes SU(2)_R\to SU(2)_{L+R}$ with a non-linear Nambu-Goldstone realization, which couples the known particles to the heavy states. We generalize the formalism developed in previous works~\cite{Pich:2016lew,Pich:2015kwa} to include colored resonances, both of bosonic and fermionic type. We study bosonic heavy states with $J^P=0^\pm$ and $J^P=1^\pm$, in singlet or triplet $SU(2)_{L+R}$ representations and in singlet or octet representations of $SU(3)_C$, and fermionic resonances with $J=\frac{1}{2}$ that are electroweak doublets and QCD triplets or singlets.
Integrating out the heavy scales, we determine the complete pattern of low-energy couplings at the lowest non-trivial order. Some specific types of (strongly- and weakly-coupled) ultraviolet completions are discussed to illustrate the generality of our approach and to make contact with current experimental searches.
\end{abstract}

\vfill

\end{titlepage}
\def\thefootnote{\arabic{footnote}}
\setcounter{footnote}{0}
\newpage

\section{Introduction}
The first years of physics runs at the LHC confirmed that the Standard Model (SM) describes the physics at the electroweak (EW) scale very well. The experimental ATLAS and CMS collaborations found a Higgs-like\footnote{We refer to this particle as ``Higgs'', even though it might not be the SM Higgs boson.} particle \cite{Aad:2012tfa,Chatrchyan:2012xdj} with couplings close to the SM Higgs expectations, seemingly completing the SM, while direct searches for new physics beyond the SM (BSM) yielded only negative results so far. The lack of new states indicates the presence of a mass gap between the electroweak scale and the scale of new physics. This separation of scales enables us to use bottom-up effective field theories (EFTs) to study the low-energy effects of heavy new physics in a systematic way.

In EFTs, the information about the ultraviolet (UV) and the infrared (IR) are separated: the UV information is entirely contained in the Wilson coefficients (or low-energy constants, LECs), and the IR information is described by the local operators. When all LECs at a given order are kept as free parameters, the approach is model-independent within a small and well-justified set of assumptions. These assumptions are usually motivated by experimental results and concern the low-energy particle content and the symmetries. When constructing an EFT for the SM, there is usually no debate about the particle content. However, the newly-discovered Higgs can be treated in two different ways. We either assume that it forms an IR doublet structure, together with the three Nambu-Goldstone bosons of the electroweak symmetry breaking, or we do not assume any specific relation between the Higgs and the Nambu-Goldstone bosons. In the latter case, the Higgs can be parametrized by a scalar singlet with independent couplings. These two cases lead to two different EFTs, with different power counting and therefore conceptually different effective expansions.

In the first case, with an $SU(2)_L$ doublet structure at the electroweak scale, the resulting EFT is an expansion in canonical dimensions that is called SM effective theory (SMEFT). It usually describes weakly coupled new physics that decouples from the SM in a certain limit. Since the doublet containing the Higgs and the EW Goldstones transforms linearly under gauge transformations, the SMEFT is also called ``linear'' EFT.

The second case, called EW effective theory (EWET), EW chiral Lagrangian (EWChL) or Higgs effective theory (HEFT), is a more general (non-linear) realization of the EW symmetry breaking, which includes the SMEFT as a particular case. It allows for rather large deviations from the Standard Model in the Higgs sector compared to the well-tested gauge-fermion sector. Such effects are usually induced by strongly-coupled physics in the UV. However, the choice of an appropriate EFT for a given UV model depends on more model-dependent details. For example, a simple extension of the SM scalar sector with an additional (heavy) singlet scalar induces at low energies the linear EFT with a Higgs doublet and an expansion in canonical dimensions as the natural EFT in most of the parameter space. In other regions of parameter space, however, the truly non-linear EFT is induced and the low-energy expansion is not given by canonical dimensions~\cite{Buchalla:2016bse}.

\begin{figure}[!h]
  \begin{center}
    \begin{tikzpicture}
      \draw[thick] {(0em,0em) rectangle (0.8\textwidth,11em)};
      \draw[->,thick] ($(1em,2em)$) -- ($(1em,9em)$) node [above] {$E$};
      \node [text width = 15em, anchor = north west, text centered] (qcd) at ($(0.5em,10em)$) {QCD $(q_{a}, G_{\mu\nu})$\\ $\updownarrow $\\ Resonance Chiral Theory\\ $(\rho, a_1,\dots)$\\ $\updownarrow $\\Chiral Perturbation Theory\\ $(f_{\pi}, \pi_{i})$ };
      \node [text width = 15em, anchor = north east, text centered] (ewXL) at ($(0.8\textwidth,11em)+(-0.5em,-1em)$) {Fundamental EW Theory (???)\\ $\updownarrow $\\ Resonance EW Theory\\ $(V, A, \dots)$\\ $\updownarrow $\\Electroweak Effective Theory\\ $(v, \varphi_{i})$};
    \end{tikzpicture}
  \end{center}
  \caption{Schematic view of the energy regimes of the different (effective) theories. A possible UV completion of the SM electroweak sector could behave somehow similarly to QCD.}
  \label{fig:scales}
\end{figure}

When working with EFTs, we can pursue two different strategies. First, in a bottom-up approach, we can fit the LECs to experimental data and look for deviations from the SM. Second, we can look at certain (classes of) UV models and analyze them in a top-down manner to study the pattern of LECs they generate. Once a deviation from the SM is observed in the data, the intuition gained in the top-down approach helps to find what type of UV model is causing it.

An interesting class of new-physics models are strongly-coupled UV completions. In this type of scenarios the electroweak scale is generated dynamically, as it happens in Quantum Chromodynamics (QCD).
The lightest states are the (pseudo-)Nambu-Goldstone bosons associated with the EW symmetry breaking, possibly including the Higgs boson. At higher energies, the resonances of the strongly-coupled interaction are excited. Since the resonance masses set the cut-off of the low-energy EFT, they cannot be described within this framework. However, as it is long known for QCD \cite{Ecker:1988te,Ecker:1989yg}, their dynamical effects on the low-energy physics are incorporated through the LECs that carry all relevant short-distance information.

The resonance states can be described within an effective Resonance Theory, where they are included {\it via} a phenomenological Lagrangian that interpolates between the UV and IR theories.
In this paper, we construct an effective Lagrangian, containing the SM states and the heavier fields, based on the symmetry group $\mG = SU(3)_C\otimes SU(2)_L \otimes SU(2)_R \otimes U(1)_X$, with $X=(\mathrm{B}- \mathrm{L})/2$, dynamically (or spontaneously) broken to $\mH =SU(3)_C\otimes SU(2)_{L+R} \otimes U(1)_X$. Thus, our main assumption is the successful pattern of EW symmetry breaking implemented in the SM. Integrating out the heavy resonance fields, one recovers the low-energy EFT with explicit values for the LECs in terms of resonance parameters. Fig.~\ref{fig:scales} illustrates the plausible analogy between UV completions of the SM and the well-known QCD case.

Previously \cite{Pich:2016lew,Pich:2015kwa}, we analyzed the impact of colorless bosonic resonances on the low-energy electroweak effective theory.
Here, we will continue to explore this direction and study the
contributions to the LECs from colored resonances, both of bosonic and fermionic type.\footnote{Some preliminary results have been already presented in Ref.~\cite{Rosell:2017kps}.}
We consider bosonic heavy states with quantum numbers $J^{P}= 0^{\pm}$ and $J^{P}= 1^{\pm}$, which are in singlet or triplet representations of the electroweak group and in singlet or octet representations of the $SU(3)_C$ group. At the lowest non-trivial order in inverse powers of the heavy masses, the only relevant fermionic resonances are states with $J=\frac{1}{2}$ that are in a doublet representation of the electroweak group.
We will discuss in some detail the new technical aspects associated with fermionic resonances, since they were not considered at all in Refs.~\cite{Pich:2016lew,Pich:2015kwa}.
The incorporation of the color degree of freedom requires an enlargement of the operator basis defined in Ref.~\cite{Pich:2016lew}, including also gluonic operators. This implies a larger number of unknown LECs. Nevertheless, the LHC data are obviously more constraining on QCD-sensitive operators. We will show, in particular, the implications of current dijet, dilepton and  diboson production data on the resonance mass scale.

The structure of this paper is as follows. In Section \ref{sec:lE-EFT}, we review the construction of the low-energy effective theory, introduce our notation, and discuss how colored fields can be incorporated. We include the resonances in Section \ref{sec:res-EFT} and derive their contributions to the LECs in Section \ref{sec:LEC}. Section~\ref{sec:pheno} discusses the current phenomenological status and illustrates how our generic EFT framework can be easily particularized to specific (classes of) UV completions and experimental searches. We conclude in Section \ref{sec:concl}. Some technical aspects are compiled in various appendices. A detailed comparison between our basis of effective operators and the basis adopted in Ref.~\cite{Buchalla:2013rka} is given in Appendix~\ref{sec:dic}, which proves their equivalence and provides a complete dictionary to translate the results obtained in the two bases. Some useful algebraic identities are listed in Appendix~\ref{app:relations}. Appendix~\ref{app:coupling} analyzes the special case of a Higgsed heavy scalar resonance with enhanced couplings proportional to its mass, and Appendix~\ref{app:fermiondiagonalization} discusses the diagonalization of the quadratic fermion Lagrangian through a redefinition of the (light and heavy) fermionic fields.

\section{The Low-Energy Effective Theory}
\label{sec:lE-EFT}

\subsection{Constructing the effective theory} \label{constructing}
We make use of the non-linear EWET Lagrangian to describe the physics at the electroweak scale \cite{Pich:2016lew,Pich:2015kwa,Rosell:2017kps,Dobado:1989ax,Dobado:1989ue,Dobado:1990zh,Dobado:1990jy,Espriu:1991vm,Herrero:1993nc,Herrero:1994iu,Feruglio:1992wf,Pich:1998xt,Bagger:1993zf,Koulovassilopoulos:1993pw,Burgess:1999ha,Wang:2006im,Grinstein:2007iv,Alonso:2012px,Buchalla:2012qq,Buchalla:2013rka,Buchalla:2013eza,Pich:2018ltt}. This is a conservative choice, as the EWET is the most general electroweak EFT, also allowing for (but not restricted to) a strongly-coupled UV completion. This is further supported by the current knowledge about the the Higgs couplings, which have still rather large experimental uncertainties of at best $\mathcal{O}(10\%)$ \cite{deBlas:2018tjm}.\footnote{More precise constraints of some Higgs couplings have been obtained in simplified one-loop analyses~\cite{deBlas:2016ojx} of the oblique $S$ and $T$ parameters, by including further assumptions like, e.g., neglecting all the $\cO(p^4)$ couplings and keeping only the (renormalized) one-loop contributions. A proper EFT study of these oblique parameters at NLO shows that the $hWW$ coupling can indeed be constrained to deviate no more than 6\% (95\% C.L.) from its SM value in some dynamical scenarios, but $\mO(10\%)$ corrections remain allowed in more general settings~\cite{Pich:2013fea,Pich:2012dv}.
}
The systematics of the construction is analogous to \cite{Pich:2016lew,Pich:2018ltt}. However, for completeness and readability we repeat the main steps here again. In addition, we will discuss how colored fermions together with the gluons of QCD are included. We construct the EFT with the following assumptions:
\begin{itemize}

\item {\bf Particle content:} We assume the particle content of the SM, including the Higgs $h$ and the three EW Goldstone bosons $\varphi^a$, the four EW gauge bosons ($W^\pm,Z,A$) and the eight gluon fields ($G^{a}$).
However, for simplicity, we will only consider a single generation of colored fermions. Occasionally, we will comment on using leptons instead of quarks, but the study of the fermion flavor structure is beyond the scope of this article. We do not assume that the Higgs forms a doublet together with the three Goldstone bosons of the electroweak symmetry breaking, {\it i.e.}, we include the Higgs as a scalar singlet with $m_h=125$ GeV.

\item {\bf Symmetries:} Although the SM possesses an $SU(2)_L\otimes U(1)_Y$ gauge symmetry, its scalar sector has an enhanced global chiral symmetry $\mG=SU(2)_L\otimes SU(2)_R$ that gets spontaneously broken down to the subgroup $\mH=SU(2)_{L+R}$.
The three electroweak Goldstone bosons parametrize the coset $\mG/\mH$ through a unitary $2\times 2$ matrix $U(\varphi)$. This pattern of
electroweak symmetry breaking (EWSB) gives rise to the masses of the gauge bosons $W^\pm$ and $Z$ and implies the successful mass relation $m_W^2 = m_Z^2 \cos^2{\theta_W}$.
The main assumption in the construction of the EWET is that the underlying BSM theory and our EWET also possess this EWSB pattern~\cite{Pich:2016lew,Pich:2015kwa,Rosell:2017kps,Appelquist:1980vg,Sikivie:1980hm,Longhitano:1980iz,Longhitano:1980tm,Feruglio:1992wf,Pich:1998xt,Buchalla:2012qq,Alonso:2012px,Buchalla:2013rka,Pich:2018ltt}:
\begin{equation}
\mG\equiv SU(2)_L\otimes SU(2)_R\quad \longrightarrow \quad \mH\equiv SU(2)_{L+R}\, .
\end{equation}
The remaining symmetry $\mH$ is called ``custodial'' symmetry \cite{Sikivie:1980hm}, because it protects the ratio of the $W$ and $Z$ masses from receiving large corrections. This ratio is related to the electroweak $T$ parameter \cite{Peskin:1990zt,Peskin:1991sw}. Given the strong experimental constraints~\cite{deBlas:2016ojx,Haller:2018nnx}, we assume that the breakings of custodial symmetry are small.
This assumption has some implications for the power counting that we discuss below. Further, we restrict our analysis to $CP$-even operators and assume conservation of baryon ($\mathrm{B}$) and lepton ($\mathrm{L}$) numbers. More precisely, the EWET Lagrangian will have the $\mathrm{B}-\mathrm{L}$ symmetry of the SM. To incorporate colored states in our EWET, we also assume a local $SU(3)_{C}$ symmetry.
\item {\bf Power counting:} The effective Lagrangian is organized as a low-energy expansion in powers of generalized momenta \cite{Pich:2016lew,weinberg,Buchalla:2012qq,Buchalla:2013rka,Buchalla:2013eza,Buchalla:2016sop,Pich:2018ltt},
\begin{eqnarray}
\mathcal{L}_{\mathrm{EWET}} &=& \sum_{\hat d\ge 2}\, \mathcal{L}_{\mathrm{EWET}}^{(\hat d)}\,, \label{EWET-Lagrangian0}
\end{eqnarray}
where the operators are not ordered according to their canonical dimensions and one must use instead the chiral dimension $\hat d$ which reflects their infrared behavior at low momenta~\cite{weinberg}. Quantum loops are renormalized order by order~\cite{Guo:2015isa,Buchalla:2017jlu,Alonso:2017tdy} in this low-energy expansion. It is interesting to spotlight two features related to this power counting, which differ from previous works~\cite{Longhitano:1980iz,Longhitano:1980tm,Buchalla:2013eza}:
\begin{enumerate}
\item Assuming that the SM fermions couple weakly to the strong sector, we assign an $\mathcal{O}(p^2)$ to fermion bilinears~\cite{Pich:2016lew}. Note that a na\"\i ve dimensional analysis would assign them an $\mathcal{O}(p)$ scaling. In practice, the additional suppression will be implicitly carried by the corresponding operator couplings as it occurs, for instance, with the $\mY$ in the Yukawa Lagrangian in Eq.~\eqref{eq:Yukawas} below.

\item Considering its phenomenological suppression and therefore assuming no strong breaking of the custodial symmetry, we assign an $\mathcal{O}(p)$ to the explicit breaking of this symmetry through the spurion field $\mT$~\cite{Pich:2016lew}, contrary to the pioneering papers studying the Higgsless EWET~\cite{Longhitano:1980iz,Longhitano:1980tm}.
This assignment leads to a more efficient ordering of operators, pushing to higher orders structures that are tightly constrained by the data, and is consistent with the SM sources of custodial symmetry breaking~\cite{Herrero:1993nc}.
\end{enumerate}
Our counting for the effective operators that will be constructed in the next subsections can be summarized as \cite{Pich:2016lew}:
\bear\label{eq:power_counting}
v\,  , \,\varphi^i\,  , \, u(\varphi/v)\, ,\, U(\varphi/v)\, , \, h\,  , \, W^i_\mu\,  ,\, B_\mu\, , \, G_\mu^a
& \sim & \cO\left(p^0\right)\, ,
\nn\\
\xi\, ,\, \bar{\xi}\, ,\,\psi\, ,\,\bar{\psi}\, , \, D_\mu U\,,\, u_\mu\, ,
\,\partial_\mu\, , \, D_\mu\, ,\, d_\mu\,, \, \nabla_\mu\,, \, \hat{W}_\mu\, , \, \hat{B}_\mu\, ,\, \hat{X}_\mu\, , \,
\hat{G}_\mu\,, & & \nn\\
m_h\, , \, m_W\, , \, m_Z\, , \, m_\psi\, ,\, g\, ,\, g'\, , g_s ,\, \mT\,,\, \mY  & \sim & \cO\left( p\right)\, ,
\nn\\
\hat{W}_{\mu\nu}\, ,\, \hat{B}_{\mu\nu}\, ,\, \hat{X}_{\mu\nu}\, ,\,
\hat{G}_{\mu\nu}\, ,\,
f_{\pm\, \mu\nu} \, ,\, c_n^{(V)}\, , \, \bar{\eta}\, \Gamma\, \zeta & \sim & \cO\left( p^2\right)\, ,
\nn\\
\partial_{\mu_1}\partial_{\mu_2} ... \partial_{\mu_n}\,\mF(h/v)  & \sim & \cO\left( p^n\right)\, ,
\label{eq:scaling}
\eear
with the Dirac spinors $\psi,\xi,\eta,\zeta$, the Dirac matrices $\Gamma$, the masses $m_{h,W,Z,\psi}$ of the corresponding fields, the gauge couplings $g,g',g_s$, the EW scale $v = \left(\sqrt{2} G_F\right)^{-1/2} = 246\:\mathrm{GeV}$, and the constants $c^{(V)}_n$ defining the Higgs potential $V(h/v)$ below. All the objects in Eq.~\eqref{eq:scaling} are defined in detail and used in the next subsections to construct the EWET Lagrangian. Thus, every effective operator will have a well-defined chiral dimension $\hat{d}$ ({\it i.e.} it will scale as $\mathcal{O}(p^{\hat{d}})$), assigned by these rules.
\end{itemize}

\subsection{Bosonic fields}
\label{sec:bosonic}

We describe the EW Goldstones in the CCWZ formalism~\cite{Coleman:1969sm,Callan:1969sn}
through the $\mG/ \mH$ coset
representative
$u(\varphi)=\exp\{ i\vec{\sigma}\,\vec{\varphi}/(2v)\}$, transforming under the chiral symmetry group
 $g\equiv (g_L^{\phantom{\dagger}},g_R^{\phantom{\dagger}})\in \mG$ as
 \begin{equation}
   \label{eq:u_transf}\nn
u(\varphi)\quad \longrightarrow\quad g_L^{\phantom{\dagger}}\,
u(\varphi)\,  g_h^\dagger \; =\; g_h^{\phantom{\dagger}} \, u(\varphi) \, g_R^\dagger\, ,
\end{equation}
\begin{equation}
\label{eq:U_u_rel}
U(\varphi)\,\equiv\,  u(\varphi)^2
\quad\longrightarrow\quad g_L^{\phantom{\dagger}}\, U(\varphi)\, g_R^\dagger \, ,
\end{equation}
with the compensating transformation $ g_h^{\phantom{\dagger}}\equiv g_h^{\phantom{\dagger}}(\varphi,g) \in SU(2)_{L+R}$.
Promoting $\mG$ to a local symmetry, we introduce the auxiliary $SU(2)_L$ and $SU(2)_R$ matrix fields, respectively
$\hat{W}_\mu$ and $\hat{B}_\mu$, and their field-strength tensors,
\begin{equation}
\hat{W}^\mu\quad\longrightarrow\quad g_L^{\phantom{\dagger}}\, \hat{W}^\mu g_L^\dagger + i\, g_L^{\phantom{\dagger}}\, \partial^\mu g_L^\dagger\, ,
\qquad\qquad
\hat{B}^\mu\quad\longrightarrow\quad g_R^{\phantom{\dagger}}\, \hat{B}^\mu g_R^\dagger + i\, g_R^{\phantom{\dagger}}\, \partial^\mu g_R^\dagger\, ,
\nn\end{equation} 
\begin{equation}
\hat{W}_{\mu\nu}  = \partial_\mu \hat{W}_\nu - \partial_\nu \hat{W}_\mu
- i\, [\hat{W}_\mu,\hat{W}_\nu]
\quad \longrightarrow\quad g_L^{\phantom{\dagger}}\, \hat{W}_{\mu\nu} \, g_L^\dagger
\, ,
\nn\end{equation} 
\begin{equation}
\hat{B}_{\mu\nu}  = \partial_\mu \hat{B}_\nu - \partial_\nu \hat{B}_\mu
- i\, [\hat{B}_\mu,\hat{B}_\nu]
\quad\longrightarrow\quad
g_R^{\phantom{\dagger}}\, \hat{B}_{\mu\nu} \, g_R^\dagger
\, ,
\nn\end{equation} 
\begin{equation}
  \label{eq:FakeTransform}
f_\pm^{\mu\nu} \, =\,
u^\dagger \hat{W}^{\mu\nu}  u \pm u\, \hat{B}^{\mu\nu} u^\dagger
\quad\longrightarrow\quad g_h^{\phantom{\dagger}}\, f_\pm^{\mu\nu}\, g_h^\dagger \, .
\end{equation}
These auxiliary fields provide the connection and covariant derivatives:
\begin{equation}
\Gamma_\mu^{L} = u^\dagger(\varphi) \left(\partial_\mu - i\,\hat{W}_\mu\right) u^{\phantom{\dagger}}(\varphi)
\, , \quad\;
\Gamma_\mu^{R} = u(\varphi) \left(\partial_\mu - i\,\hat{B}_\mu\right) u^{\dagger}(\varphi)
\, , \quad\;
\Gamma_\mu =
\Frac{1}{2} \left(\Gamma_\mu^{L} +\Gamma_\mu^{R}\right) ,
\nn\end{equation}
\begin{equation}
u_\mu  \, =\, i\,
\left( \Gamma_\mu^R - \Gamma_\mu^L\right)
= i\, u\, (D_\mu U)^\dagger u \, =\, -i\, u^\dagger D_\mu U\, u^\dagger
\, =\, u_\mu^\dagger
\quad\longrightarrow\quad g_h^{\phantom{\dagger}}\, u_\mu\, g_h^\dagger \, ,
\nn\end{equation}
\begin{equation}
D_\mu U \, =\, \partial_\mu U -  i\, \hat{W}_\mu  U + i\, U \hat{B}_\mu
\quad\longrightarrow\quad g_L^{\phantom{\dagger}}\, (D_\mu U) \, g_R^\dagger
\, ,
\nn\end{equation}
\begin{equation}
  \label{eq:CovDev}
\nabla_\mu \mX \, =\, \partial_\mu \mX \, +\, [\Gamma_\mu , \mX ]
\quad\longrightarrow\quad g_h^{\phantom{\dagger}} \, (\nabla_\mu \mX)\, g_h^\dagger\, ,
\end{equation}
with any tensor $\mX$ transforming like $\mX\to g_h \mX g_h^\dagger$.
The parity and charge-conjugation properties and the hermitian conjugation of all these bosonic tensors can be found in App.~A of Ref.\cite{Pich:2016lew}.

The identification \cite{Pich:2012jv}
\bel{eq:SMgauge}
\hat{W}^\mu \, =\, -g\;\frac{\vec{\sigma}}{2}\, \vec{W}^\mu \, ,
\qquad\qquad\qquad\quad
\hat{B}^\mu\, =\, -g'\;\frac{\sigma_3}{2}\, B^\mu\, ,
\ee
explicitly breaks the chiral symmetry group $\mG$ while preserving
the $SU(2)_L\otimes U(1)_Y$ gauge symmetry, as in the SM.

Combining the previous covariant tensors and taking traces, one can easily build effective operators that are invariant under $\mG$.
Taking into account that the Higgs field is a singlet under
$SU(2)_L\otimes SU(2)_R$, we can multiply these invariant operators
by arbitrary analytical functions of $h$ \cite{Grinstein:2007iv}.
Then, we can write the bosonic part of
the leading order (LO) effective Lagrangian $\mathcal{L}_{\mathrm{EWET}}^{(2)}$ as
\bel{eq:L2}
\mathcal{L}_{\mathrm{EWET}}^{(2)}\Big|^{\mathrm{Bosonic}} \, =\, \frac{1}{2}\,
\partial_\mu h\,\partial^\mu h
\, -\,\frac{1}{2}\, m_h^2\, h^2 \, -\, V(h/v)
\, +\,
\frac{v^2}{4}\,\mF_u(h/v)\;\langle u_\mu u^\mu\rangle_2 \, +\, \mL_{YM}\, ,
\ee
with $\mL_{YM}$ the Yang-Mills Lagrangian of the SM gauge bosons,
\be\label{eq:Fhu_V}
  V(h/v)\, = \, v^4\;\sum_{n=3} c^{(V)}_n \left(\frac{h}{v}\right)^n\, ,
\qquad\qquad
\mF_u(h/v)\, = \, 1\, +\, \sum_{n=1} c^{(u)}_n \left(\frac{h}{v}\right)^n\, ,
\ee
and $\langle A\rangle_k$ denoting the $SU(k)$-trace of a matrix  $A$. One recovers the SM scalar Lagrangian for
 $c^{(V)}_3 = \frac{1}{2}\, m_h^2/v^2$, $c^{(V)}_4 = \frac{1}{8}\, m_h^2/v^2$, $c^{(V)}_{n>4} = 0$,
$c^{(u)}_1 = 2$, $c^{(u)}_2 = 1$ and $c^{(u)}_{n>2} = 0$. Although the symmetry allows us to multiply the kinetic term of the Higgs by an arbitrary
function $\mF_h(h/v)$,
this function can be always reduced to $\mF_h=1$ through
an appropriate Higgs field redefinition. Two equivalent, but complementary, explicit derivations are given in
App. B of Ref.~\cite{Pich:2016lew} and
App. A of Ref.~\cite{Buchalla:2013rka}.

We can incorporate the explicit breaking of custodial symmetry by means of a right-handed spurion:
\bel{eq:T_covariant}
\mT_R\quad\longrightarrow\quad g_R^{\phantom{\dagger}}\, \mT_R\, g_R^\dagger\, ,
\qquad\qquad  \mT\, =\, u\, \mT_R\, u^\dagger \quad\longrightarrow\quad g_h^{\phantom{\dagger}} \mT g_h^\dagger\, .
\ee
Making the identification
\begin{equation}
\label{eq:T_R-value}
   \mT_R    \, =\, -g'\;\frac{\sigma_3}{2}\, ,
\end{equation}
one obtains the custodial symmetry breaking operators induced
through quantum loops with internal $B_\mu$ lines. Note that $\mT_R$ is the only custodial-symmetry-violating spurion consistent with the SM gauge symmetries~\cite{Buchalla:2014eca}.

\subsection{Fermionic fields}
When colored SM fermions are incorporated to the EWET, the symmetry group must be extended to $\mG=SU(3)_C \otimes SU(2)_L\otimes SU(2)_R\otimes U(1)_{X}$ with $X=(\mathrm{B}-\mathrm{L})/2$, being $\mathrm{B}$ and $\mathrm{L}$ the baryon and lepton quantum numbers, respectively~\cite{Hirn:2005fr}. This $U(1)_X$ component does not affect the SM bosons and the expressions in the previous subsection remain unaffected.
The left and right chiralities of the SM quarks form $SU(2)_L$ and $SU(2)_R$ doublets, respectively:
\bear
\psi_L = \left( \begin{array}{c} t_L \\ b_L \end{array}\right)
\; &\longrightarrow &\; g_C \,g_X\, g_L\;  \psi_L
\, ,
\qquad\quad\;\,
\psi_R  =\left( \begin{array}{c} t_R \\ b_R \end{array}\right)
\quad\longrightarrow\quad g_C \, g_X\, g_R\;  \psi_R  \, ,
\label{eq.psi-transformation}
\nn\\
\xi_{L} \,\equiv\,  u^\dagger \, \psi_L
\; &\longrightarrow &\; g_C \,g_X\, g_h\;  \xi_L
\, ,
\qquad\qquad\quad
\xi_{R} \,\equiv\,   u \,\psi_R
\quad\longrightarrow\quad g_C \,g_X\, g_h\;  \xi_R
\, ,
\eear
with $\psi_{L,R}=P_{L,R}\,\psi$ transforming under
$(g_C,g_L,g_R,g_X)\in SU(3)_C\otimes SU(2)_L\otimes SU(2)_R\otimes U(1)_X$,
the compensating transformation $g_h\in SU(2)_{L+R}$ introduced
in the previous subsection in the CCWZ representation of the EW Goldstones
and $P_{L,R} =\frac{1}{2}\, (1\mp \gamma_5)$. Leptons are similarly organized except that they are singlets under $SU(3)_C$. From now on, we will focus our analysis on the quark doublet.
The corresponding covariant derivatives of these fermion doublets are given by
\bear \label{e_covG}
D_\mu^L\psi_L\, &\! = &\! \,\left(\partial_\mu - i\,\hat G_\mu - i\,\hat{W}_\mu - i\,
x_{\psi}\,\hat{X}_\mu \,
\right) \psi_L
\quad\longrightarrow\quad g_C \,g_X\, g_L\;  D_\mu^L\psi_L\, ,
\nn\\
D_\mu^R\psi_R\, &\! = &\! \,\left(\partial_\mu - i\,\hat G_\mu - i\,\hat{B}_\mu - i\,
x_{\psi}\,\hat{X}_\mu
\right) \psi_R
\quad\longrightarrow\quad g_C \,g_X\, g_R\;  D_\mu^R\psi_R\, ,
\nn\\
d_\mu^{L} \xi_{L}   &\! =&\!
u^\dagger D_\mu^L\psi_L
\quad\longrightarrow\quad   g_C \, g_X \, g_h\; d_\mu^{L} \xi_{L}
\, ,
\qquad
d_\mu^{R} \xi_{R}   \! =\!
u D_\mu^R \psi_R
\quad\longrightarrow\quad   g_C \, g_X \, g_h\; d_\mu^{R} \xi_{R}
\, ,
\nn\\
d_\mu \xi  &\! =&\!  d_\mu^R\xi_R + d_\mu^L\xi_L \quad\longrightarrow\quad   g_C \, g_X \, g_h\; d_\mu \xi
\, ,
\eear
where $x_{\psi}$ is the corresponding $U(1)_{X}$ charge of the fermion $\psi$.
The fields $\hat{G}_\mu$ and $\hat{X}_\mu$ are introduced to keep the covariance under local $SU(3)_C$ and $U(1)_X$ transformations, respectively. These fields and their field-strength tensors transform under $\mG$ like
\begin{equation}
\hat G_\mu \;\; \longrightarrow\;\; g_C\, \hat G_\mu\, g_C^\dagger \, + \, i \,g_C\,\partial_\mu g_C^\dagger\,,
\quad\qquad
\hat G_{\mu\nu}\, =\,
 \partial_\mu  \hat{G}_\nu  -\partial_\nu   \hat{G}_\mu  -i\, [\hat{G}_{\mu},\hat{G}_\nu]
 \;\;\longrightarrow\;\; g_C\, \hat{G}_{\mu\nu}\, g_C^\dagger\, ,
 \nn
\end{equation}
\begin{equation}
\label{eq:Xtransform}
\hat{X}^\mu\quad\longrightarrow\quad \hat{X}^\mu
+ i\, g_X^{\phantom{\dagger}}\, \partial^\mu g_X^\dagger\, ,
\qquad\qquad
\hat{X}_{\mu\nu}\, =\,\partial_\mu\hat{X}_\nu -\partial_\nu \hat{X}_\mu  \quad\longrightarrow\quad \hat{X}_{\mu\nu}\, .
\end{equation}

The SM gauge interactions are recovered by setting $\hat{W}_\mu$, $\hat{B}_\mu$, $\hat{X}_\mu$ and $\hat{G}_ \mu$
to the values given in Eq.~(\ref{eq:SMgauge}) in the previous subsection and
\bel{eq:Xmu-fixing}
\hat G_\mu\, =\, g_s\, G^a_\mu\, T^a\,,
\qquad \qquad\qquad
\hat{X}_\mu  \, =\, \,-\, g'\, B_\mu \, .
\ee
These assignments explicitly break $\mG$ while preserving
the SM gauge symmetry $SU(3)_C\otimes SU(2)_L\otimes U(1)_Y\subset \mG$.
The QCD gluons $G_\mu^a$ ensure that the covariant derivatives transform indeed covariantly
under $SU(3)_C$ gauge transformations: $g_s$ is the strong coupling
and $T^a= \frac{1}{2}\,\lambda^a\,$ ($a=1,\ldots,\,8$) are the $SU(3)_C$ generators in the fundamental representation,
with normalization $\bra T^aT^b\ket_3=\delta^{ab}/2$. Further details on the QCD algebra are given in Appendix~\ref{app:QCD-algebra}.

The fermion masses are incorporated through the Yukawa Lagrangian that explicitly breaks the symmetry group $\mG$.
To account for this type of symmetry breaking one introduces right-handed spurion fields transforming as
\begin{equation}
\label{eq:Yspurion}
\mY_R\quad\longrightarrow\quad g_R^{\phantom{\dagger}}\,\mY_R\, g_R^\dagger\, ,
\qquad\qquad\qquad
\mY\, =\, u\,\mY_R\, u^\dagger
\quad\longrightarrow\quad g_h^{\phantom{\dagger}}\,\mY\, g_h^\dagger
\, .
\end{equation}
The Yukawa Lagrangian then takes the form
\begin{equation}
\label{eq:Yukawas}
\mathcal{L}_{\mathrm{EWET}}^{(2)}\Big|^{\mathrm{Yukawa}}\, =\, -\sum_{\xi}\left(v\; \bar\xi_L\, \mY\, \xi_R \, +\, \mathrm{h.c.}\right)
\, =\, -\sum_{\psi}\left(v\;\bar\psi_L\, U(\varphi)\,\mY_R\,\psi_R\, +\, \mathrm{h.c.} \,\right) ,
\end{equation}
which is formally invariant under $\mG$ transformations. The explicit symmetry breaking
in the SM Lagrangian
is recovered when the spurion field adopts the value \cite{Buchalla:2013rka,Appelquist:1984rr,Bagan:1998vu}
\bear
\label{eq:SM_Yspurion_h}
\mY_{R}\, =\, \hat{Y}_t(h/v)\,\mP_+ + \hat Y_b(h/v)\,\mP_-\, ,
\,\,\,\,
\mP_\pm \,\equiv\,\frac{1}{2}\,\left( I_2\pm\sigma_3\right)\, ,
\,\,\,\,
\hat{Y}_{t,b}(h/v)\, =\, \sum_{n=0}\, \hat{Y}_{t,b}^{(n)}\, \left(\frac{h}{v}\right)^n\, ,
\eear
where $\hat{Y}_{\psi}^{(0)}=\hat{Y}_{\psi}^{(1)}=m_\psi/v$ and
$\hat{Y}_{\psi}^{(n\geq 2)}=0$ in the SM.

\subsection{Leading-order Lagrangian}

After having introduced all the required notation for the bosonic and fermionic fields in the previous sections, we summarize the LO low-energy Lagrangian as~\cite{Pich:2016lew,Pich:2018ltt,Feruglio:1992wf,Bagger:1993zf,Koulovassilopoulos:1993pw,Burgess:1999ha,Wang:2006im,Grinstein:2007iv,Alonso:2012px,Buchalla:2012qq,Buchalla:2013rka,Buchalla:2013eza}
\begin{align}
  \begin{aligned}
    \label{eq:L.LO}
   \mathcal{L}_{\mathrm{EWET}}^{(2)} &=\, \sum_{\xi} \left[ i\,\bar\xi \gamma^\mu d_\mu \xi -v \left( \; \bar\xi_L\, \mY\, \xi_R \, +\, \mathrm{h.c.} \right) \right] \\
     & -\frac{1}{2g^{2}}\langle \hat{W}_{\mu\nu}\hat{W}^{\mu\nu}\rangle_{2}-\frac{1}{2g^{'2}}\langle \hat{B}_{\mu\nu}\hat{B}^{\mu\nu}\rangle_{2}-\frac{1}{2g_{s}^{2}}\langle \hat{G}_{\mu\nu}\hat{G}^{\mu\nu}\rangle_{3} \\
    & +\frac{1}{2}\partial_\mu h\,\partial^\mu h \, -\,\frac{1}{2}\, m_h^2\, h^2 \, -\, V(h/v) \, +\, \frac{v^2}{4}\,\mF_u(h/v)\;\langle u_\mu u^\mu\rangle_{2}\, .
  \end{aligned}
\end{align}

\subsection{Next-to-leading-order Lagrangian}

In order to construct the next-to-leading-order (NLO) EWET Lagrangian, $\mathcal{L}_{\mathrm{EWET}}^{(4)}$, the fermionic fields are combined into generic bilinears $J_\Gamma$ and $J_\Gamma^8 $ with well-defined Lorentz transformation properties, which can be further used to build Lagrangian operators with an even number of fermion fields. Note that the inclusion of gluons and colored quarks implies changes in the notation, compared to Ref.~\cite{Pich:2016lew}, since different transformation properties under $SU(3)_C$ are required:
\bear
{\rm Color\ singlet:} &\qquad\qquad & J_\Gamma \,\equiv\, \bar \xi_i\, \Gamma \, \xi_j\; \delta_{ij} \, ,
 \nn\\
 {\rm Color\ octet:} &\qquad\qquad &  J_\Gamma^8 \,\equiv\, (J_{\Gamma}^{8,a})\, T^a  \,\equiv\, (\bar \xi_i\, \Gamma \, \xi_j\, T^{a}_{ij})\; T^a  \, .  \label{bilinears1}
\eear
These bilinears are singlets under $U(1)_X$ and transform under $SU(2)_L \otimes SU(2)_R$ like
\begin{equation}
(J_\Gamma)_{mn}  \,=\,  \bar{\xi}^n_i \Gamma\xi^m_j\; \delta_{ij}
\quad\longrightarrow\quad (g_{h})_{mr} \, (J_\Gamma)_{rs} \, (g_{h}^\dagger)_{sn} \, , \label{bilinears2}
\end{equation}
where $\Gamma = \left\{ I, i\gamma_5, \gamma^\mu, \gamma^\mu\gamma_5, \sigma^{\mu\nu}\right\}$
is the usual basis of Dirac matrices. An analogous expression
is assumed for $J_\Gamma^8$, being these indices left implicit
in the $SU(2)$ trace expressions. In Eqs.~(\ref{bilinears1}) and~(\ref{bilinears2}),
$a=1,\dots,8$ and $i,j=1,2,3$ are $SU(3)$ indices, while $m,n,r,s=1,2$ refer to $SU(2)$ indices.

Obviously, the inclusion of color in the EWET raises the number of operators, compared to Ref.~\cite{Pich:2016lew}.
At $\mO(p^4)$, one needs to consider one additional bosonic operator,
\begin{equation}
\label{eq:mO12}
\mO_{12}\, =\, \bra \hat G_{\mu\nu}\,\hat G^{\mu\nu} \ket_3 \, ,
\end{equation}
and one additional two-fermion operator,
\be
\mO^{\psi^2}_8\, =\, \bra \hat G_{\mu\nu}\, J_{T}^{8\, \mu\nu} \ket_{2,3} \, .
\ee
There are no more new operators of these types due to the fact that color indices must be closed. The set of $\mO(p^4)$ four-fermion operators, however, grows from 6 independent operators to 12 when including color. We applied different relations among the operators to minimize the number of structures. Particularly, using $SU(3)$ relations and Fierz identities (see Apps.~\ref{app:QCD-algebra} and~\ref{app:fierz}, respectively),
we were able to relate operators involving the color octet current to operators involving the color singlet current.

\begin{table}[t!]
	\begin{center}
		\renewcommand{\arraystretch}{1.8}
		\begin{tabular}{|c||c|c|c|}
			\hline
			$i$ & ${\cal O}_i$ &  ${\cal O}^{\psi^2}_i$ & ${\cal O}^{\psi^4}_i$\\

                        \hline
			\hline
			$1$  & 	$\Frac{1}{4}\,\bra {f}_+^{\mu\nu} {f}_{+\, \mu\nu}- {f}_-^{\mu\nu} {f}_{-\, \mu\nu}\ket_2$ & $\bra J_S \ket_2 \bra u_\mu u^\mu \ket_2$  &$\bra J_{S} J_{S} \ket_2 $
			\\ [1ex]
			\hline
			$2$  & 	$ \Frac{1}{2} \,\bra {f}_+^{\mu\nu} {f}_{+\, \mu\nu} + {f}_-^{\mu\nu} {f}_{-\, \mu\nu}\ket_2$& $ i \,  \bra J_T^{\mu\nu} \left[ u_\mu, u_\nu \right] \ket_2$  &$\bra J_{P} J_{P} \ket_2 $
			\\ [1ex]
			\hline
			$3$  &	$\Frac{i}{2} \,\bra {f}_+^{\mu\nu} [u_\mu, u_\nu] \ket_2$ &  $\bra J_T^{\mu \nu} f_{+ \,\mu\nu} \ket_2 $ & $\bra J_{S} \ket_2 \bra  J_{S} \ket_2 $
			\\ [1ex]
			\hline
			$4$  & 	$\bra u_\mu u_\nu\ket_2 \, \bra u^\mu u^\nu\ket_2 $ &$\hat{X}_{\mu\nu} \bra J_T^{\mu \nu} \ket_2 $ & $\bra J_{P} \ket_2 \bra  J_{P} \ket_2 $
			\\ [1ex]
			\hline
			$5$  & $  \bra u_\mu u^\mu\ket_2 \, \bra u_\nu u^\nu\ket_2$ & $\displaystyle\frac{\partial_\mu h}{v} \, \bra u^\mu J_P \ket_2 $ & $\bra J_V^\mu J_{V,\mu}^{\phantom{\mu}}\ket_2 $
			\\ [1ex]
			\hline
			$6$ & $\Frac{(\partial_\mu h)(\partial^\mu h)}{v^2}\,\bra u_\nu u^\nu \ket_2$ &$\bra J_A^\mu \ket_2 \bra u_\mu \mathcal{T} \ket_2 $ & $\bra J_A^\mu J_{A,\mu}^{\phantom{\mu}}\ket_2 $
			\\ [1ex]
			\hline
			$7$  & 	$\Frac{(\partial_\mu h)(\partial_\nu h)}{v^2} \,\bra u^\mu u^\nu \ket_2$ &$\Frac{(\partial_\mu h) (\partial^\mu h)}{v^2} \bra J_S\ket_2 $& $\bra J_V^\mu\ket_2 \bra J_{V,\mu}^{\phantom{\mu}}\ket_2 $
			\\ [1ex]
			\hline
			$8$ &  $\Frac{(\partial_\mu h)(\partial^\mu h)(\partial_\nu h)(\partial^\nu h)}{v^4}$ &$\bra \hat G_{\mu\nu} J_{T}^{8\,\mu\nu} \ket_{2,3} $  & $\bra J_A^\mu\ket_2 \bra J_{A,\mu}^{\phantom{\mu}}\ket_2 $
			\\ [1ex]
			\hline
			$9$ & $\Frac{(\partial_\mu h)}{v}\,\bra f_-^{\mu\nu}u_\nu \ket_2$ & --- &$\bra J^{\mu\nu}_{T} J_{T\,\mu\nu}^{\phantom{\mu}} \ket_2 $
			\\ [1ex]
			\hline
			$10$ & $\bra \mT u_\mu\ket_2\, \bra \mT u^\mu\ket_2$  & --- &$\bra J^{\mu\nu}_{T} \ket_2 \bra J_{T\,\mu\nu}^{\phantom{\mu}} \ket_2 $
			\\ [1ex]
			\hline
			$11$ & $ \hat{X}_{\mu\nu} \hat{X}^{\mu\nu}$ & --- & ---
			\\ [1ex]
			\hline
			$12$ & $\bra \hat G_{\mu\nu}\,\hat G^{\mu\nu} \ket_3 $ & --- & ---
			\\ [1ex]
			\hline
		\end{tabular}
	\end{center}
	\caption{\small
		$CP$-invariant and $P$-even operators of the $\cO(p^4)$ EWET Lagrangian. The left column shows bosonic, the central two-fermion, and the right column four-fermion operators. All Wilson coefficients are Higgs-dependent functions. }
	\label{tab:P-even-Op4}
\end{table}
%

  \begin{table}[!t] 
   \begin{center}
     \renewcommand{\arraystretch}{1.8}
     \begin{tabular}{|c||c|c|c|}
       \hline
       $i$ & $\widetilde{\cal O}_i$& $\widetilde{\cal O}^{\psi^2}_i$ & $\widetilde{\cal O}^{\psi^4}_i$  \\ \hline\hline
       1   & $\Frac{i}{2} \,\bra {f}_-^{\mu\nu} [u_\mu, u_\nu] \ket_2$&  $\bra J_T^{\mu \nu} f_{- \,\mu\nu} \ket_2 $ &   $\bra J_V^\mu J_{A,\mu}^{\phantom{\mu}}\ket_2 $\\  [1ex] \hline
       2  & $\bra {f}_+^{\mu\nu} {f}_{-\, \mu\nu} \ket_2 $&   $\displaystyle\frac{\partial_\mu h}{v} \, \bra u_\nu J^{\mu\nu}_T \ket_2 $ &  $\bra J_V^\mu\ket_2 \bra J_{A,\mu}^{\phantom{\mu}}\ket_2 $\\  [1ex] \hline
       3 &  $\Frac{(\partial_\mu h)}{v}\,\bra f_+^{\mu\nu}u_\nu \ket_2$  & $\bra J_V^\mu \ket_2 \bra u_\mu \mathcal{T} \ket_2 $  & ---	\\  [1ex] \hline	
     \end{tabular}
     \caption{\small
       $CP$-invariant and $P$-odd operators of the $\cO(p^4)$ EWET Lagrangian. The left column shows bosonic, the central two-fermion, and the right column four-fermion operators. All Wilson coefficients are Higgs-dependent functions. }
     \label{tab:P-odd-Op4}
   \end{center}
 \end{table}

The final list of independent $P$-even and $P$-odd operators is given in Tables~\ref{tab:P-even-Op4} and \ref{tab:P-odd-Op4}, respectively, and can be summarized in the following NLO Lagrangian:
\begin{eqnarray}
  \label{eq:L-NLO}
\mathcal{L}_{\mathrm{EWET}}^{(4)} & =&
\sum_{i=1}^{12} \mF_i(h/v)\; \mO_i \, +\, \sum_{i=1}^{3}\widetilde\mF_i(h/v)\; \widetilde \mO_i  \nonumber
\, +\,  \sum_{i=1}^{  8  } \mF_i^{\psi^2}(h/v)\; \mO_i^{\psi^2} \, +\, \sum_{i=1}^{  3   } \widetilde\mF_i^{\psi^2}(h/v)\; \widetilde \mO_i^{\psi^2}
\nonumber \\ &&
\, +\, \sum_{i=1}^{10}\mF_i^{\psi^4}(h/v)\; \mO_i^{\psi^4} \, +\, \sum_{i=1}^{2}\widetilde\mF_i^{\psi^4}(h/v)\; \widetilde \mO_i^{\psi^4} \, .
\end{eqnarray}

In order to crosscheck the completeness of the considered operator basis, we have compared our results with the NLO basis of Ref.~\cite{Buchalla:2013rka}. We give the detailed comparison in terms of a dictionary in the Appendix \ref{sec:dic}, together with other useful operator relations in Appendix \ref{app:relations}. We found agreement in the overall number of operators in both bases.

\section{The Resonance Effective Theory}
\label{sec:res-EFT}

\subsection{The power counting}
\label{sec:res-pc}
When writing a Lagrangian of resonances interacting with the light fields, we leave the realms of the low-energy EFT, as the heavy masses are at or above the cutoff. The power counting we discussed in Section~\ref{constructing} is therefore not directly applicable. However, we can still treat this Lagrangian in a consistent phenomenological way, which properly interpolates between the UV and IR regimes, generating the correct low-energy predictions~\cite{Ecker:1988te,Ecker:1989yg}. For that purpose, let us organize the Lagrangian schematically by the number of resonance fields
in each operator \cite{Pich:2013fea}:
\begin{equation}
  \label{eq:res-pc.1}
  \mathcal{L}\, =\, \mathcal{L}_{\text{non-res}} + \sum_{R,i}c_{i}\,
  \chi_{R}^{(i)} \,
  R + \sum_{R,R',i}d_{i}\, \chi_{R,R'}^{(i)}\,
  R \, R'+ \cdots \, ,
\end{equation}
where $\mathcal{L}_{\text{non-res}}$ and the chiral structures $\chi^{(i)}_R,\, \chi^{(i)}_{R,R'},\dots$ only contain light degrees of freedom (dof).
At energies below the resonance masses, we can integrate out the resonance fields, recovering in this way the low-energy EWET.

At tree-level and for bosonic resonances, the Lagrangian~\eqn{eq:res-pc.1} gives low-energy equation-of-motion (EoM) solutions of the form
\begin{equation}
  \label{eq:res-pc.2}
  R\,\sim\, \sum_{i}\frac{c_{i}\,\chi_{R}^{(i)}}{M_{R}^{2}}
  + \mathcal{O}(v^{5}/M_{R}^{4})\, ,
\end{equation}
where $M_R$ denotes the resonance mass.
Unless $c_{i}\sim M_{R}^{2}/ v^{2}$, which is a special case that we discuss in Appendix~\ref{app:coupling}, the contributions from operators with higher number of resonances are further suppressed by additional powers of $v^{2}/M_{R}^{2}$. At NLO, $\mathcal{O}(v^{2}/M_{R}^{2})$, it is therefore sufficient to consider Eq.~\eqref{eq:res-pc.1} up to operators with a single resonance.

Inserting the expression~\eqref{eq:res-pc.2} back in Eq.~\eqref{eq:res-pc.1} and keeping terms up to $\mathcal{O}(v^{2}/M_{R}^{2})$, the resonance Lagrangian transforms into a sum of chiral-invariant structures of the form $(\sum_{i} c_{i}  \,\chi_{R}^{(i)})^2/M_R^2 $. Matching this result with the low-energy EWET Lagrangian discussed in Section \ref{sec:lE-EFT}, we can identify  the different $c_i c_j /M_{R}^{2}$ terms with the corresponding pre-factors $\mF_j$ of the NLO EWET operators, with their given normalization factors properly taken into account.
In order to pin down all possible contributions to the
$\mathcal{O}(p^{4})$ LECs, we only need to consider tensors $\chi_{R}^{(i)}$ in Eq.~\eqref{eq:res-pc.1} of $\mO(p^{2})$ or lower. We discuss the detailed form of Eq.~\eqref{eq:res-pc.2} below.

For fermion resonances $\Psi$, the low-energy EoM solution in Eq.~\eqref{eq:res-pc.2} gets modified into $\Psi\sim \sum_i c_i\, \chi_\Psi^{(i)}/M_{\Psi}$ which, when inserted back in~\eqref{eq:res-pc.1},
generates EWET structures that only contain a single heavy-mass factor: $(\sum_i c_i\, \chi_\Psi^{(i)})^2/M_{\Psi}$. However, the resulting fermionic contributions to the LECs, $c_i c_j/M_{\Psi}$,  are also of $\cO(p^4)$ because, according to our power-counting rules, the EWET fermionic operators always come in combination with weak coupling factors in $c_i c_j$ that provide an additional suppression.

\subsection{Colored bosonic resonances}

Considering also colored fields in our analysis implies new tree-level interactions with resonances and thus additional contributions to the LECs of the EWET. Furthermore, we consider colorful heavy objects that are in an octet representation of the $SU(3)_C$ group, while all the heavy objects considered in Ref.~\cite{Pich:2016lew} correspond to the singlet representation. In consequence, several types of heavy resonance states can be distinguished and they read as
\begin{align}
R_1^1\quad &\! \longrightarrow\quad R_{1}^1 \, , &\!
\qquad   &\quad\;\, \partial_\mu R^1_1 \, , 
\nn\\
R^1_3 \quad &\! \longrightarrow\quad g_h^{\phantom{\dagger}}\, R^1_3\, g_h^\dagger\, ,&\!
\qquad \nabla_\mu R^1_3 & = \partial_{\mu} R^1_3 + [\Gamma_\mu, R^1_3]\,, &\!
\nn\\
R^8_{1}\quad  &\! \longrightarrow\quad g_C\,R^8_{1}\,g_C^{\dagger} \, , &\!
\qquad \hat\partial_\mu R^8_1 &  = \partial_{\mu} R^8_1 + i\, [\hat G_\mu, R^8_1]\,, &\!
\nn\\
R^8_3 \quad  &\! \longrightarrow\quad g_C\,g_h^{\phantom{\dagger}}\, R^8_3\, g_h^\dagger\,g_C^{\dagger}\, ,  &\!
\qquad \hat \nabla_\mu R^8_3 & = \partial_{\mu} R^8_3 +  i\, [\hat G_\mu, R^8_3] + [\Gamma_\mu, R^8_3]\,, &\!
\label{eq.R-transform}
\end{align}
where the dimension of the resonance representation is indicated, here and in the following, with upper and lower indices in the scheme $R_{SU(2)}^{SU(3)}$. As usual, $R$ stands for any of the four possible $J^{PC}$ bosonic states with quantum numbers $0^{++}$ (S), $0^{-+}$ (P), $1^{--}$ (V) and $1^{++}$ (A). For instance, $P^8_3$ refers to a pseudoscalar heavy multiplet that is an $SU(2)$ triplet and an $SU(3)$ octet. Equation \eqref{eq.R-transform} also incorporates the covariant derivatives for the colored resonances.
The normalization used for the n-plets of resonances is
\begin{align}
R^{n}_3 &\!= \frac{1}{\sqrt{2}}\,\sum_{i=1}^3\,\sigma_{i}\,R^n_{3,i}\,, &\! \,\, &\! \mbox{with}  &\! \quad  \bra R^{n}_3\,R^{m}_3 \ket_2 &\!=
 \sum_{i=1}^3   R^{n}_{3,i}\,R^{m}_{3,i}\,, &  
\quad n,m=1,8\,;
\nn\\
R_n^8 &\!=  \sum_{a=1}^8\,T^a\, R_n^{8,a}\,, &\! \,\, &\! \mbox{with} &\! \quad  \bra R^8_{n}\,R^8_{m} \ket_3 &\!=
 \sum_{a=1}^8 \frac{1}{2} \,R^{8,a}_{n}\,R^{8,a}_{m}\,, & 
\quad n,m=1,3\, ,
\end{align}
with $\bra \sigma_i\sigma_j\ket_2 = 2\delta_{ij}$ and $\bra T^a T^b\ket_3 = \delta^{ab}/2$.

In order to find the imprints of these heavy states in the LECs of the EWET at NLO, it is enough to analyze the tree-level exchange of resonances. Including only interactions linear in the resonance fields, the spin-$0$ Lagrangians read
\begin{align}
\mL_{R^1_1} \, & = \, \Frac{1}{2}  \left(  \partial^\mu R_1^1\,  \partial_\mu R^1_1 \, -\, M_{R^1_1}^2\, (R_1^1)^2 \right)  \; +\;
R^1_1\, \chi_{R^1_1}^{\phantom{\mu}} \,,  
\nn\\
\mL_{R^1_3}\, & =\, \Frac{1}{2}\bra \nabla^\mu R^1_3\,  \nabla_\mu R^1_3 \, -\, M_{R^1_3}^2\, (R^1_3)^2\ket_2 \; +\; \bra R^1_3\, \chi_{R^1_3}\ket_2 \, , 
\nn\\
\mL_{R^8_1}\, & =\,  \bra  \hat\partial^\mu R^8_1\,  \hat\partial_\mu R^8_1 \, -\, M_{R_1^8}^{2}\, (R_1^8)^2 \ket_3  \; +\;
\bra R_1^8\, \chi_{R_1^8} \ket_3\,, 
\nn\\
\mL_{R^8_3}\, & =\, \bra \hat\nabla^\mu R^8_3\,  \hat\nabla_\mu R^8_3 \, -\, M_{R^8_3}^2\, (R^8_3)^2 \ket_{2,3} \; +\; \bra R^8_3\, \chi_{R^8_3} \ket_{2,3} \,, 
\label{eq:L0}
\end{align}
with $R^m_n = S^m_n, P^m_n$.
We have included all tree-level interactions in the different $\chi_R$ tensors. The $O(p^2)$ interacting structures for the scalar and pseudoscalar resonances are:\footnote{Note that we have slightly changed the notation with respect to Ref.~\cite{Pich:2016lew}.}
\begin{align}
\chi_{S_1^1} \, &\! = \,
\lambda_{hS_1} \, v \, h^2  \, +\,
\Frac{c_{d}}{\sqrt{2}}\, \bra u_\mu u^\mu \ket_2 \, +\,
\Frac{c^{S_1^1}}{\sqrt{2}}\, \bra J_S \ket_2  \, ,&\!
\qquad\quad
\chi_{P_1^1}\, &\! =\,
\Frac{c^{P_1^1}}{\sqrt{2}}\,  \bra  J_P\ket_2 \, , &\!
\nn\\
\chi_{S^1_3} \,&\! =\, c^{S^1_3}\,  J_S \, ,&\!
\qquad\quad
\chi_{P^1_3}\, &\! =\,
c^{P^1_3}\, J_P \, +\,
d_P\, \Frac{(\partial_\mu h)}{v}\, u^\mu \, , &\!
\nn\\
\chi_{S_1^{8}}\, &\! =\,  \Frac{c^{S_1^8}}{\sqrt{2}}\,\bra J^8_{S} \ket_2 \,,    &\!
\qquad\quad
\chi_{P_1^{8}}\, &\! =\,  \Frac{c^{P_1^8}}{\sqrt{2}}\,\bra J^8_{P} \ket_2 \,,    &\!
\nn\\
\chi_{S^8_3} \,&\! =\,  c^{S^8_3}\,J^8_{S}   \, , &\!
\qquad\quad
\chi_{P^8_3}\, &\! =\,   c^{P^8_3}\,J^8_{P}  \,.&\!
\label{eq:L0tensor}
\end{align}

For the massive spin-1 fields there is some freedom in the selection of resonance formalism: we can use either the four-vector Proca description $\hat{R}^\mu$ or the rank-2 antisymmetric tensor $R^{\mu \nu}$ \cite{Ecker:1988te,Ecker:1989yg}. By using a change of variables within the path integral formulation, and once a good short-distance behavior is required, we demonstrated the equivalence of both formalisms in Ref.~\cite{Pich:2016lew}: they lead to the same predictions for the LECs of the EWET at NLO. Although both descriptions are fully equivalent, they involve different Lagrangians. Depending on the particular phenomenological application, one formalism can be more efficient predicting the LECs than the other, in the sense that direct contributions from $\mathcal{L}_{\text{non-res}}$ (local operators without resonances) in Eq.~\eqref{eq:res-pc.1} are absent.

The antisymmetric tensor involves interactions of the type $R_{\mu\nu} \chi_R^{\mu\nu}$ and $R_{\mu\nu}(\nabla^\mu \chi_R^\nu -\nabla^\nu\chi_R^\mu)$ while, owing to its different Lorentz structure, the couplings of the Proca field can be written in two different ways:
$\hat R_\mu \chi^\mu_{\hat R}$  and $(\nabla_\mu \hat R_\nu - \nabla_\nu \hat R_\mu) \,\chi^{\mu\nu}_{\hat R}$. There is a one-to-one correspondence among vertices in both formalisms, although with different chiral dimensions. The different momentum dependence of the two spin-1 representations is compensated by different (but related) local Lagrangians $\mathcal{L}_{\text{non-res}}$ that adjust a proper UV behavior.
In Ref.~\cite{Pich:2016lew} we demonstrated the following important result that simplifies the calculation of the LECs:

{\it The sum of tree-level resonance-exchange contributions from $\cO(p^2)$ $\chi^\mu_{\hat{R}}$ (Proca) and $\chi^{\mu \nu}_R$ (antisymmetric) structures gives the complete (non-redundant and correct) set of predictions for the $\cO(p^4)$ EWET LECs, without any additional contributions from local $\mathcal{L}_{\text{non-res}}$ operators.}

In the following we just take advantage of this useful theorem, referring to Ref.~\cite{Pich:2016lew} for its detailed proof.
The relevant spin-$1$ Proca resonance Lagrangians are:
\begin{align}
	\mL_{\hat R_1^1} \, &\! =  -\Frac{1}{4}  \left(  \hat R^1_{1\,\mu\nu}\,  \hat R_1^{1\,\mu\nu}  - 2 M_{R_1^1}^2\,\hat R^1_{1\,\mu}\, \hat R_1^{1\,\mu} \right)   +
	\hat R^1_{1\,\mu}\, \chi_{\hat R_1^1}^{\mu} \,, 
	\nn\\
	\mL_{\hat R^1_3}\, &\! = -\Frac{1}{4}  \bra  \hat R^1_{3\,\mu\nu}\,  \hat R^{1\,\mu\nu}_3  - 2 M_{R^1_3}^2\,\hat R^1_{3\,\mu}\, \hat R^{1\,\mu}_3 \ket_2   +
	\bra \hat R_{3\,\mu}^1 \, \chi_{\hat R^1_3}^{\mu}\ket_2 \,, 
	\nn\\
	\mL_{\hat R^8_1}\, &\! =  -\Frac{1}{2}  \bra  \hat R_{1\,\mu\nu}^8\,  \hat R_1^{8\,\mu\nu}  - 2 M_{R_1^8}^2\,\hat R_{1\,\mu}^8\, \hat R_1^{8\,\mu} \ket_3   +
	\bra \hat R_{1\,\mu}^8\, \chi_{\hat R_1^8}^{\mu} \ket_3 \,, 
	\nn\\
	\mL_{\hat R^8_3}\, &\! = -\Frac{1}{2}  \bra  \hat R_{3\,\mu\nu}^8\,  \hat R^{8\,\mu\nu}_3  - 2 M_{R^8_3}^2\,\hat R_{3\,\mu}^8\, \hat R^{8\,\mu}_3 \ket_{2,3}   +
	\bra \hat R_{3\,\mu}^8\, \chi_{\hat R^8_3}^{\mu} \ket_{2,3} \,, 
	\label{eq:L1proca}
\end{align}
where $\hat R^m_n = \hat V^m_n, \hat A^m_n$ and $\hat R_{\mu\nu}$ stands for the corresponding resonance strength tensor ({\it e.g.}, $\hat R^8_{3\,\mu\nu}=\hat \nabla_\mu \hat R^8_{3\,\nu} - \hat \nabla_\nu \hat R^8_{3\,\mu}$). The $\chi_{\hat R}^\mu$ chiral structures take the form:
\begin{align}
\lefteqn{\hskip -.7cm
\chi_{\hat V^1_1}^{\mu} \, = \,   \widetilde{c}_{\mathcal{T}}\,  \bra u^\mu \mathcal{T} \ket_2 \,+\,
\Frac{c^{{\hat{V}}^1_1}}{\sqrt{2}} \, \bra J^\mu_V\ket_2
\, +\, \Frac{   \widetilde c^{{\hat{V}}^1_1}  }{\sqrt{2}}
\, \bra J^\mu_A\ket_2 \, ,}&&&&
\nn\\[5pt]
\lefteqn{\hskip -.7cm
\chi_{\hat A_1^1}^{\mu}\, =\,  c_{\mathcal{T}} \, \bra u^\mu \mathcal{T} \ket_2 \,+\,
\Frac{  c^{{\hat{A}}^1_1}  }{\sqrt{2}} \, \bra J^\mu_A\ket_2
\, +\,
\Frac{  \widetilde c^{{\hat{A}}^1_1}  }{\sqrt{2}}
\, \bra J^\mu_V\ket_2 \, ,}&&&&
\nn\\[5pt]
\chi_{\hat V^1_3}^{\mu} \,& =\,  c^{\hat{V}^1_3}\, J^\mu_V  \,+\, \widetilde c^{\hat{V}^1_3}\, J^\mu_A       \, ,&
\qquad
\chi_{\hat A^1_3}^{\mu}\, & =\, c^{\hat{A}^1_3}\, J^\mu_A  \,+\, \widetilde c^{\hat{A}^1_3}\, J^\mu_V       \, , &
\nn\\[5pt]
\chi_{\hat V_1^{8}}^\mu\, & =\,  \Frac{c^{{\hat{V}}_1^8}}{\sqrt{2}}
\, \bra J^{8\,\mu}_{V}\ket_2  \,+\,
\Frac{   \widetilde c^{{\hat{V}}_1^8}  }{\sqrt{2}}
\, \bra J^{8\,\mu}_{A}\ket_2       \,,&
\qquad
\chi_{\hat A_1^{8}}^\mu\, & =\, \Frac{  c^{{\hat{A}}_1^8}  }{\sqrt{2}}
\, \bra J^{8\,\mu}_{A}\ket_2  \,+\,
\Frac{  \widetilde c^{{\hat{A}}_1^8}  }{\sqrt{2}}
\, \bra J^{8\,\mu}_{V} \ket_2         \,,&
\nn\\[5pt]
\chi_{\hat V^{8}_3}^\mu \,& =\,  c^{\hat{V}^8_3}\, J^{8\,\mu}_{V}  \,+\, \widetilde c^{\hat{V}^8_3}\, J^{8\,\mu}_{A}       \,, &
\qquad
\chi_{\hat A^{8}_3}^\mu\, & =\,  c^{\hat{A}^8_3}\, J^{8\,\mu}_{A}  \,+\, \widetilde c^{\hat{A}^8_3}\, J^{8\,\mu}_{V}        \,.&
\label{eq:L1procatensor}
\end{align}
The antisymmetric Lagrangians read:
\begin{align}
\mL_{R_1^1} \, &\! =  -\Frac{1}{2}  \left(  \partial^\lambda R^1_{1\,\lambda\mu}\,   \partial_\sigma R_1^{1\,\sigma\mu}  - \frac{1}{2} M_{R_1^1}^2\, R_{1\,\mu\nu}^{1}\,  R_1^{1\,\mu\nu} \right)\,   +R^1_{1\,\mu\nu}\,\chi_{ R_1^1}^{\mu\nu}\,,
\nn\\[5pt]
\mL_{ R^1_3}\, &\! = -\Frac{1}{2} \, \bra   \nabla^\lambda R^1_{3\,\lambda\mu}\,  \nabla_\sigma R^{1\,\sigma\mu}_3  - \frac{1}{2} M_{R^1_3}^2\, R^1_{3\,\mu\nu}\,  R^{1\,\mu\nu}_3 \ket_2  \, +  \bra R^1_{3\,\mu\nu}\,\chi_{ R^1_3}^{\mu\nu}\ket_2 \,,
\nn\\[5pt]
\mL_{ R^8_1}\, &\! =  -\,  \bra \hat\partial^\lambda R_{1\,\lambda\mu}^8\,   \hat\partial_\sigma R_1^{8\,\sigma\mu}  - \frac{1}{2} M_{R_1^8}^2\, R_{1\,\mu\nu}^8\,  R_1^{8\,\mu\nu} \ket_3  \, + \bra R_{1\,\mu\nu}^8\,\chi_{ R_1^8}^{\mu\nu} \ket_3 \,,
\nn\\[5pt]
\mL_{ R^8_3}\, &\! = - \, \bra  \hat\nabla^\lambda R_{3\,\lambda\mu}^8\, \hat\nabla_\sigma  R^{8\,\sigma\mu}_3 - \frac{1}{2} M_{R^8_3}^2\, R_{3\,\mu\nu}^8\,  R^{8\,\mu\nu}_3 \ket_{2,3}   \, + \bra R_{3\,\mu\nu}^8\,\chi_{ R^8_3}^{\mu\nu} \ket_{2,3} \, ,
\label{eq:L1anti}
\end{align}
with $R^m_n = V^m_n, A^m_n$ and the tensor structures:
\begin{align}
\chi_{V^1_1}^{\mu\nu} & =\,
 F_{X}\, \hat X^{\mu\nu}
\, +\, \Frac{C^{V^1_1}_{0}}{\sqrt{2}}\,  \bra J_T^{\mu\nu} \ket_2
\, , &\!
\qquad
 \chi_{A^1_1}^{\mu\nu}  & =\,
 \widetilde{F}_{X}\, \hat X^{\mu\nu}
\, +\, \Frac{\widetilde{C}^{A^1_1}_{0}}{\sqrt{2}}\,  \bra J_T^{\mu\nu} \ket_2
\, , &\!
\nn\\[5pt]
\lefteqn{\hskip -.7cm
\chi_{V^1_3}^{\mu\nu}  \, =\,
\Frac{F_V}{2\sqrt{2}}\,  f_+^{\mu\nu}\, +\,
\Frac{i\, G_V}{2\sqrt{2}}\, [u^\mu, u^\nu]
\, +\, \Frac{\widetilde{F}_V }{2\sqrt{2}}\, f_-^{\mu\nu} \, + \,
\Frac{ \widetilde{\lambda}_1^{hV} }{\sqrt{2}}\left[
 (\partial^\mu h)\, u^\nu-(\partial^\nu h)\, u^\mu \right]
 \, +\, C_{0}^{V^1_3}\, J_T^{\mu\nu} \, ,}&&&&
\nn\\[5pt]
\lefteqn{\hskip -.7cm
\chi_{A^1_3}^{\mu\nu}  \, =\,
\Frac{F_A}{2\sqrt{2}}\,  f_-^{\mu\nu} \, +\,
\Frac{ \lambda_1^{hA} }{\sqrt{2}} \left[
(\partial^\mu h)\, u^\nu-(\partial^\nu h)\, u^\mu \right]
\,+ \,
\Frac{\widetilde{F}_A}{2\sqrt{2}}\, f_+^{\mu\nu}\, +\,
  \Frac{i\, \widetilde{G}_A}{2\sqrt{2}}\, [u^{\mu}, u^{\nu}]
  \, +\,  \widetilde{C}_{0}^{A^1_3}\,  J_{T}^{\mu\nu}\, ,} &&&&
\nn\\[5pt]
\chi^{\mu\nu}_{V^8_1} & =\,  \Frac{C_0^{V_1^8}}{\sqrt{2}}\, \bra J^{8\,\mu\nu}_{T} \ket_2  \,+\, C_G\,\hat G^{\mu\nu}\,, &\!
\qquad
 \chi^{\mu\nu}_{A^8_1} & =\,  \Frac{\widetilde C_0^{A_1^8}}{\sqrt{2}}\, \bra J^{8\,\mu\nu}_{T} \ket_2   \,+\, \widetilde C_G\,\hat G^{\mu\nu}\,, &\!
\nn\\[5pt]
\chi^{\mu\nu}_{V^8_3} & =\,  C_0^{V^8_3}\, J^{8\, \mu\nu}_{T}\,,   &\!
\qquad
  \chi^{\mu\nu}_{A^8_3} & =\,   \widetilde C_0^{A^8_3}\, J^{8\,\mu\nu}_{T} \,.    &\!
\label{eq:L1antitensor}
\end{align}


\subsection{Fermionic doublet resonances}
\label{sec:ferm.res}

We are going to construct the interactions between the SM fields and one fermionic resonance~$\Psi$. The resonance has to be an electroweak doublet, as with other representations it is not possible to construct invariant operators with just a single $\Psi$ field and the low-energy SM fermions at the order we consider.  Under the complete symmetry group, $\mG = SU(3)_C\otimes SU(2)_L\otimes SU(2)_R\otimes  U(1)_{X}$ with $X={\rm (B-L)}/2$, the resonance $\Psi$ transforms as
\bear
\Psi\quad \longrightarrow \quad g_C\, g_X\, g_h \Psi \, ,
\eear
with $g_C$ and $g_{X}$ the appropriate color and $U(1)_X$ transformations for the corresponding representation of $\Psi$.

After performing several simplifications, the most compact and diagonalized form of the fermionic resonance Lagrangian reads
\be
\mL_\Psi \, =\, \bar{\Psi}(i\slashed{D}-M_{\Psi})\Psi
+  \left( \overline\Psi \chi_\Psi + \overline{\chi}_\Psi \Psi \right)\, ,
\label{eq:Psi-L}
\ee
with the $\mO(p^2)$ fermionic chiral tensors
	\bear
	\chi_\Psi
	&=&  u_\mu \gamma^\mu (\lambda_1\gamma_5+\wl_1) \xi\, -\, i \Frac{(\partial_\mu h)}{v} \, \gamma^\mu (\lambda_2+\wl_2\gamma_5)\xi
	\, + \, (\lambda_0 +\widetilde{\lambda}_0 \gamma_5 )\mT \xi   \, ,
	\nn\\
	\overline{\chi}_\Psi &=&
	\bar\xi\gamma^\mu (\lambda_1\gamma_5+\wl_1)  u_\mu
	\, +\, i \bar\xi\gamma^\mu (\lambda_2+\wl_2\gamma_5)\, \Frac{(\partial_\mu h)}{v}
	\, + \, \bar{\xi} \mT(\lambda_0 - \widetilde{\lambda}_0\gamma_5 ) \, .
	\label{eq:Psi-interaction}
	\eear
Notice that the resonance Lagrangian in Eq.~\eqref{eq:Psi-L} is only invariant if $\xi$ and $\Psi$ have the same color and ${\rm B-L}$ quantum numbers. This means that if $\Psi$ is a quark-type doublet (triplet in color and ${\rm B-L}=1/3$) then it will only couple to quarks and only the EWET LECs $\mF_j^{\psi^2}$ and $\widetilde\mF_j^{\psi^2}$ with $\psi^2\sim \bar{q}q$ will receive contributions. If $\Psi$ is a lepton-type doublet (singlet in color and ${\rm B-L}=-1$), it will only couple to (neutral or charged) leptons and only the EWET LECs $\mF_j^{\psi^2}$ and $\widetilde\mF_j^{\psi^2}$ with $\psi^2\sim \bar{\ell} \ell $ will arise at low energies. Any other $SU(3)_C\otimes U(1)_X$ representation of $\Psi$ will not generate contributions to the low-energy theory at the order we study.

Since the inclusion of fermionic resonances in the EWET was not discussed in previous references, we detail in the Appendix~\ref{app:fermiondiagonalization} the bare unsimplified fermionic doublet resonance Lagrangian and how to obtain its final version in Eq.~\eqref{eq:Psi-L}.

\section{Imprints of the Heavy States in the EWET}
\label{sec:LEC}
Since there is no direct experimental evidence of the presence of heavy resonances, one can only glimpse their effects through the imprints they leave in the LECs of the EWET, at energies lower  than the resonance masses. The resonance contributions to the LECs lead to specific deviations in observables with respect to the SM expectations. The presence of some interacting heavy states might be then inferred from the precise pattern of deviations they induce.

Once the resonance Lagrangian is set, the heavy states are integrated out from the action and the outcome can be organized in powers of momenta over the resonance masses.
This procedure is standard in the context of EFTs and more details can be found for example in Refs.~\cite{Pich:2016lew,Burgess:2007pt}. \\

\subsection{Bosonic resonances}

Replacing the resonance fields in the interaction Lagrangians \eqref{eq:L0}, \eqref{eq:L1proca} and \eqref{eq:L1anti} by their classical EoM solutions, truncated at leading order, one obtains the tree-level resonance-exchange contributions to the $\mO(p^4)$ EWET Lagrangian. The scalar singlet resonance $S_{1}^{1}$ also induces contributions that renormalize parameters of the LO Lagrangian $\mathcal{L}_{\mathrm{EWET}}^{(2)}$ in Eq.~\eqref{eq:L.LO}. These additional singlet contributions have been already discussed in Ref.~\cite{Pich:2016lew}, as colored bosonic resonances do not induce operators of $\mathcal{L}_{\mathrm{EWET}}^{(2)}$. Therefore, we do not repeat here the discussion again, although some additional details can be found in App.~\ref{app:coupling}.

The resulting $\mO(p^4)$ low-energy Lagrangians from spin-$0$ scalar and pseudoscalar resonance exchanges are:
\begin{align}
\Delta \mL_{R^1_1}^{\cO(p^4)} & =\, \Frac{1}{2M_{R^1_1}^2} \,
( \chi_{R^1_1})^2,
\nn\\
\Delta \mL_{R^1_3}^{\cO(p^4)} & =\, \Frac{1}{2M_{R^1_3}^2}
\left(  \bra  \chi_{R^1_3}\, \chi_{R^1_3}\ket_2 \,-\, \frac{1}{2}
\bra \chi_{R^1_3}\ket_2\, \bra \chi_{R^1_3}\ket_2\right), 
\nn\\
\Delta \mL_{R_1^8}^{\cO(p^4)} & =\,  \Frac{1}{ 4 M_{R^8_1}^2}\,  \bra  \chi_{R_1^{8}}\, \chi_{R_1^{8}}\ket_3    \, , 
\nn\\
\Delta \mL_{R^8_3}^{\cO(p^4)} & =\,  \Frac{1}{4 M_{R^8_3}^2}
\left(  \bra  \chi_{R^{8}_3}\, \chi_{R^{8}_3}\ket_{2,3}\, -\, \Frac{1}{2} \bra \bra \chi_{R^8_3} \ket_2 \, \bra \chi_{R^8_3} \ket_2 \ket_3 \right)  \,, 
\label{integration0}
\end{align}
with $R^m_n = S^m_n,\,P^m_n$ and the interaction tensors $\chi_R$ listed in Eq.~(\ref{eq:L0tensor}).  Contributions coming from the product of two separated $SU(3)$ traces are not listed since they vanish at $\mO(p^4)$ (closed color indices imply that these traces are proportional to $\bra T^a\ket_3\bra T^a\ket_3 = 0)$.

For the spin-$1$ resonances, we combine results from the Proca and antisymmetric formalisms together, as indicated before, which ensures the completeness of the obtained set of LECs. Through the exchange of Proca fields, the $\chi^\mu_{\hat R} \hat R_\mu$ interactions in Eq.~\eqref{eq:L1proca} generate at low energies the following $O(p^4)$ Lagrangians:
\begin{align}
	\Delta \mL_{\hat R^1_1}^{\cO(p^4)} & =\, -\Frac{1}{2M_{R^1_1}^2} \,
	( \chi_{\hat R_1^{1}}^\mu)^2 \,, 
	\nn\\
	\Delta \mL_{\hat R^1_3}^{\cO(p^4)} & =\, -\Frac{1}{2M_{R^1_3}^2}
	\left(  \bra \chi^{\mu}_{\hat R^1_3}\, \chi_{\hat R^1_3\,\mu}\ket_2 \,-\, \frac{1}{2}
	\bra \chi^{\mu}_{\hat R^1_3} \ket_2 \,\bra \chi_{\hat R^1_3\,\mu} \ket_2 \right)\,, 
	\nn\\
	\Delta \mL_{\hat R_1^8}^{\cO(p^4)} & =\,  -\Frac{1}{4 M_{R^{8}_1}^2}
	\, \bra  \chi^\mu_{\hat R_1^{8}}\, \chi_{\hat R_{1}^{8}\,\mu}\ket_3\,,  
	\nn\\
	\Delta \mL_{\hat R^8_3}^{\cO(p^4)} & =\,  -\Frac{1}{4 M_{R^8_3}^2}
	\left(  \bra  \chi^\mu_{\hat R^{8}_3}\, \chi_{\hat R^{8}_3\,\mu}\ket_{2,3}\, -\, \Frac{1}{2} \bra \bra \chi^\mu_{\hat R^8_3} \ket_2 \,\bra \chi_{\hat R^8_3\,\mu} \ket_2 \ket_3 \right)\,,  
	\label{integration1proca}
\end{align}
with $\hat R^m_n = \hat V^m_n,\,\hat A^m_n$ and $\chi^\mu_{\hat R}$ the tensor structures in Eq.~\eqref{eq:L1procatensor}. The $\chi^{\mu\nu}_R R_{\mu\nu}$ interactions in Eq.~\eqn{eq:L1anti} give rise, through the exchange of antisymmetric tensor fields, to the $O(p^4)$ Lagrangians:
\begin{align}
\Delta \mL_{ R^1_1}^{\cO(p^4)} & =\, -\Frac{1}{M_{R^1_1}^2} \,
(  \chi_{ R^1_1}^{\mu\nu})^2 \,,
\nn\\
\Delta \mL_{ R^1_3}^{\cO(p^4)} & =\, -\Frac{1}{M_{R^1_3}^2}
\left(  \bra  \chi^{\mu\nu}_{ R^1_3}\, \chi_{ {R^1_3}\,\mu\nu}^{\phantom{\mu\nu}}\ket_2 \,-\, \frac{1}{2}
\bra \chi^{\mu\nu}_{ R^1_3} \ket_2\,\bra \chi_{{ R^1_3}\,\mu\nu}^{\phantom{\mu\nu}} \ket_2 \right) \,,
\nn\\
\Delta \mL_{ R_1^8}^{\cO(p^4)} & =\, -\Frac{1}{2 M_{R^{8}_1}^2}
 \bra  \chi^{\mu\nu}_{ R_1^{8}}\, \chi_{ R_{1}^{8}\,\mu\nu}^{\phantom{\mu\nu}}\ket_3\,,
\nn\\
\Delta \mL_{ R^8_3}^{\cO(p^4)} & =\, -\Frac{1}{2 M_{R^8_3}^2}
\left(  \bra  \chi^{\mu\nu}_{ R^8_3}\, \chi_{ {R^8_3}\,\mu\nu}^{\phantom{\mu\nu}} \ket_{2,3}\, -\, \Frac{1}{2} \bra \bra \chi^{\mu\nu}_{ R^8_3} \ket_2 \, \bra \chi_{{ R^8_3}\,\mu\nu}^{\phantom{\mu\nu}} \ket_2 \ket_3 \right) \,,
\label{integration1anti}
\end{align}
where $\chi_R^{\mu\nu}$ refer to the tensor structures of Eq.~\eqref{eq:L1antitensor}.

\begin{table}[tb]  
	\begin{center}
		\renewcommand{\arraystretch}{2.6}
		\begin{tabular}{|c|c|c||c|c|}
                  \hline
                  $i$ & $\Delta\mF_i$ & $\Delta\widetilde{\mF}_i$& $i$ & $\Delta\mF_i$ \\
                  \hline\hline
                  1  & $- \Frac{F_V^2-\widetilde{F}_V^2}{4M_{V^1_3}^2} + \Frac{F_A^2-\widetilde{F}_A^2}{4M_{A^1_3}^2} $  & $- \Frac{\widetilde{F}_VG_V}{2M_{V^1_3}^2} - \Frac{F_A\widetilde{G}_A}{2M_{A^1_3}^2}$	& 7  & $\Frac{ d_P^2}{2 M_{P^1_3}^2}+\Frac{\lambda_1^{hA\,\, 2}v^2}{M_{A^1_3}^2} +  \Frac{\widetilde{\lambda}_1^{hV\,\, 2}v^2}{M_{V^1_3}^2} $\\[1ex] \hline
                  2  & $- \Frac{F_V^2+{\widetilde{F}_V}^2}{8M_{V^1_3}^2} - \Frac{F_A^2+{\widetilde{F}_A}^2}{8M_{A^1_3}^2}$  & $- \Frac{F_V \widetilde{F}_V}{4M_{V^1_3}^2} - \Frac{F_A \widetilde{F}_A}{4M_{A^1_3}^2}$ & 8  & 0	\\[1ex] \hline
                  3  & $-  \Frac{F_VG_V}{2M_{V^1_3}^2} - \Frac{\widetilde{F}_A\widetilde{G}_A}{2M_{A^1_3}^2}$  & $- \Frac{F_V \widetilde{\lambda}_1^{hV} v}{M_{V^1_3}^2} - \Frac{\widetilde{F}_A \lambda_1^{hA} v}{M_{A^1_3}^2}$& 9  & $  - \Frac{F_A \lambda_1^{hA} v}{M_{A^1_3}^2} - \Frac{\widetilde{F}_V \widetilde{\lambda}_1^{hV} v}{M_{V^1_3}^2}$	\\[1ex] \hline
                  4  & $\Frac{G_V^2}{4M_{V^1_3}^2} + \Frac{{\widetilde{G}_A}^2}{4M_{A^1_3}^2} $  & --- & 10 & $-\displaystyle\frac{\widetilde{c}_{\mathcal{T}}^2}{2M_{V^1_1}^2}-\displaystyle\frac{c_{\mathcal{T}}^2}{2M_{A^1_1}^2}$	\\[1ex] \hline
                  5  & $\Frac{c_{d}^2}{4M_{S^1_1}^2}-\Frac{G_V^2}{4M_{V^1_3}^2} - \Frac{{\widetilde{G}_A}^2}{4M_{A^1_3}^2}$  & --- &11 & $- \Frac{F_{X}^2}{M_{V^1_1}^2} - \Frac{\widetilde{F}_{X}^2}{M_{A^1_1}^2} $	\\[1ex] \hline
                  6  & $ - \Frac{\widetilde{\lambda}_1^{hV\,\, 2}v^2}{M_{V^1_3}^2} - \Frac{\lambda_1^{hA\,\, 2}v^2}{M_{A^1_3}^2}$  & --- &12 & $  - \Frac{(C_G)^2}{2 M_{V^8_1}^2} - \Frac{(\widetilde{C}_G)^2}{2 M_{A^8_1}^2}  $ 	\\[1ex] \hline
\end{tabular}
\caption{{\small
Contributions to the bosonic $\cO(p^4)$ LECs from heavy scalar, pseudoscalar, vector, and axial-vector exchanges.}}
\label{tab:result:bos}
\end{center}
\end{table}

Projecting these expressions into the $\mO(p^4)$ operator basis, one gets the corresponding contributions to the EWET LECs. They are compiled in Tables \ref{tab:result:bos}, \ref{tab:result:2ferm.even}, \ref{tab:result:4ferm.even}, and \ref{tab:result:ferm.odd} together with the results from fermionic resonance exchanges. Table~\ref{tab:result:bos} shows the LECs of bosonic operators, Table~\ref{tab:result:2ferm.even} the $P$-even two-fermion LECs, Table~\ref{tab:result:4ferm.even} the $P$-even four-fermion ones, and Table~\ref{tab:result:ferm.odd} the LECS of $P$-odd fermion operators. We have applied several Fierz identities to express the resulting four-fermion operators in terms of our operator basis. Interestingly, the spin-$0$ resonances only contribute to $P$-even operators.

\begin{table}[tb]
  \begin{center}
    \renewcommand{\arraystretch}{2.4}
    \begin{tabular}{|c|c||c|c|}
      \hline
      $i$ &  $\Delta\mF^{\psi^2}_i$ &      $i$ &  $\Delta\mF^{\psi^2}_i$\\  \hline\hline
      1   &  $\displaystyle\frac{c_d c^{S^1_1}}{2M_{S^1_1}^2}+\Frac{\wl_1^2-\lambda_1^2}{2 M_\Psi}$ &       5   &  $\displaystyle\frac{d_P c^{P^1_3}}{M_{P^1_3}^2}+\Frac{    2(\lambda_1\lambda_2 -\wl_1\wl_2)   }{M_\Psi}$ \\[1ex] \hline
      2   &  $-\displaystyle\frac{G_V C_0^{V^1_3}}{\sqrt{2}M_{V^1_3}^2} \!-\! \displaystyle\frac{\widetilde{G}_A \widetilde{C}_0^{A^1_3}}{\sqrt{2}M_{A^1_3}^2}-\Frac{\wl_1^2-\lambda_1^2}{2 M_\Psi} $ &      6   &  $-\displaystyle\frac{\widetilde{c}_{\mathcal{T}} \widetilde{c}^{\hat{V}^1_1}    }{\sqrt{2} M_{V^1_1}^2} - \displaystyle\frac{c_{\mathcal{T}} c^{\hat{A}^1_1}    }{\sqrt{2} M_{A^1_1}^2}+ \Frac{(\lambda_0\lambda_1 +\wl_0  \wl_1)}{M_\Psi}$     \\[1ex] \hline
      3   &  $-\displaystyle\frac{F_V C_0^{V^1_3}}{\sqrt{2}M_{V^1_3}^2} -  \displaystyle\frac{\widetilde{F}_A \widetilde{C}_0^{A^1_3}}{\sqrt{2}M_{A^1_3}^2} $  &      7   & $\displaystyle \Frac{\lambda_2^2 -\wl_2^2}{M_\Psi}$    \\[1ex] \hline
      4   &  $-\displaystyle\frac{\sqrt{2}F_{X} C_0^{V^1_1}}{M_{V^1_1}^2} -\displaystyle\frac{\sqrt{2}\widetilde{F}_{X} \widetilde{C}_0^{A^1_1}}{M_{A^1_1}^2} $ & 8   &  $ -\Frac{C_G C_0^{V^8_1}}{\sqrt{2} M^2_{V^8_1}} - \Frac{\widetilde C_G \widetilde C_0^{A^8_1}}{\sqrt{2} M^2_{A^8_1}}  $ \\[2ex] \hline
\end{tabular}
\end{center}
\caption{\small
Contributions to the $\cO(p^4)$ LECs of two-fermion $P$-even operators from heavy scalar, pseudoscalar, vector, axial-vector and fermionic exchanges.}
\label{tab:result:2ferm.even}
\end{table}

\begin{table}[pt] 
\begin{center}
\renewcommand{\arraystretch}{2.5}
\begin{tabular}{|c||c|}
\hline
$i$  & $\Delta\mF_i^{\psi^4}$
\\
\hline\hline
& $ -\Frac{(c^{\hat V^8_3})^2}{32 M^2_{V^8_3}} + \Frac{(\widetilde c^{\hat{V}^8_3})^2}{32 M^2_{V^8_3}}  + \Frac{(c^{\hat V^8_1})^2}{32 M^2_{V^8_1}}   - \Frac{(\widetilde c^{\hat V^8_1})^2}{32 M^2_{V^8_1}} +\Frac{(c^{\hat A^8_3})^2}{32 M^2_{A^8_3}} - \Frac{(\widetilde c^{\hat A^8_3})^2}{32 M^2_{A^8_3}} $
\\
1 & $ - \Frac{(c^{\hat A^8_1})^2}{32 M^2_{A^8_1}}   + \Frac{(\widetilde c^{\hat A^8_1})^2}{32 M^2_{A^8_1}} - \Frac{3 (C_0^{V^8_3})^2}{16 M^2_{V^8_3}}  + \Frac{3 (C_0^{V^8_1})^2}{16 M^2_{V^8_1}}  - \Frac{3 (\widetilde C_0^{A^8_3})^2}{16 M^2_{A^8_3}}  +  \Frac{3 (\widetilde C_0^{A^8_1})^2}{16 M^2_{A^8_1}}+\Frac{(c^{S^1_3})^2}{2M_{S^1_3}^2}$
\\
& $ +\displaystyle \frac{1}{128}\! \left( - \Frac{5(c^{S^8_3})^2}{3 M^2_{S^8_3}}  - \Frac{(c^{S_1^8})^2}{ M^2_{S_1^8}} -  \Frac{(c^{P^8_3})^2}{ M^2_{P^8_3}}   + \Frac{(c^{P_1^8})^2}{ M^2_{P_1^8}} \right) $
\\[1ex]\hline
& $ -\Frac{(c^{\hat V^8_3})^2}{32 M^2_{V^8_3}} + \Frac{(\widetilde c^{\hat V^8_3})^2}{32 M^2_{V^8_3}}  + \Frac{(c^{\hat V^8_1})^2}{32 M^2_{V^8_1}}   - \Frac{(\widetilde c^{\hat V^8_1})^2}{32 M^2_{V^8_1}}- \Frac{(c^{\hat A^8_1})^2}{32 M^2_{A^8_1}}  + \Frac{(\widetilde c^{\hat A^8_1})^2}{32 M^2_{A^8_1}} $
\\
2 & $ +\Frac{(c^{\hat A^8_3})^2}{32 M^2_{A^8_3}} - \Frac{(\widetilde c^{\hat A^8_3})^2}{32 M^2_{A^8_3}} +\Frac{3 (C_0^{V^8_3})^2}{16 M^2_{V^8_3}}  - \Frac{3 (C_0^{V^8_1})^2}{16 M^2_{V^8_1}}  + \Frac{3 (\widetilde C_0^{A^8_3})^2}{16 M^2_{A^8_3}}  - \Frac{3 (\widetilde C_0^{A^8_1})^2}{16 M^2_{A^8_1}}+\Frac{(c^{P^1_3})^2}{2M_{P^1_3}^2} $
\\
& $ + \displaystyle\frac{1}{128}\! \left( - \Frac{(c^{S^8_3})^2}{ M^2_{S^8_3}}+ \Frac{(c^{S_1^8})^2}{ M^2_{S_1^8}}-  \Frac{5(c^{P^8_3})^2}{3 M^2_{P^8_3}}- \Frac{(c^{P_1^8})^2}{ M^2_{P_1^8}} \right) $ 	
\\[1ex]\hline
3 &  $ \Frac{(c^{\hat V^8_3})^2}{16 M^2_{V^8_3}}  - \Frac{( \widetilde c^{\hat V^8_3})^2}{16 M^2_{V^8_3}}  -\Frac{(c^{\hat A^8_3})^2}{16 M^2_{A^8_3}} + \Frac{(\widetilde c^{\hat A^8_3})^2}{16 M^2_{A^8_3}} +  \Frac{3 (C_0^{V^8_3})^2}{8 M^2_{V^8_3}} + \Frac{3 (\widetilde C_0^{A^8_3})^2}{8 M^2_{A^8_3}}$
\\
& $- \Frac{(c^{S^1_3})^2}{4M_{S^1_3}^2}+\Frac{(c^{S^1_1})^2}{4M_{S^1_1}^2} + \displaystyle\frac{1}{64}\! \left(- \Frac{(c^{S^8_3})^2}{ 3 M^2_{S^8_3}}- \Frac{2(c^{S_1^8})^2}{3 M^2_{S_1^8}}+ \Frac{(c^{P^8_3})^2}{M^2_{P^8_3}} \right)  $
\\[1ex]\hline
4 & $\Frac{(c^{\hat V^8_3})^2}{16 M^2_{V^8_3}}  - \Frac{( \widetilde c^{\hat V^8_3})^2}{16 M^2_{V^8_3}}  -\Frac{(c^{\hat A^8_3})^2}{16 M^2_{A^8_3}} + \Frac{(\widetilde c^{\hat A^8_3})^2}{16 M^2_{A^8_3}}  - \Frac{3 (C_0^{V^8_3})^2}{8 M^2_{V^8_3}} - \Frac{3 (\widetilde C_0^{A^8_3})^2}{8 M^2_{A^8_3}}$
\\
& $-\Frac{(c^{P^1_3})^2}{4M_{P^1_3}^2}+\Frac{(c^{P^1_1})^2}{4M_{P^1_1}^2}+ \displaystyle \frac{1}{64}\! \left(  \Frac{(c^{S^8_3})^2}{ M^2_{S^8_3}}- \Frac{(c^{P^8_3})^2}{3 M^2_{P^8_3}}- \Frac{2(c^{P_1^8})^2}{3 M^2_{P_1^8}} \right)  $
\\[1ex]\hline
5 & $\displaystyle \frac{1}{128}\! \left( \Frac{(c^{S^8_3})^2}{ M^2_{S^8_3}}- \Frac{(c^{S_1^8})^2}{ M^2_{S_1^8}}+ \Frac{(c^{P^8_3})^2}{ M^2_{P^8_3}}- \Frac{(c^{P_1^8})^2}{ M^2_{P_1^8}} \right)-\Frac{(c^{\hat{V}^1_3})^2}{2M_{V^1_3}^2} -\Frac{(\widetilde{c}^{\hat{A}^1_3})^2}{2M_{A^1_3}^2}$
\\
& $+\displaystyle \frac{1}{64}\! \left( \Frac{7(c^{\hat V^8_3})^2}{3 M^2_{V^8_3}} + \Frac{(\widetilde c^{\hat V^8_3})^2}{ M^2_{V^8_3}} - \Frac{(c^{\hat V^8_1})^2}{ M^2_{V^8_1}} - \Frac{(\widetilde c^{\hat V^8_1})^2}{ M^2_{V^8_1}}  +\Frac{(c^{\hat A^8_3})^2}{ M^2_{A^8_3}}  + \Frac{7(\widetilde c^{\hat A^8_3})^2}{3 M^2_{A^8_3}} - \Frac{(c^{\hat A^8_1})^2}{ M^2_{A^8_1}} - \Frac{(\widetilde c^{\hat A^8_1})^2}{ M^2_{A^8_1}} \right) $
\\[1ex]\hline
6 & $- \Frac{(\widetilde{c}^{\hat{V}^1_3})^2}{2M_{V^1_3}^2} -\Frac{({c}^{\hat{A}^1_3})^2}{2M_{A^1_3}^2}+ \displaystyle \frac{1}{128}\! \left( -\Frac{(c^{S^8_3})^2}{ M^2_{S^8_3}}+ \Frac{(c^{S_1^8})^2}{M^2_{S_1^8}}- \Frac{(c^{P^8_3})^2}{ M^2_{P^8_3}}+ \Frac{(c^{P_1^8})^2}{ M^2_{P_1^8}} \right)$
\\
& $+\displaystyle \frac{1}{64}\! \left( \Frac{(c^{\hat V^8_3})^2}{ M^2_{V^8_3}} + \Frac{7(\widetilde c^{\hat V^8_3})^2}{3 M^2_{V^8_3}} - \Frac{(c^{\hat V^8_1})^2}{ M^2_{V^8_1}} - \Frac{(\widetilde c^{\hat V^8_1})^2}{ M^2_{V^8_1}} +\Frac{7(c^{\hat A^8_3})^2}{3 M^2_{A^8_3}}  + \Frac{(\widetilde c^{\hat A^8_3})^2}{ M^2_{A^8_3}} - \Frac{(c^{\hat A^8_1})^2}{ M^2_{A^8_1}} - \Frac{(\widetilde c^{\hat A^8_1})^2}{ M^2_{A^8_1}} \right) $
\\[2ex]\hline
\end{tabular}
\caption{\small
Contributions to the $\cO(p^4)$ LECs of four-fermion $P$-even operators from heavy scalar, pseudoscalar, vector and axial-vector exchanges.}
\label{tab:result:4ferm.even}
\end{center}
\end{table}
\begin{table}[t] 
\begin{center}
\renewcommand{\arraystretch}{2.5}
\begin{tabular}{|c||c|}
\hline
$i$  & $\Delta\mF_i^{\psi^4}$
\\
\hline\hline
7 & $-\Frac{(c^{S^8_3})^2}{64 M^2_{S^8_3}}- \Frac{(c^{P^8_3})^2}{64 M^2_{P^8_3}} +\Frac{({c}^{\hat{V}^1_3})^2}{4M_{V^1_3}^2} +\Frac{(\widetilde{c}^{\hat{A}^1_3})^2}{4M_{A^1_3}^2} -\Frac{({c}^{\hat{V}^1_1})^2}{4M_{V^1_1}^2} -\Frac{(\widetilde{c}^{\hat{A}^1_1})^2}{4M_{A^1_1}^2}$
\\
& $ -\Frac{(c^{\hat V^8_3})^2}{24 M^2_{V^8_3}}  - \Frac{(\widetilde c^{\hat V^8_3})^2}{32 M^2_{V^8_3}}   + \Frac{( c^{\hat V^8_1})^2}{96 M^2_{V^8_1}} - \Frac{(c^{\hat A^8_3})^2}{32 M^2_{A^8_3}}  - \Frac{(\widetilde c^{\hat A^8_3})^2}{24 M^2_{A^8_3}}  + \Frac{( \widetilde c^{\hat A^8_1})^2}{96 M^2_{A^8_1}} $
\\[1ex]\hline
8 & $\Frac{(c^{S^8_3})^2}{64 M^2_{S^8_3}}+ \Frac{(c_1^{P^8_3})^2}{64 M^2_{P^8_3}}+ \Frac{(\widetilde{c}^{\hat{V}^1_3})^2}{4M_{V^1_3}^2} +\Frac{({c}^{\hat{A}^1_3})^2}{4M_{A^1_3}^2} -\Frac{(\widetilde{c}^{\hat{V}^1_1})^2}{4M_{V^1_1}^2}  -\Frac{({c}^{\hat{A}^1_1})^2}{4M_{A^1_1}^2}$
\\
&  $- \Frac{( c^{\hat V^8_3})^2}{32 M^2_{V^8_3}}   -\Frac{(\widetilde c^{\hat V^8_3})^2}{24 M^2_{V^8_3}}    + \Frac{(\widetilde c^{\hat V^8_1})^2}{96 M^2_{V^8_1}} - \Frac{(c^{\hat A^8_3})^2}{24 M^2_{A^8_3}} - \Frac{( \widetilde c^{\hat A^8_3})^2}{32 M^2_{A^8_3}}  +  \Frac{(c^{\hat A^8_1})^2}{96 M^2_{A^8_1}} $
\\[1ex]\hline
9 & $-\displaystyle\frac{(C_0^{V^1_3})^2}{M_{V^1_3}^2} - \displaystyle\frac{(\widetilde{C}_0^{A^1_3})^2}{M_{A^1_3}^2}   + \Frac{ 7( C_0^{V^8_3})^2}{96 M^2_{V^8_3}}   - \Frac{ (C_0^{V^8_1})^2}{32 M^2_{V^8_1}}  + \Frac{7 (\widetilde C_0^{A^8_3})^2}{96 M^2_{A^8_3}}  -  \Frac{ (\widetilde C_0^{A^8_1})^2}{32 M^2_{A^8_1}}$
\\
& $+ \displaystyle \frac{1}{256}\! \left( \Frac{(c^{S^8_3})^2}{ M^2_{S^8_3}} - \Frac{(c^{S_1^8})^2}{ M^2_{S_1^8}} - \Frac{(c^{P^8_3})^2}{ M^2_{P^8_3}} +  \Frac{(c^{P_1^8})^2}{ M^2_{P_1^8}} \right) $
\\[1ex]\hline
10 & $\displaystyle\frac{(C_0^{V^1_3})^2}{2M_{V^1_3}^2} - \displaystyle\frac{(C_0^{V^1_1})^2}{2M_{V^1_1}^2}  + \displaystyle\frac{(\widetilde{C}_0^{A^1_3})^2}{2M_{A^1_3}^2} - \displaystyle\frac{(\widetilde{C}_0^{A^1_1})^2}{2M_{A^1_1}^2} - \Frac{ (C_0^{V^8_3})^2}{12 M^2_{V^8_3}} + \Frac{ (C_0^{V^8_1})^2}{48 M^2_{V^8_1}} - \Frac{ (\widetilde C_0^{A^8_3})^2}{12 M^2_{A^8_3}} +  \Frac{ (\widetilde C_0^{A^8_1})^2}{48 M^2_{A^8_1}}$
\\
& $-\Frac{(c^{S^8_3})^2}{128 M^2_{S^8_3}}+ \Frac{(c^{P^8_3})^2}{128 M^2_{P^8_3}}$
\\[2ex]\hline
\end{tabular}
\addtocounter{table}{-1}
\caption{\small (continued)
Contributions to the $\cO(p^4)$ LECs of four-fermion $P$-even operators from heavy scalar, pseudoscalar, vector and axial-vector exchanges.}
\end{center}
\end{table}

\begin{table}[!tbh]  
	\begin{center}
		\renewcommand{\arraystretch}{2.4}
		\begin{tabular}{|c|c|c|}
			\hline
			$i$ &  $\Delta\widetilde{\mF}^{\psi^2}_i$ &$\Delta\widetilde{\mF}_i^{\psi^4}$
			\\  \hline\hline
			1 &  $-\Frac{\widetilde{F}_V C_0^{V^1_3}}{\sqrt{2}M_{V^1_3}^2}  -  \Frac{F_A \widetilde{C}_0^{A^1_3}}{\sqrt{2}M_{A^1_3}^2} $ & $ -\Frac{c^{\hat{V}^1_3} \widetilde{c}^{\hat{V}^1_3}} {M_{V^1_3}^2} -\Frac{ c^{\hat{A}^1_3}\widetilde{c}^{\hat{A}^1_3}}{M_{A^1_3}^2}$ \\
                  & & $\Frac{5 c^{\hat V^8_3} \widetilde c^{\hat V^8_3}}{48 M^2_{V^8_3}} -\Frac{ c^{\hat V^8_1} \widetilde c^{\hat V^8_1}}{16 M^2_{V^8_1}} + \Frac{5 c^{\hat A^8_3} \widetilde c^{\hat A^8_3}}{48 M^2_{A^8_3}} - \Frac{ c^{\hat A^8_1} \widetilde c^{\hat A^8_1}}{16 M^2_{A^8_1}} $ \\[1ex] \hline
			2 &  $-\Frac{2\sqrt{2}\,v\,\widetilde{\lambda}_1^{hV}C_0^{V^1_3}}{M_{V^1_3}^2} - \Frac{2\sqrt{2}\,v\,\lambda_1^{hA}\widetilde{C}_0^{A^1_3}}{M_{A^1_3}^2}$ &$ \Frac{c^{\hat{V}^1_3}\widetilde{c}^{\hat{V}^1_3}}{2M_{V^1_3}^2} -\Frac{c^{\hat{V}^1_1}\widetilde{c}^{\hat{V}^1_1}}{2M_{V^1_1}^2}+\Frac{c^{\hat{A}^1_3}\widetilde{c}^{\hat{A}^1_3}}{2M_{A^1_3}^2} -\Frac{c^{\hat{A}^1_1}\widetilde{c}^{\hat{A}^1_1}}{2M_{A^1_1}^2}$\\
                  &$+\displaystyle\Frac{   2(\wl_1\lambda_2 -\lambda_1\wl_2)   }{M_\Psi}$ & $-\Frac{7 c^{\hat V^8_3} \widetilde c^{\hat V^8_3}}{48 M^2_{V^8_3}} +\Frac{ c^{\hat V^8_1} \widetilde c^{\hat V^8_1}}{48 M^2_{V^8_1}} - \Frac{7 c^{\hat A^8_3} \widetilde c^{\hat A^8_3}}{48 M^2_{A^8_3}} + \Frac{ c^{\hat A^8_1} \widetilde c^{\hat A^8_1}}{48 M^2_{A^8_1}} $\\[1ex] \hline
			3 &$-\Frac{\widetilde{c}_{\mathcal{T}} c^{\hat{V}^1_1}    }{\sqrt{2} M_{V^1_1}^2} - \Frac{c_{\mathcal{T}} \widetilde{c}^{\hat{A}^1_1}    }{\sqrt{2} M_{A^1_1}^2}+\displaystyle  \Frac{(\lambda_0\wl_1 +\wl_0\lambda_1)}{M_\Psi}$ & ---\\[2ex] \hline
		\end{tabular}
	\end{center}
	\caption{\small
Contributions to the $\cO(p^4)$ LECs of two- and four-fermion $P$-odd operators from heavy scalar, pseudoscalar, vector, axial-vector and fermionic exchanges.}
\label{tab:result:ferm.odd}
\end{table}
%

\subsection{Fermionic resonances}

Similarly, we solve the EoM for the fermionic resonance $\Psi$ and expand it for $p\ll M_\Psi$, up to $\cO(p^2)$. We obtain
\be
\Psi^{\cO(p^2)} =  \Frac{1}{M_\Psi} \,\chi_\Psi\, ,
\qquad\qquad\qquad
\overline{\Psi}^{\cO(p^2)} = \Frac{1}{M_\Psi} \,\overline{\chi}_\Psi\, .
\ee
Substituting this EoM solution in the resonance Lagrangian~(\ref{eq:Psi-L}),
we get the contribution from the $\Psi$ exchanges to the low-energy EWET at $\cO(p^4)$:
\bear
\Delta\mL_{\Psi}^{\cO(p^4)}&=&
\Frac{1}{M_\Psi} \;\overline{\chi}_\Psi\,\chi_\Psi\,
\nn\\
 &=&  \,  \Frac{g^{\prime\, 2}(\lambda_0^2-\wl_0^2 )}{4 M_\Psi}\; \bar{\xi}\xi\,
+\sum_{j=1,2,5,6,7}\Delta\mF_j^{\psi^2}\mO_j^{\psi^2}
+\sum_{j=2,3}\Delta\widetilde{\mF}_j^{\psi^2}\widetilde{\mO}_j^{\psi^2}
\, ,
\label{eq:Psi-contribution}
\eear
which expressed in terms of the previously developed EFT basis yields the corresponding LECs. The remaining NLO operators do not receive contributions from the doublet fermionic resonance~$\Psi$. The results are shown in detail in Tables~\ref{tab:result:2ferm.even} and \ref{tab:result:ferm.odd}.

The first term in the second line of~(\ref{eq:Psi-contribution})
has the form of a SM fermion mass Lagrangian. From pure na\"ive dimensional analysis, the fermion bilinear would scale in the low-energy counting as $\cO(p)$ while the factor  $g^{\prime2}$, coming from the custodial breaking tensor $\mT$ in the interaction vertex~(\ref{eq:Psi-interaction}), scales like $\cO(p^2)$. In addition, our assumption that the SM fermions carry an additional weak coupling suppression
(encoded in the operator coupling) implies that $\lambda_0^2$ and $\wl_0^2$ scale like $\cO(p)$. Hence, this contribution is formally of $\mO(p^4)$ although, at the practical level,
it adds up to the Yukawa mass terms in the LO Lagrangian and will not be discussed separately any more.

We would like to make one final remark on the distinction between
fermionic resonances with quark ($\Psi_{\rm quark}$) and lepton ($\Psi_{\rm lepton}$) quantum numbers, when they are considered at the same time. Unless $\Psi_{\rm quark}$ and $\Psi_{\rm lepton}$ belong to a common multiplet of some unified group ({\it e.g.} the $SU(4)$ Pati-Salam group~\cite{Pati:1974yy}), one must distinguish between quark and leptonic EFT operators, being the effective couplings of these two sectors independent parameters.

\section{Phenomenology}
\label{sec:pheno}
In this section we discuss some phenomenological signatures of the electroweak resonances, at energies lower than their mass scale. The more generic (model-independent) aspects are analyzed first. Afterwards, we connect our general approach to three specific examples with very contrasting initial assumptions, encompassing both weakly and strongly coupled models. This reflects the great power and versatility of the EWET, being able to properly describe any high-energy scenario provided the assumed pattern of EWSB is satisfied. Note that for some of the models, where the Higgs boson is embedded in an electroweak doublet, the SMEFT also gives a convenient description.
\subsection{Current EWET phenomenology}
\label{sec:EWET-pheno}
We have seen in Section~\ref{sec:LEC} that most of the resonances contribute only to the NLO (and higher-order) LECs of the EWET. The scalar singlet $S_{1}^{1}$  and the fermionic doublet $\Psi$ are the only exceptions that also contribute to LO structures. However, all resonance contributions are of $\mO(p^4)$ and suppressed by inverse powers of the resonance mass scale ($1/M_R^2$ for bosonic resonances and $1/M_R$ for fermionic ones), also the contributions to LO operators. Thus, even though the $S_{1}^{1}$ state could manifest in current measurements of the Higgs couplings, its effects are expected to be suppressed. This is consistent with the recent fit to LHC data~\cite{deBlas:2018tjm}, which does not see any deviation from the SM predictions. However, the present uncertainties are still sizable, of $\mathcal{O}(10\%)$.

Another possibility to access the LO couplings is double-Higgs production, which is sensitive to the triple Higgs coupling~\cite{Grober:2016wmf}.
If a discrepancy with the SM would be identified in a LO coupling, the mass scale of some of the resonances could also be estimated using unitarization techniques \cite{Dobado:1989gr,Dobado:1990jy,Dobado:1990am,Dobado:1999xb,Filipuzzi:2012bv,Delgado:2014dxa,Delgado:2015kxa,Delgado:2017cat}, albeit with some degree of model dependence originating in the resonance couplings neglected in those analyses.

The fermion resonance doublet $\Psi$ also contributes to the LO EWET Lagrangian:
a similar shift $\Delta m_\psi= g^{'\, 2} (\widetilde{\lambda}_0^2-\lambda_0^2)/M_\Psi$ is generated for the mass of both the $t$ and $b$ components of the SM doublet $\psi$. If $\Delta m_\psi$ was of the order of the top mass, it seems unlikely that the rest of SM fermion masses would remain much smaller than $m_t$. Thus, one expects a very suppressed correction $|\Delta m_\psi|\ll m_t$, in agreement with our assumptions, which classify these fermion contributions as NLO in the EWET.

The oblique $S$ and $T$ parameters are sensitive to vector and axial-vector triplet resonances that contribute
to $S$ already at LO~\cite{Peskin:1990zt,Peskin:1991sw}. Their effects were studied in Refs.~\cite{Pich:2012jv,Pich:2012dv,Pich:2013fea},
including NLO corrections. Current experimental bounds~\cite{deBlas:2016ojx,Haller:2018nnx}
on these parameters and the requirement of a good high-energy behavior of the UV theory push the masses of the vector and axial-vector resonances
to the TeV range~\cite{Pich:2013fea,Pich:2015kwa}.
Scalar and fermion resonances can generate additional NLO corrections to the gauge-boson self-energies, which have not been studied yet.

Apart from modifications of the SM-like couplings, one expects new interaction vertices not present in the SM Lagrangian. An important class of these novel contributions are four-fermion operators. Standard dijet and dilepton studies at LHC~\cite{Aaboud:2017yvp,Sirunyan:2017ygf,ATLAS:2014cra,CMS:2014aea} and LEP~\cite{Schael:2013ita} have looked for four-fermion operators containing light leptons and/or quarks. Their theoretical interpretation depends on whether one considers a diagonal flavor structure of the theory, with similar couplings and LECs for all generations, or a particular suppression of BSM interactions in the first and second families. In these experimental searches, it has been customary to express the four-fermion LECs in terms of a suppression scale $\Lambda$ defined through $|\mF^{\psi^4}_j| = 2\pi/\Lambda^2$. Currently, the tightest (95\% CL) lower limits on these contact interactions are:\footnote{We note that Refs.\cite{Schael:2013ita,ATLAS:2014cra} consider a different LEC normalization ($|\mF^{\psi^4}_j| = 4\pi/\Lambda^2$) for some analyses. Nevertheless, as we are just interested in showing the order of magnitude of the current lower bounds on the new-physics scale, the values of $\Lambda$ quoted in the present article are simply those given in Refs.~\cite{Schael:2013ita,ATLAS:2014cra}, without any modification.}
\begin{enumerate}
\item From dijet production:
\begin{itemize}
\item[--] $\Lambda\geq 21.8$~TeV from ATLAS~\cite{Aaboud:2017yvp},
\item[--] $\Lambda\geq 18.6$~TeV from CMS~\cite{Sirunyan:2017ygf},
\item[--] $\Lambda\geq 16.2$~TeV from LEP~\cite{Schael:2013ita}.
\end{itemize}
\item From dilepton production:
\begin{itemize}
\item[--] $\Lambda\geq 26.3$~TeV from ATLAS~\cite{ATLAS:2014cra},
\item[--] $\Lambda\geq 19.0$~TeV from CMS~\cite{CMS:2014aea},
\item[--] $\Lambda\geq 24.6$~TeV from LEP~\cite{Schael:2013ita}.
\end{itemize}
\end{enumerate}

Since typically these analyses check the contact four-fermion operators one by one, setting the remaining NLO vertices to zero,
the exact constraints vary from operator to operator and between production channels. Nonetheless, they all lead in general to bounds of the order $\Lambda\gsim \cO( 10$~TeV$)$. For further limits on leptonic four-fermion operators see also Ref.~\cite{Falkowski:2017pss}.

The lower bounds above refer to first and second generation four-fermion operators. Nevertheless, in recent years there
have been also some studies on four-fermion operators including top and bottom quarks (see Ref.~\cite{AguilarSaavedra:2018nen} for a wider review of these results):\footnote{For illustration, we have taken the most stringent bounds in every case. In order to ease the comparison, the four-fermion couplings $c_j$ given in Refs.~\cite{AguilarSaavedra:2018nen,Zhang:2017mls,Buckley:2015lku,Greljo:2017vvb,Isidori:2013ez,Jung:2018lfu} are related to the scale $\Lambda$ presented here through $|c_j| = 2\pi/\Lambda^2$.}
\begin{enumerate}
\item From high-energy collider studies:
\begin{itemize}
\item[--] $\Lambda\geq 1.5$~TeV from multi-top production at LHC and Tevatron~\cite{Zhang:2017mls}, 
\item[--] $\Lambda\geq 2.3$~TeV from $t$ and $t\bar{t}$ production at LHC and Tevatron~\cite{Buckley:2015lku},
\item[--] $\Lambda\geq 4.7$~TeV from dilepton production at LHC~\cite{Greljo:2017vvb}. 
\end{itemize}
\item From low-energy studies:
\begin{itemize}
\item[--] $\Lambda\geq 14.5$~TeV from $B_s-\overline{B}_s$ mixing~\cite{Isidori:2013ez}, 
\item[--] $\Lambda\geq 3.3$~TeV from semileptonic B decays~\cite{Jung:2018lfu}. 
\end{itemize}
\end{enumerate}

Thus, although pure na\"ive dimensional analysis would assign them a chiral dimension $\hat{d}=2$, dilepton and dijet studies suggest a stronger suppression of the four-fermion effective operators. This provides a strong phenomenological confirmation of our fermion power-counting in the EWET, and the assigned formal scaling $\bar\xi\Gamma\eta \sim \cO(p^2)$. An additional weak coupling is implicitly assumed to be contained in the (non-SM) couplings of the fermionic interaction Lagrangian, such that the suppression scale is much larger than $\mO(1$~TeV$)$.
In the next subsection, we will come back to the four-fermion operators within a more specific theoretical framework: the heavy vector-triplet (simplified) model.

\subsection{The Heavy Vector Triplet (simplified) model}
\label{sec:HVT}
LHC diboson ($WW$, $WZ$, $ZZ$, $Wh$ and $Zh$) production analyses~\cite{Aaboud:2016lwx,Khachatryan:2014hpa,Aad:2015owa,Aaboud:2017ahz,Aaboud:2017cxo,Sirunyan:2017wto,Aaboud:2016okv,Sirunyan:2018iff,Sirunyan:2017nrt,Sirunyan:2017acf,Liu:2018pkg,Sirunyan:2018ivv,Sirunyan:2018hsl,ATLAS:2018tpf}
confront the search for new physics from a different perspective.
Instead of looking for a smooth increase of the signal over the SM background, these studies look for narrow bumps in the diboson invariant mass spectrum. The absence of any positive signals has set lower limits on the mass of a possible $SU(2)$-triplet and colorless spin-1 resonance $V_{3}^{1}$ (see~\cite{Dorigo:2018cbl} and references therein). Nevertheless, these limits heavily depend on the resonance width and quantum numbers assumed in the analysis.

Particular models and benchmark points are considered as a
common basis to interpret the experimental data, among them the spin-1 Heavy Vector Triplet model B (HVT-B) as one of the most popular ones~\cite{Pappadopulo:2014qza}.
The HVT-A variant has a coupling to fermions similar to the HVT-B one, but a much smaller resonance coupling to dibosons~\cite{Pappadopulo:2014qza}. In general, HVT-A always leads to looser exclusion bounds and, hence, we will focus on the HVT-B scenario in what follows.
In the narrow-width approximation the diboson production cross section
is given by
\bear
\sigma(pp\to V\to {\rm diboson}) &=& \sum_{i,j\,\in\, p}\,
\Frac{48\pi^2 \gamma_{ij} }{(2 S_i+1)(2S_j+1) C_i C_j}
\; \Frac{dL_{ij}}{d\hat{s}}\bigg|_{\hat{s}=M_V^2} \, ,
\eear
with $i,j\in p=\{q,\bar{q}',W,Z,\dots\}$, $\frac{dL_{ij}}{d\hat{s}}$
the corresponding parton luminosity, $S_{k}$ and $C_k$
the spin and color factor of the parton $k$, respectively, while
\be
\gamma_{ij} \, =\, \Frac{\Gamma_{V\to ij}}{M_V} \,
\times \, \mB_{V\to {\rm dibos}}
\ee
incorporates the partial width $\Gamma_{V\to ij}$ into the partons $ij$, the resonance mass $M_V$ and its branching ratio $\mB_{V\to {\rm dibos}}$
into the diboson final state of the particular experimental data at hand
(diboson $\in\{ W^+W^-, Zh\}$ or $\{W^\pm Z, W^\pm h\}$ for $V^0$ and $V^\pm$ production, respectively).
In the HVT-B model, the branching ratios into two weak bosons and into a Higgs and a gauge boson are roughly 50\% each, the branching ratios into fermions being less than 1\%.

It is important to remark that experimental diboson studies implicitly assume that Drell-Yan ($q\bar{q}'\to V \to $ diboson) is the dominant $V$ production channel. Most works rely on the HVT benchmark
for the interpretation of vector triplet signals~\cite{Dorigo:2018cbl}.
Although the HVT-B variant has a fermionic branching ratio two orders of magnitude smaller than the bosonic one, the $q\bar{q}'$ luminosity is from four to six orders of magnitude larger than the diboson one, strongly suppressing the vector-boson-fusion production channel even with an enhanced partial width~\cite{Pappadopulo:2014qza}.
Thus, unless the BSM light-quark couplings are extremely suppressed,
$\mB(V\to q\bar{q}')\ll 10^{-4}$,
Drell-Yan  production will dominate:
\begin{equation}
\sigma(pp\to V\to {\rm diboson}) \,\simeq \,
\sum_{q,\bar{q}'}\, \Frac{48\pi^2 \gamma_{q\bar{q}'} }{4 N_C^2}
\; \Frac{dL_{q,\bar{q}'}}{d\hat{s}}\bigg|_{\hat{s}=M_V^2} \, ,
\end{equation}
with $q$ and $\bar{q}'$ summed over the different flavors ($u,d,\dots$).
Works relying on vector-boson-fusion production for a vector triplet lead
to much smaller cross sections~\cite{Delgado:2017cls} than those usually predicted with the HVT benchmark~\cite{Pappadopulo:2014qza,Dorigo:2018cbl}.

In the HVT model, the partial decay width of the vector resonance into quarks is given by
$\Gamma_{V^0 \to q\bar{q}}/M_V
= (g^2 g_V^{-1} c_F)^2 \times N_C/(96\pi)$
with $g^2= 4 m_W^2/v^2$ and $c_F\simeq 1$
(for a spin-1 triplet, $\Gamma_{V^\pm \to q\bar{q}'}=2\Gamma_{V^0 \to q\bar{q}}$)~\cite{Pappadopulo:2014qza}.
In the present article, we consider a much more general fermionic structure where the triplet partial width is related to the resonance couplings $c^{\hat{V}_3^1}$,
$\widetilde{c}^{\hat{V}_3^1}$ and $C_0^{V_3^1}$ through
\bear
c_{\rm eff}^2\,\,\equiv \,\, (c^{\hat{V}_3^1})^2 +(\widetilde{c}^{\hat{V}_3^1})^2
 + \frac{1}{2}(C_0^{V_3^1})^2  &=&
\Frac{ 24\pi}{N_C}\, \Frac{ \gamma_{q\bar{q}} }{\mB_{V^0\to{\rm dibos.}}}
\,\,\, \geq\,\,\,  \Frac{ 24\pi}{N_C}\, \gamma_{q\bar{q}}
\,\,\equiv\,\, (c_{\rm eff}^{\rm bound})^2
\, .
\label{eq:R-coupling-bound}
\eear
The identities $c_{\rm eff} = \frac{1}{2}\, g^2 g_V^{-1} c_F$ and $(c_{\rm eff}^{\rm bound})^2 =c_{\rm eff}^2 \times \mB_{V^0\to{\rm dibos.}}$ provide a simple relation between the two theoretical frameworks.
We note that $\gamma_{q\bar{q}}$ here refers to the neutral $V^0$ production.

The experimental analyses consider benchmark points with constant $\gamma_{q\bar{q}}$. In particular, HVT-B$_{g_V=3}$ studies have $\gamma_{q\bar{q}}= 1.0\times 10^{-4}$
(together with $c_F\simeq 1$ and $\mB_{V\to{\rm dibos.}}\simeq 50\%$).
The $V$ mass is varied and values
that yield a larger cross section than the experimentally observed
one become excluded. In general, the HVT-B$_{g_V=3}$
analyses have excluded masses $M_V\leq  M_V^{\rm bound}$ at the 95\% CL~\cite{ATLAS:2018tpf}: the most stringent bound, $M_V^{\rm bound}=4.15$~TeV, has been given by the recent $W^+W^-/W^\pm Z/ZZ$
production analysis~\cite{ATLAS:2018tpf}, where the diboson system was reconstructed
using large-radius jets.
At constant $\gamma_{q\bar{q}}$,
lower resonance masses yield larger production cross sections and are excluded, whereas larger values of $\gamma_{q\bar{q}}$ give
higher cross sections for fixed $M_V$ and are, therefore, also excluded.
This implies the 95\% CL exclusion limit $c_{\rm eff}
\geq c_{\rm eff}^{\rm bound}=5.0\times 10^{-2}$, for $M_V\leq M_V^{\rm bound}=4.15$~TeV~\cite{ATLAS:2018tpf},
on the resonance coupling combination provided by Eq.~(\ref{eq:R-coupling-bound}).
Note that, for a given value of $\gamma_{q\bar{q}}$ (which at the end encodes all the new physics in standard diboson analyses),
the limit on this $c_{\rm eff}$ combination becomes stronger if one assumes a model with particular values for the diboson branching
ratio. {\it E.g.}, in the HVT-B model, $\mB_{V\to{\rm dibos}}= 50\%$ would actually imply $c_{\rm eff}\geq \sqrt{2} \, c_{\rm eff}^{\rm bound}$.
These resonance couplings determine the $V_3^1$ contribution to the low-energy four-fermion LECs through the relations (see Table~\ref{tab:result:4ferm.even}):
\begin{equation}
\mF_7^{\psi^4} +\mF_8^{\psi^4} + \frac{\mF_{10}^{\psi^4}}{4} \,\,=\,\,
-\Frac{1}{2}\left( \mF_5^{\psi^4} +\mF_6^{\psi^4} + \frac{\mF_9^{\psi^4}}{4}\right)
\,\, =\,\, \Frac{ c_{\rm eff}^2 }{4 M_V^2}
\,\,=\,\,  \Frac{6\pi \Gamma_{V^0\to q\bar{q}} }{N_C M_V^3}
\, .
\label{eq:LEC-bound}
\end{equation}
Analogous expressions have been derived in the past for $\cO(p^4)$ LECs in Chiral Perturbation Theory, in terms of the ratio of the vector triplet partial width
and its mass~\cite{Guo:2007ff,Ledwig:2014cla}. Those results are based on
the same basic assumption we employ: an appropriate behavior
of the amplitudes at high energies, dictated by unitarity
(in addition to the standard requirements of analyticity
and crossing of a well-defined field theory).

In order to ease the comparison with four-fermion interaction studies
at LHC~\cite{Aaboud:2017yvp,Sirunyan:2017ygf,ATLAS:2014cra,CMS:2014aea}
and LEP~\cite{Schael:2013ita}, it is convenient to rewrite the previous combination of LECs in terms of the $\Lambda$ scale suppression,
\bear
2\pi \Lambda^{-2} \,\,\equiv \,\, \mF_7^{\psi^4}+\mF_8^{\psi^4}+\frac{\mF_{10}^{\psi^4}}{4}
\qquad \Longrightarrow \qquad \Lambda^2\,=\, \Frac{ 8\pi  M_V^2}{c_{\rm eff}^2}
\, .
\label{eq:Lambda-scale}
\eear
%

\begin{figure}[!t]
\centering
\includegraphics[scale=1]{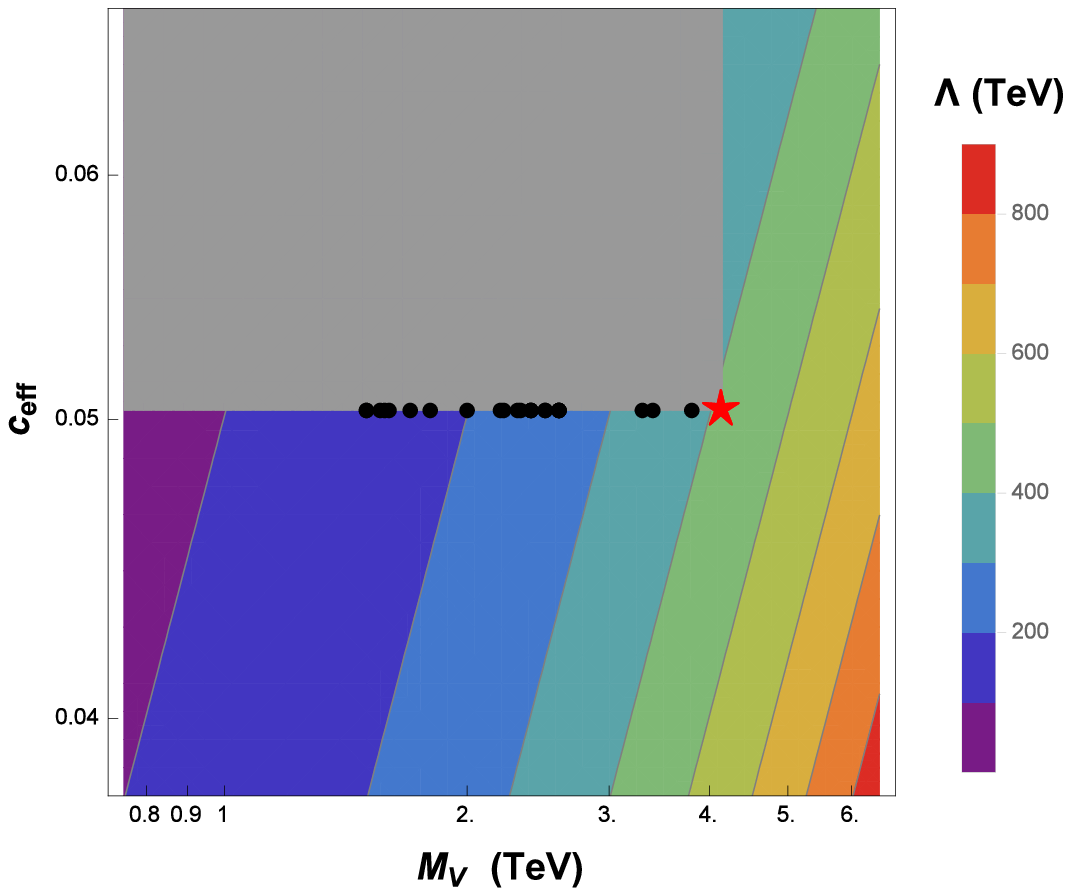}
\caption{{\small
Exclusion regions based on diboson production experimental analyses~\cite{Aaboud:2016lwx,Khachatryan:2014hpa,Aad:2015owa,Aaboud:2017ahz,Aaboud:2017cxo,Sirunyan:2017wto,Aaboud:2016okv,Sirunyan:2018iff,Sirunyan:2017nrt,Sirunyan:2017acf,Liu:2018pkg,Sirunyan:2018ivv,Sirunyan:2018hsl,ATLAS:2018tpf,Dorigo:2018cbl} (shaded gray region). The black dots represent the values $(M_V^{\rm bound},c_{\rm eff}^{\rm bound})$
obtained in the various HVT-B$_{g_V=3}$ studies, representing the red star the most stringent constraint $M_V^{\rm bound}=4.15$~TeV~\cite{ATLAS:2018tpf}. The relevant LEC combination is expressed in terms of the scale $\Lambda$ defined by $\mF_7+\mF_8+\mF_{10}/4 \,= c_{\rm eff}^2/(4 M_V^2) = 2\pi \Lambda^{-2}$, with the value for each point given by the color legend on the right-hand side. More details are given in the text.}}\label{fig:exclusion}
\end{figure}

Each particular $\gamma_{q\bar{q}}$ benchmark-point study leads to the exclusion of vector triplet resonances with $c_{\rm eff}\geq c_{\rm eff}^{\rm bound}$ if $M_V\leq M_V^{\rm bound}$.
The black dots in Fig.~\ref{fig:exclusion} provide the values $(M_V^{\rm bound},c_{\rm eff}^{\rm bound})$ for each HVT-B$_{g_V=3}$ analysis~\cite{Aaboud:2016lwx,Khachatryan:2014hpa,Aad:2015owa,Aaboud:2017ahz,Aaboud:2017cxo,Sirunyan:2017wto,Aaboud:2016okv,Sirunyan:2018iff,Sirunyan:2017nrt,Sirunyan:2017acf,Liu:2018pkg,Sirunyan:2018ivv,Sirunyan:2018hsl,ATLAS:2018tpf,Dorigo:2018cbl}. The most stringent exclusion bound~\cite{Sirunyan:2017acf} is provided by the red star in Fig.~\ref{fig:exclusion}
$(M_V^{\rm bound}=4.15$~TeV$,c_{\rm eff}^{\rm bound}= 5.0\times 10^{-2} )$, with $\gamma_{q\bar{q}}=10^{-4}$.
It yields the widest 95\% CL exclusion region, shown in Fig.~\ref{fig:exclusion} as the upper-left shaded gray region. At this red-star point, the LEC scale suppression is found to be $\Lambda= 410$~TeV.

Thus, these standard diboson searches lead to an important extension of the $(M_V,c_{\rm eff})$ exclusion regions listed in the previous section~\ref{sec:EWET-pheno}, derived through effective four-fermion operator analyses which typically excluded $(M_V,c_{\rm eff})$ points with $\Lambda\lsim 10$~TeV. These points would be in the
upper-left corner of Fig.~\ref{fig:exclusion}, being, indeed, out of the range of the plot presented here.

Since the spin-1 triplet studies have been mainly focused on just two $\gamma_{q\bar{q}}$ benchmark points (corresponding to HVT-A$_{g_V=1}$ and HVT-B$_{g_V=3}$), there are still several exclusion regions unexplored: smaller values of $\Lambda$ are still allowed for
1) a light spin-1 resonance if $\gamma_{q\bar{q}}$  (or $c_{\rm eff}$) is small enough, and
2) a large resonance parameter $\gamma_{q\bar{q}}$ (or $c_{\rm eff}$) if $M_V$ is large enough.

It is remarkable that direct resonance searches ---based on specific benchmarks--- lead to the exclusion of $(M_V,c_{\rm eff})$ values with a $\Lambda$ which is one order of magnitude more stringent than dedicated four-fermion studies, leading to the shaded gray rectangular exclusion region in Fig.~\ref{fig:exclusion}.
However, these diboson searches are not truly exhaustive for $\Lambda$.
Four-fermion analyses at LHC and LEP lead to looser bounds, but an exclusion limit is given by a constant $\Lambda$ that corresponds to a diagonal line in Fig.~\ref{fig:exclusion}; all points on the top or left of this line are excluded. Therefore, although these experimental analyses do not reach such strong bounds as the diboson ones, they cover ranges of the resonance parameter space that scape to the latter:
bounds from contact interaction studies serve to exclude values of $(M_V,c_{\rm eff})$ that are not tested by diboson searches, providing
complementary information.
 Thus, a reanalysis of these diboson studies for different values of $\gamma_{q\bar{q}}=N_C (c_{\rm eff}^{\rm bound})^2 /(24\pi)$ is advised,
as it would definitely enlarge the exclusion range shown in Fig.~\ref{fig:exclusion}.

The study in this Subsection only refers to the available diboson analyses performed by other groups,
which considered a narrow width approximation. In order to extract information on our LECs from those
works~\cite{Aaboud:2016lwx,Khachatryan:2014hpa,Aad:2015owa,Aaboud:2017ahz,Aaboud:2017cxo,Sirunyan:2017wto,Aaboud:2016okv,Sirunyan:2018iff,Sirunyan:2017nrt,Sirunyan:2017acf,Liu:2018pkg,Sirunyan:2018ivv,Sirunyan:2018hsl,ATLAS:2018tpf,Dorigo:2018cbl,Pappadopulo:2014qza},
we followed the approximations used therein.
These assumptions agree with those in the present work,
as we are neglecting one-loop corrections and,
in particular, the impact from resonance widths
in the intermediate propagators.
In the range of resonance couplings discussed in~\cite{Aaboud:2016lwx,Khachatryan:2014hpa,Aad:2015owa,Aaboud:2017ahz,Aaboud:2017cxo,Sirunyan:2017wto,Aaboud:2016okv,Sirunyan:2018iff,Sirunyan:2017nrt,Sirunyan:2017acf,Liu:2018pkg,Sirunyan:2018ivv,Sirunyan:2018hsl,ATLAS:2018tpf,Dorigo:2018cbl,Pappadopulo:2014qza},
resonance widths are small enough to safely use the narrow-width approximation.
If (much) larger resonance couplings than those depicted in Fig.~\ref{fig:exclusion} are studied in future collider
analyses, finite-width corrections should be properly accounted and loop effects must be properly incorporated.

\subsection{Composite Higgs models}
Composite Higgs models (CHMs)~\cite{Kaplan:1983fs,Kaplan:1983sm,Georgi:1984ef,Georgi:1984af,Dugan:1984hq,Agashe:2004rs,Contino:2006qr,Marzocca:2012zn,Ferretti:2014qta,Barnard:2013zea,Ferretti:2013kya,Cacciapaglia:2014uja,Vecchi:2015fma,Belyaev:2016ftv,Liu:2016idz} are appealing UV completions,
able to accommodate a light Higgs particle. The Higgs arises as a pseudo-Nambu-Goldstone boson (pNGB)  (similarly to the QCD pions)
and is, therefore, naturally lighter than the other new-physics states. The spectrum of the heavier resonances, however, is model dependent. Nevertheless, a broad class of these models can be described by the electroweak resonance theory and the EWET, as we outline below.
Subsets of resonance states have been analyzed previously in the context of CHMs~\cite{Vignaroli:2011um,Contino:2011np,Marzocca:2012zn,Belyaev:2016ftv,Buschmann:2017ucg,Yepes:2017pjr,Yepes:2018dlw}, as well as compositeness of the electroweak gauge bosons~\cite{Liu:2016idz}. There are also lattice studies \cite{DeGrand:2015zxa,Appelquist:2016viq,Bennett:2017kga,Holdom:2017wpj,Ayyar:2017qdf} that bring a deeper understanding on this type of scenarios.

The minimal incarnation of the CHMs is the Minimal Composite Higgs Model (MCHM)~\cite{Agashe:2004rs,Contino:2006qr,Contino:2010rs}. It is based on the coset $SO(5)/SO(4)$ and therefore implies four real Goldstone Bosons.
Even though they transform as a fundamental representation of the $SO(4)$, {\it i.e.}, there is a $SU(2)_{L}$ Higgs doublet $\Phi$ in the spectrum, it is betterdescribed by the EWET. The reason is that this coset construction,
reflecting the strongly-coupled UV completion,
induces in the leading Lagrangian~\eqref{eq:L.LO}
a function of the form $\mathcal{F}_{u} = f^{2}/v^{2} \sin^{2}{(\theta+(h/f))}$,
with $f$ the $SO(5)/SO(4)$ spontaneous symmetry breaking scale and
$\theta = \arcsin{(v/f)}$ the misalignment angle ~\cite{Contino:2011np}.

As pointed out by~\cite{Alonso:2016btr,Alonso:2016oah}, sometimes a non-linearly realized EWET
can be rewritten in terms of a linear SMEFT realization.
This is the case of the MCHM~\cite{Agashe:2004rs,Contino:2011np,Alonso:2016btr,Alonso:2016oah}:
the non-linear transformation $h/f=\arcsin(\sqrt{2 \Phi^\dagger \Phi}/f) - \theta$
yields an equivalent SMEFT scalar Lagrangian
$\Delta\mL= D_\mu\Phi^\dagger D^\mu \Phi
- \frac{c_H}{\Lambda^2} |\Phi|^2 \Box |\Phi|^2- V(|\Phi|)$
up to NLO in the $1/\Lambda^2$ expansion, with
$c_H/\Lambda^2=1/(2 f^2)$ and the expectation value
$\bra \Phi^\dagger\Phi\ket =v^2/2$~\cite{Alonso:2016btr,Alonso:2016oah,Sanz-Cillero:2017jhb}.
Note, however, that this SMEFT rearrangement  is linked to a $v^2/f^2$
expansion in the observables. SMEFT perturbation theory may then eventually
break down (or poorly converge) for small enough values of $f$, while these potentially
large corrections are always well accounted in the EWET approach as far as energies
remain below the lightest resonance mass $M_R$. In general, this scale $f$ is not
the cut-off (the lightest resonance mass) but rather the decay constant,
which can be significantly smaller than $M_R$. Notice that current $\cO(10\%)$ Higgs coupling
experimental uncertainties still allow for MCHM values of $f$ as low as 0.6~TeV
based on the relation $g_{hWW}/g_{hWW}^{\rm SM}=\sqrt{1-v^2/f^2}$~\cite{deBlas:2018tjm,Contino:2010rs}.
The $\mathcal{F}_{u}$ operator of the $SO(5)/SO(4)$ MCHM Lagrangian,
 as well as other operators,
are then best written in the non-linear EWET (see \cite{Buchalla:2014eca}
for an explicit matching of the operators). 
On the other hand, when $v\ll f$, the SMEFT representation provides a more economic and efficient approach.

The MCHM has a non-renormalizable holographic (partial) UV-completion
in five spacetime dimensions~\cite{Contino:2010rs}. There are other
CHMs beyond the minimal coset that
have four-dimensional UV completions~\cite{Cacciapaglia:2014uja} based on strong dynamics.

This underlying strong dynamics is inspired by QCD, but cannot be a scaled-up version of it. A ``wishlist'' of required properties for phenomenologically-viable models was compiled in Ref.~\cite{Ferretti:2013kya}.
They should have a simple, asymptotically-free hypercolor gauge group; a global symmetry group that contains the SM custodial and gauge groups; a pNGB state that has the quantum numbers of the SM Higgs; as well as fermionic bound states (``baryons'') with the right quantum numbers to couple to the top quark. The latter, called top partners, are usually assumed to generate the fermion masses {\it via} the partial compositeness paradigm~\cite{Kaplan:1991dc}. Within partial compositeness, the top partner couples linearly to the top quark, generating its mass through mixing effects. This mixing introduces factors of weak couplings, in complete agreement with our EWET power counting.

Since only the Higgs particle has been observed so far, the other new-physics states are assumed to have masses above the electroweak scale, and our resonance formalism can be used to describe them, especially given the strongly-coupled nature of the underlying theory. There are various effects that contribute to the masses of the pNGBs in these scenarios~\cite{Ferretti:2016upr}: loops of the SM gauge bosons (the SM is a subgroup of the global symmetry group), loops of the top quark, and the masses of the hyperfermions. The explicit spectrum of the resonances is of course model-dependent, however, there are some common features that are shared by all these models~\cite{Cacciapaglia:2015eqa,Belyaev:2016ftv}.
\begin{itemize}
\item Usually, having both, a Higgs candidate and a top partner in the low-energy spectrum, requires more than one irreducible representation of hyperfermions in the UV~\cite{Ferretti:2013kya,Ferretti:2016upr}. This induces $U(1)$ hyperfermion number symmetries in the UV that give rise to pseudoscalars singlet Goldstone Bosons at in the low-energy spectrum. One linear combination of them is anomalous and gets a large mass (similar to the $\eta'$ in QCD), while an orthogonal linear combination gets its mass from the explicit symmetry breaking by the underlying hyperfermion masses. Generic couplings of these states can be described with $P_{1}^{1}$ in our notation.
\item The need for a top partner generally induces colored pNGBs. Among them, there is always a real color octet~\cite{Belyaev:2016ftv}. Its mass comes from the explicit symmetry breaking of the hyperfermion masses, as well as from QCD effects. It is expected to be slightly heavier than the singlet pseudoscalar and can be described by $P_{1}^{8}$.
\item The top partners get their masses from the hyperfermion masses as well as the dynamical symmetry breaking of the condensate. Their mass would then be around the mass of the pseudoscalar octet. They correspond to our fermionic resonance $\Psi$.
\item Vector and axial-vector resonances would get their masses from dynamical and explicit symmetry breaking. Depending on the UV structure, there might also be the possibility for a UV-symmetry invariant mass term, making them heavier than the other states. These states can be described by our  $V_{3}^{1}, A_{3}^{1}, A_{1}^{1}$ and $V_{1}^{8}$.
\end{itemize}
We have already discussed dedicated searches for heavy vector triplets, within composite Higgs models or from other origins, in the previous subsection.

\subsection{Weakly-coupled scenarios}

The discussion presented so far is not limited to strongly-coupled UV completions.
 In weakly-coupled scenarios, the SMEFT linear realization framework tends to be more economic from the practical point of view.
Still, the non-linear EWET approach can be also safely applied, even if it usually leads to longer and more tedious computations in these cases.
Thus, we want to remark that the weakly-coupled models studied in this Subsection are not the main motivation for this work.
Their discussion just shows how general the present approach is:
it can be applied without any {\it a priori} consideration on the strength of the underlying interactions,
avoiding any unwanted bias in the phenomenological analysis.

There are many renormalizable, weakly-coupled models of new physics with additional heavy fields that have the same quantum numbers as we consider for the resonances. Writing a phenomenological Lagrangian, linear in those heavy fields, and integrating them out to study their low-energy implications \cite{Jiang:2016czg,delAguila:2000rc,delAguila:2008pw,delAguila:2010mx,deBlas:2014mba,deBlas:2017xtg} is completely analogous to our formal study within the non-linear formalism. The closest analysis to our approach is Ref.~\cite{deBlas:2017xtg}, which is based on a series of previous works \cite{delAguila:2000rc,delAguila:2008pw,delAguila:2010mx,deBlas:2014mba}.

Obviously, for those UV completions incorporating an $SU(2)_L$ Higgs doublet, the linear SMEFT provides a more efficient low-energy description. Nevertheless, any model of this type can always be written in non-linear notation, using field redefinitions. The simplest example is provided by the SM itself. Combining the four scalar fields contained in the SM Higgs doublet $\Phi$ into the $2\times 2$ matrix \cite{Appelquist:1980vg}
\bel{eq:Sigma_def}
\Sigma(x) \,\equiv\, (\Phi^c,\Phi)\, =\,  \left( \bat
\Phi^{0*} & \Phi^+ \\ -\Phi^- & \Phi^0 \ea\right)
\, =\, \frac{1}{\sqrt{2}}\, [v + h(x)]\; U(\varphi(x))\, ,
\ee
makes explicit the invariance of the SM scalar Lagrangian under global $SU(2)_L\otimes SU(2)_R$ transformations:
\bel{eq:Sigma_transf}
\Sigma\;\to\; g_L^{\phantom{\dagger}}\,\Sigma\, g_R^\dagger\, ,
\qquad\qquad
g_{L,R}\,\in\, SU(2)_{L,R}\, ,
\ee
and its spontaneous breaking to the custodial $SU(2)_{L+R}$ subgroup \cite{Sikivie:1980hm}. Moreover, writing the SM Lagrangian in terms of $h(x)$ and the matrix $U(\varphi)$ that parametrizes the electroweak Nambu-Goldstone fields, one trivially obtains the non-linear EWET Lagrangian with specific values for all its couplings~\cite{Pich:2018ltt,Pich:1998xt}.
In particular, $c^{(V)}_{n>4}$, $c^{(u)}_{n>2}$, $\hat Y^{(n\ge 2)}_\psi$ and all LECs of $\mO(p^{n>2})$ turn out to be identically zero, reflecting the renormalizability of the SM Lagrangian, implied by the $SU(2)_L$ doublet structure of the Higgs multiplet.

When studying the effects of massive states, from weakly-coupled UV completions, on the linear SMEFT, the heavy fields usually transform in a
well-defined way under the symmetry group
$SU(3)_{C}\otimes SU(2)_{L}\otimes U(1)_{Y}$ (see \cite{Dawson:2017vgm} for a classification in the scalar sector). Nevertheless, using the coset representative $u(\varphi)$, they can always be rewritten in terms of fields with definite transformation properties under the custodial subgroup $SU(2)_{L+R}$, such as those indicated in Eq.~\eqn{eq.R-transform}.

Since our analysis only contains terms linear in the heavy fields and operators with more than one resonance lead to higher-order contributions, only singlets and doublets of $SU(2)_{L}$ could contribute to the LECs at $\mO(p^4)$.
Comparing our EWET results with the SMEFT approach, one can na\"\i vely identify the list of generated dimension six operators and the list of operators with chiral dimension four, when the tensors $\chi_{R}^{(i)}$ have canonical dimension three and chiral dimension two.
This is the case for the fermionic bilinears $J_{\Gamma}$, but not necessarily for other $\chi_{R}^{(i)}$ that involve the Higgs or the Goldstone bosons.
This, however, precisely reflects the consequences of a possible strongly-coupled nature of the electroweak symmetry breaking. Objects of chiral dimension two, for example $\langle u_{\mu}u_{\nu}\rangle_{2}$ or $(\partial_{\mu} h) u_{\nu}$, translate to objects with canonical dimension four or more, indicating a further suppression in the weakly-coupled case.\footnote{In the same way, some bosonic $\cO(p^4)$ EWET operators like, {\it e.g.}, $\bra u_\mu u^\mu\ket_2\, \bra u_\nu u^\nu\ket_2$, translate into objects with canonical dimension eight or higher, indicating a similar additional suppression in the case of weakly-coupled theories.}

As an explicit example, let us consider a simple enlargement of the SM scalar sector with a color-octet and $SU(2)_L$-doublet scalar multiplet $\mathcal{S}$ of hypercharge $Y=\frac{1}{2}$. The scalar field $\mathcal{S}$ and its charge conjugate $\mathcal{S}^c = i \sigma_2 \mathcal{S}^*$ can be combined together into a bidoublet $\Xi \equiv \left( \mathcal{S}^c ,\mathcal{S}\right)$, transforming as
$g_L^{\phantom{\dagger}}\,\Xi \, g_R^\dagger$, which allows one to build easily the most general, renormalizable scalar potential
$V(\Sigma, \Xi )$, invariant under global $SU(2)_L\otimes SU(2)_R$ transformations \cite{Manohar:2006ga}. Imposing also custodial symmetry on the fermionic couplings of the color-octet scalar multiplet ({\it i.e.}, $y_u=y_d\equiv y$), its Yukawa interactions can be written in the form
\begin{equation}
\label{eq:XiYukawa}
\mL_Y^8\, =\, - y\;\left(  \bar q_L \mathcal{S}^c u_R +  \bar q_L \mathcal{S} d_R \right) + \mathrm{h.c}.
\, =\, -y\; \bar q_L \Xi q_R + \mathrm{h.c}.
\, =\, -y\; \bar\xi_L u^\dagger \Xi u^\dagger \xi_R + \mathrm{h.c}. \,,
\end{equation}
which makes the connection with our $S^8_1$ and $P^8_3$ fields clear:
\begin{equation}
\label{eq:Xi.to.SP}
u^\dagger \Xi u^\dagger\, =\, \frac{1}{\sqrt{2}}\, S^8_1 + i \, P^8_3\,,
\end{equation}
where the new physics interactions are assumed to be invariant under $\Xi\stackrel{CP}{\longrightarrow} \Xi'$ in the present EWET construction,
leading to the relation $y\Xi'=y^*\Xi^\dagger$. Since we are further considering that the color octet $\Xi$ can be decomposed in a combination of $CP$ eigenstates,
there are in fact two possible transformations for $\Xi$:
Eq.~\eqref{eq:Xi.to.SP} provides the case  $\Xi\stackrel{CP}{\longrightarrow} \Xi'= \,+\, \Xi^\dagger$
and implies a real value
for $y$.\footnote{There is a second case, with $\Xi\stackrel{CP}{\longrightarrow} \Xi'= \,-\, \Xi^\dagger$. The coupling constant $y$ is then purely imaginary and Eq.~\eqref{eq:Xi.to.SP} is now modified to $u^\dagger \Xi u^\dagger = \frac{1}{\sqrt{2}} P^8_1 - i  S^8_3$. We have explicitly checked the complete agreement between our LEC predictions and the results in Refs.~\cite{deBlas:2017xtg,deBlas:2014mba} for this assignment, although, for illustration, we just provide the outcomes for the $\Xi\stackrel{CP}{\longrightarrow} \Xi'= \,+\, \Xi^\dagger$ case.}
Note, however, that other works~\cite{Manohar:2006ga,deBlas:2017xtg,deBlas:2014mba} study the general complex $y$ case and its impact on $CP$ violation.
The Yukawa interaction of Eq.~\eqref{eq:XiYukawa} contributes to our single-resonance Lagrangian of Eq.~\eqref{eq:L0}, in the following way:
\begin{equation}
\label{eq:MW:1}
\mL_Y^8\, =\, -\sqrt{2} y\; \langle S_{1}^{8}\langle J^{8}_{S}\rangle_{2}\rangle_{3}  -   2  y\; \langle P^{8}_{3} J^{8}_{P}\rangle_{2,3}\, ,
\end{equation}
which determines the octet-scalar couplings,
\begin{equation}
\label{eq:MW:2}
c^{S^8_1}\, =\, c^{P^8_3}\, =\,  - 2y   \, .
\end{equation}
Assuming custodial symmetry, the electroweak doublet structure of the $J=0$ multiplet $\Xi$ implies identical couplings for its $SU(2)_{L+R}$ triplet and singlet components, as well as $M_{S^8_1}=M_{P^8_3}=M_{\Xi}$. Inserting these values in Table~\ref{tab:result:4ferm.even}, one gets the low-energy structure generated by $\Xi$ exchange:
\begin{equation}
\label{eq:MW-results2}
  -\mF_1^{\psi^4}\,=\,- 3 \mF_2^{\psi^4}\,=\, 3 \mF_3^{\psi^4}\,=\,-3\mF_4^{\psi^4} \, =\,- \mF_7^{\psi^4}\, =\, \mF_8^{\psi^4}\, =\, -2 \mF_9^{\psi^4} \,=\, 2 \mF_{10}^{\psi^4}\, =\,  \frac{y^2}{16 M_{\Xi}^2}\, ,
\end{equation}
where the contributions to the remaining LECs vanish. This result is also in agreement with those compiled in Refs.~\cite{deBlas:2017xtg,deBlas:2014mba} ($\mathcal{S}$ is denoted as $\Phi$ therein).

The custodial-breaking parts of the Yukawa interactions could be analyzed in a similar way, in terms of the spurion $\mT$. However, they give rise to higher-order structures, not included in our analysis, because of the additional power suppression implicitly carried by $\mT$.
Owing to the color-octet nature of the $\Xi$ field, the scalar potential does not contain interactions linear in $\Xi$ and, therefore, cannot contribute either at $\mO(p^4)$.

The collider constraints on the scalar-octet $\mathcal{S}$ have been critically re-analyzed in Ref.~\cite{Hayreter:2017wra}, in the context of the Manohar-Wise model~\cite{Manohar:2006ga} that incorporates the three fermion generations with flavor-aligned Yukawa interactions \cite{Pich:2009sp}.
Although mass limits around 1~TeV are obtained in the most sensitive regions of parameter space, lower scalar masses cannot be yet excluded with current data. The existence of such additional colored particles would also alter the running of $\alpha_{s}$ above their mass scale, with different implications \cite{Sannino:2015sel}. The CMS collaboration has measured the running up to $Q\approx 1.4~\text{TeV}$\cite{CMS:2014mna,Khachatryan:2016mlc}, finding consistency within errors with the SM expectations.

 \section{Conclusions}
 \label{sec:concl}

The EWET is the most general EFT framework incorporating the known particle states and the SM symmetries. It is based on the successful pattern of electroweak symmetry breaking $SU(3)_C\otimes SU(2)_L\otimes SU(2)_R\otimes U(1)_X \to SU(3)_C\otimes SU(2)_{L+R}\otimes U(1)_X$, with $X=(\mathrm{B}-\mathrm{L})/2$, and it does not make any assumption concerning the electroweak transformation properties of the recently-discovered Higgs particle. The Higgs boson is parametrized as a light neutral scalar $h$, singlet under the electroweak group, and a non-linear realization of the Nambu-Goldstone bosons is adopted without any `a priori' relation with the Higgs field. The EWET contains all operators allowed by the assumed symmetries and field content, organized according to their infrared behavior in an expansion in powers of derivatives over some high-energy scale.

In the absence of any experimental evidence for additional non-SM particles at the electroweak scale, EFT methods are the appropriate tool to investigate the existence of hypothetical heavy states above the energies currently explored. Since the scale of new physics appears to be separated from the electroweak scale by a wide energy gap, the only accessible information about the heavy fields is encoded in the imprints they leave on the LECs of the EWET.

In this paper, we have built an effective Lagrangian that couples the light particles to generic heavy states. We have generalized the formalism developed in Refs.~\cite{Pich:2016lew,Pich:2015kwa} in order to incorporate colored resonances, including also fermionic states that were not considered in our previous works. Integrating out the heavy scales, we have determined at the lowest non-trivial order the complete pattern of EWET LECs generated by different types of resonances. We have considered bosonic states with $J^P=0^\pm$ and $J^P=1^\pm$, in singlet or triplet $SU(2)_{L+R}$ representations and in singlet or octet representations of $SU(3)_C$, and fermionic resonances with $J=\frac{1}{2}$ that are electroweak doublets and color triplets or singlets.

Our results, summarized in Tables~\ref{tab:result:bos}, \ref{tab:result:2ferm.even}, \ref{tab:result:4ferm.even} and \ref{tab:result:ferm.odd},
exhibit a rich set of low-energy signals with different types of resonances potentially contributing to the same LECs. Experimental evidence for a non-zero value of some particular coupling would certainly select a set of possible resonance quantum numbers. However, one would need a clear pattern of measured LECs in order to neatly identify the culprit state through its low-energy fingerprints. Whenever the needed data would be available, the comprehensive information contained in these tables will prove very useful to infer the kind of high-scale physics responsible for any observed anomalies.

Given the limitations of current data samples, the experimental analyses are usually performed within the context of benchmark models or simplified EFT-like approaches. We have illustrated with a few examples how our general EFT formalism can be easily particularized to these simpler scenarios and the corresponding model-dependent bounds worked out. While pragmatic phenomenological analyses are unavoidable, their implicit assumptions should be carefully scrutinized once any new-physics signal arises. The general formalism developed here provides the necessary tools to confront the data information in a model-independent way.

\section*{Acknowledgements}
This work has been supported in part by the Spanish National Research Agency (AEI) and ERDF funds from the European Commission (FPA2014-53631-C2-1-P, FPA2016-75654-C2-1-P, FPA2017-84445-P); by the Spanish Centro de Excelencia Severo Ochoa Program (SEV-2014-0398); by the Generalitat Valenciana (PROMETEO/2017/053); by the Universidad Cardenal Herrera-CEU (INDI16/10 and INDI17/11); by La Caixa (Ph.D. grant for Spanish universities); and by the STSM Grant from COST Action CA16108.
C.K. acknowledges the support of the Alexander von Humboldt Foundation. This manuscript has been authored by Fermi Research Alliance, LLC under Contract No. DE-AC02-07CH11359 with the U.S. Department of Energy, Office of Science, Office of High Energy Physics.

\appendix
\section{The Electroweak Chiral Dictionary}
\label{sec:dic}

As a cross-check of the $\mO(p^4)$ operator basis listed in Tables \ref{tab:P-even-Op4} and \ref{tab:P-odd-Op4}, we compare it to the different basis chosen in Ref.~\cite{Buchalla:2013rka}. For future reference, we provide the necessary translation dictionary here. When more than one generation of quarks, or quarks and leptons, are present simultaneously, the dictionary changes slightly.
We comment on this at the end of this section. To increase the readability, we refer to the basis adopted in this paper and in Ref.~\cite{Pich:2016lew} as VLC basis and to the basis used in Refs.~\cite{Buchalla:2012qq,Buchalla:2013rka,Krause:2016uhw}
as MUC basis.\footnote{We use the IATA airport codes of the cities as abbreviations.}

The main difference between the MUC and VLC bases is the choice of group variables to represent the electroweak Nambu-Goldstone bosons.
VLC adopts the $SU(2)_L\otimes SU(2)_R/SU(2)_{L+R}$
coset representative $(u_L,u_R)$ with the canonical choice $u_L^{\phantom{\dagger}}=u_R^\dagger=u(\varphi)$, which transforms as $u\to g_L^{\phantom{\dagger}} ug_h^\dagger=g_h^{\phantom{\dagger}} u g_R^\dagger$ under the action of the chiral group element  $(g_L,g_R)\in SU(2)_L\otimes SU(2)_R$. MUC chooses instead $U(\varphi) = u_L^{\phantom{\dagger}} u_R^\dagger = u^2$, with $U\to g_L U g_R^\dagger$, getting rid of the compensating $SU(2)_{L+R}$ transformation $g_h$.
Thus, the building blocks of these two bases transform differently under $\mG$. The VLC choice is more suitable for the inclusion of resonances \cite{Ecker:1988te,Pich:2016lew},
since they transform covariantly under $SU(2)_{L+R}$.

Furthermore, the operators in the MUC basis are listed without any assumptions on $CP$ properties and without specifying any particular suppression for the violations of custodial symmetry.
If custodial symmetry is only broken by weak interactions,
like hypercharge and Yukawa couplings, operators that violate custodial symmetry will come with additional factors of weak couplings.
This increases the chiral dimension and, therefore, the order at which these operators must appear in the EFT with respect to their expected order from pure na\"\i ve dimensional analysis arguments. The custodial suppression of some MUC operators was already discussed in Ref.~\cite{Krause:2016uhw}. Therefore, we start by listing all $CP$-even combinations of operators of the MUC basis that are not custodially suppressed beyond $\mathcal{O}(p^{4})$.
We collect in Appendix~\ref{app:relations} some algebraic identities that are useful to relate the operators within the two bases.

\subsection{Bosonic operators}
From the complete list of bosonic operators in the MUC basis,
we only need the operators that are $CP$-even.
Furthermore, we assume that the custodial-breaking
spurion $\tau_{L}$ in the MUC basis comes with an explicit weak, custodial-breaking coupling (such as $g'$).\footnote{$\tau_{L}$ is related to the VLC custodial-breaking spurion through $\mT=-g' u^\dagger \tau_L u$.} If there is no $B_{\mu\nu}$ (and therefore $g'$) in the operator, it will be further suppressed beyond $\mathcal{O}(p^{4})$. The corresponding set contains the following 12 operators, where we use the naming convention of Ref.~\cite{Buchalla:2013rka}:
\begin{equation}
  \label{eq:dic9}
  \mathcal{O}_{\beta}, \,\mathcal{O}_{D1}, \,\mathcal{O}_{D2}, \,\mathcal{O}_{D7}, \,\mathcal{O}_{D8}, \,\mathcal{O}_{D11}, \,\mathcal{O}_{Xh1}, \,\mathcal{O}_{Xh2}, \,\mathcal{O}_{Xh3}, \,\mathcal{O}_{XU1}, \,\mathcal{O}_{XU7}, \,\mathcal{O}_{XU8}.
\end{equation}
In the VLC basis, there are 15 operators involving only bosonic fields: $\mathcal{O}_{1-12}$ and $\mathcal{\tilde O}_{1-3}$. We find the relations between the two sets of operators given in Table \ref{tab:dic:bosonic}. Just by numbers, there are three operators too much in the VLC basis. $\mathcal{O}_{11}$ is the kinetic term for the auxilliary field, it becomes redundant once we go to the SM. We discuss the two remaining operators, $\mathcal{\tilde O}_{3}$ and $\mathcal{O}_{9}$, after the fermion bilinear operators.

{\renewcommand{\arraystretch}{1.5}
\begin{table}[!ht]
    \centering
    \begin{tabular}[t]{|c|c||c|c|}
\hline
VLC & MUC & VLC & MUC \\
\hline
\hline
$\mathcal{O}_{1}$ & $\mathcal{O}_{XU1}$ & $\mathcal{O}_{6}$ & $\mathcal{O}_{D7}$ \\
\hline
$\mathcal{\tilde O}_{1}$ & $\frac{1}{2}\mathcal{O}_{XU7} - \frac{1}{2}\mathcal{O}_{XU8}$ & $\mathcal{O}_{7}$ & $\mathcal{O}_{D8}$ \\
\hline
$\mathcal{O}_{2}$ & $\mathcal{O}_{Xh2} + \frac{1}{2}\mathcal{O}_{Xh1}$ & $\mathcal{O}_{8}$ & $\mathcal{O}_{D11}$ \\
\hline
$\mathcal{\tilde O}_{2}$ &$\mathcal{O}_{Xh2} - \frac{1}{2}\mathcal{O}_{Xh1}$ & $\mathcal{O}_{9}$ & to be discussed below \\
\hline
$\mathcal{O}_{3}$ & $-\frac{1}{2}\mathcal{O}_{XU7} - \frac{1}{2}\mathcal{O}_{XU8}$ & $\mathcal{O}_{10}$ & $\frac{1}{v^{2}}\mathcal{O}_{\beta}$ \\
\hline
$\mathcal{\tilde O}_{3}$ & to be discussed below & $\mathcal{O}_{11}$ & redundant when $\hat X^{\mu} \rightarrow - g' B^{\mu}$\\
\hline
$\mathcal{O}_{4}$ & $\mathcal{O}_{D2}$ & $\mathcal{O}_{12}$ & $\mathcal{O}_{Xh3}$ \\
\hline
$\mathcal{O}_{5}$ & $\mathcal{O}_{D1}$ & & \\
\hline
    \end{tabular}
    \caption{Dictionary of bosonic operators.}
    \label{tab:dic:bosonic}
  \end{table}
}

\subsection{Fermion bilinears}

In the MUC basis, the fermion-bilinear operators are defined such that they reduce to operators with a single $Z$ or $W^{\pm}$ in the unitary gauge. Since this violates custodial symmetry, we first have to find the linear combinations that are not custodially suppressed. We use for the vector fermion bilinears
\begin{align}
  \begin{aligned}
      \label{eq:dic10}
      \tilde{\mathcal{O}}_{\psi V1}^{L}\, &\equiv\, \bar\psi_{L}\gamma_{\mu}L^{\mu}\psi_{L}\, =\, -(2\mathcal{O}_{\psi V2}+\mathcal{O}_{\psi V3}+\mathcal{O}_{\psi V3}^{\dagger} )\, ,\\
      \tilde{\mathcal{O}}_{\psi V2}^{L}\, &\equiv\, \bar\psi_{L}\gamma_{\mu}\{\tau_{L},L^{\mu}\}\psi_{L}\, =\, -\mathcal{O}_{\psi V1}\, ,\\
      \tilde{\mathcal{O}}_{\psi V1}^{R}\, &\equiv\, \bar\psi_{R}\gamma_{\mu}U^{\dagger}L^{\mu}U\psi_{R}\, =\, -(\mathcal{O}_{\psi V4}-\mathcal{O}_{\psi V5}+\mathcal{O}_{\psi V6}+\mathcal{O}_{\psi V6}^{\dagger} )\, ,\\
      \tilde{\mathcal{O}}_{\psi V2}^{R}\, &\equiv\, \bar\psi_{R}\gamma_{\mu}U^{\dagger}\{\tau_{L},L^{\mu}\}U\psi_{R}\, =\, -\tfrac{1}{2}(\mathcal{O}_{\psi V4}+\mathcal{O}_{\psi V5})\, .\\
  \end{aligned}
\end{align}
The scalar fermion bilinears are
\begin{align}
  \begin{aligned}
      \label{eq:dic11}
      \tilde{\mathcal{O}}_{\psi S1}\, &\equiv\, \bar\psi_{L}U\psi_{R} \tfrac{(\partial_{\mu}h)}{v} \tfrac{(\partial^{\mu}h)}{v}\, =\, \mathcal{O}_{\psi S14}+\mathcal{O}_{\psi S15}\, ,\\
      \tilde{\mathcal{O}}_{\psi S2}\, &\equiv\, \bar\psi_{L}L_{\mu}U\psi_{R} \tfrac{(\partial^{\mu}h)}{v}\,  =\, \mathcal{O}_{\psi S10}-\mathcal{O}_{\psi S11}+\mathcal{O}_{\psi S12}+\mathcal{O}_{\psi S13}\, ,\\
      \tilde{\mathcal{O}}_{\psi S3}\, &\equiv\, \bar\psi_{L}L_{\mu}L^{\mu}U\psi_{R}\,  =\,\tfrac{1}{2}(\mathcal{O}_{\psi S1}+\mathcal{O}_{\psi S2})\, .
  \end{aligned}
\end{align}
The tensor fermion bilinears are
\begin{align}
  \begin{aligned}
      \label{eq:dic12}
      \tilde{\mathcal{O}}_{\psi T1}\, &\equiv\, \bar\psi_{L}\sigma_{\mu\nu}L^{\mu}L^{\nu}U\psi_{R}\,  =\, \mathcal{O}_{\psi T1}-\mathcal{O}_{\psi T2}+2\mathcal{O}_{\psi T3}-2\mathcal{O}_{\psi T4}\, ,\\
      \tilde{\mathcal{O}}_{\psi T2}\, &\equiv\, \bar\psi_{L}\sigma_{\mu\nu}L^{\mu}U\psi_{R} \tfrac{(\partial^{\nu}h)}{v}\,  =\,\mathcal{O}_{\psi T7}-\mathcal{O}_{\psi T8}+\mathcal{O}_{\psi T9}+\mathcal{O}_{\psi T10}\, .
  \end{aligned}
\end{align}
And the dipole operators are
\begin{align}
  \begin{aligned}
      \label{eq:dic13}
      \tilde{\mathcal{O}}_{\psi X1}\, &\equiv\, g' \bar\psi_{L}\sigma_{\mu\nu}U\psi_{R} B^{\mu\nu}\, =\,\mathcal{O}_{\psi X1}+\mathcal{O}_{\psi X2}\, ,\\
      \tilde{\mathcal{O}}_{\psi X2}\, &\equiv\, g' \bar\psi_{L}\sigma_{\mu\nu}\tau_{L}U\psi_{R} B^{\mu\nu}\, =\,\mathcal{O}_{\psi X1}-\mathcal{O}_{\psi X2}\, ,\\
      \tilde{\mathcal{O}}_{\psi X3}\, &\equiv\, g \bar\psi_{L}\sigma_{\mu\nu}W^{\mu\nu}U\psi_{R}\, =\,\mathcal{O}_{\psi X3}-\mathcal{O}_{\psi X4}+\mathcal{O}_{\psi X5}+\mathcal{O}_{\psi X6}\,,\\
      \tilde{\mathcal{O}}_{\psi X4}\, &\equiv\, g_{s} \bar \psi_{L} \sigma_{\mu\nu}G^{\mu\nu}U\psi_{R}\, =\,\mathcal{O}_{\psi X7}+\mathcal{O}_{\psi X8}\, .
  \end{aligned}
\end{align}
These are in total 13 operators in the MUC basis, while in the VLC basis, there are only 11 operators, $\mathcal{O}^{\psi^{2}}_{1-8}$ and $\tilde{\mathcal{O}}^{\psi^{2}}_{1-3}$. We find the dictionary of Table \ref{tab:dic:bilinear}. Just based on the numbers, the VLC basis has two operators less than the MUC basis. The two `missing' operators are $\tilde{\mathcal{O}}_{\psi V1}^{L}$ and $\tilde{\mathcal{O}}_{\psi V1}^{R}$. They are related to the bosonic operators $\mathcal{\tilde O}_{3}$ and $\mathcal{O}_{9}$ that we discussed before.

{\renewcommand{\arraystretch}{1.5}
\begin{table}[!ht]
    \centering
    \begin{tabular}[t]{|c|c||c|c|}
\hline
VLC & MUC& VLC & MUC \\
\hline
\hline
$\mathcal{O}^{\psi^{2}}_{1}$ & $2(\tilde{\mathcal{O}}_{\psi S3}+\mathrm{h.c.})$ & $\mathcal{O}^{\psi^{2}}_{4}$ & $-(\tilde{\mathcal{O}}_{\psi X1}+\mathrm{h.c.}) $\\
\hline
$\mathcal{\tilde O}^{\psi^{2}}_{1}$ & $(\tilde{\mathcal{O}}_{\psi X2}+\mathrm{h.c.})-(\tilde{\mathcal{O}}_{\psi X3}+\mathrm{h.c.})$ & $\mathcal{O}^{\psi^{2}}_{5}$ & $i\tilde{\mathcal{O}}_{\psi S2}+\mathrm{h.c.} $ \\
\hline
$\mathcal{O}^{\psi^{2}}_{2}$ & $2 i(\tilde{\mathcal{O}}_{\psi T1}+\mathrm{h.c.}) $ & $\mathcal{O}^{\psi^{2}}_{6}$ & $\tfrac{1}{2}(-\tilde{\mathcal{O}}_{\psi V2}^{L}+\tilde{\mathcal{O}}_{\psi V2}^{R}) $\\
\hline
$\mathcal{\tilde O}^{\psi^{2}}_{2}$ &$-(\tilde{\mathcal{O}}_{\psi T2}+\mathrm{h.c.}) $ & $\mathcal{O}^{\psi^{2}}_{7}$ & $\tilde{\mathcal{O}}_{\psi S1}+\mathrm{h.c.} $\\
\hline
$\mathcal{O}^{\psi^{2}}_{3}$ & $-(\tilde{\mathcal{O}}_{\psi X2}+\mathrm{h.c.})-(\tilde{\mathcal{O}}_{\psi X3}+\mathrm{h.c.}) $ & $\mathcal{O}^{\psi^{2}}_{8} $ & $\tfrac{1}{2}\tilde{\mathcal{O}}_{\psi X4}$\\
\hline
$\mathcal{\tilde O}^{\psi^{2}}_{3}$ & $\tfrac{1}{2}(\tilde{\mathcal{O}}_{\psi V2}^{L}+\tilde{\mathcal{O}}_{\psi V2}^{R}) $ & & \\
\hline
    \end{tabular}
    \caption{Dictionary of fermionic bilinear operators.}
    \label{tab:dic:bilinear}
  \end{table}
}

In total, the numbers of operators in the bosonic and the fermionic-bilinear section of the bases coincide (apart from the kinetic term of $\hat X_{\mu}$ for obvious reasons). The only thing left to show is how the MUC operators $\tilde{\mathcal{O}}_{\psi V1}^{L}$ and $\tilde{\mathcal{O}}_{\psi V1}^{R}$  are related to $\mathcal{\tilde O}_{3}$ and $\mathcal{O}_{9}$ of VLC. We start from the VLC operators and integrate by parts. For simplicity, we will not write down the Higgs-dependent function that multiplies the operator and the terms that it induces when the integration by parts is performed:
\begin{align}
  \begin{aligned}
      \label{eq:dic14}
      \tfrac{(\partial_{\mu}h)}{v} \langle f^{\mu\nu}_{\pm}u_{\nu}\rangle_{2}\, &\rightarrow\, -\langle (u^{\dagger}D_{\mu}\hat W^{\mu\nu}u \pm u D_{\mu}\hat B^{\mu\nu}u^{\dagger} ) u_{\nu}\rangle_{2} - \frac{i}{2}\,\langle f^{\mu\nu}_{\mp}[u_{\mu},u_{\nu}]\rangle_{2} + \frac{1}{2}\,\langle f^{\mu\nu}_{\pm}f_{- \, \mu\nu}\rangle_{2}\\
      &=\,\frac{g^{2}}{4}\, \langle u_{\mu}(J^{\mu}_{V}-J^{\mu}_{A})\rangle_{2} \pm \frac{g'^{2}}{4}\, \langle u_{\mu}(J^{\mu}_{V}+J^{\mu}_{A})\rangle_{2} -\frac{v^{2}}{4}\, (g^{2}\pm g'^{2})\,\langle u_{\mu}u^{\mu}\rangle_{2} \\
      &\phantom{=\;} - \frac{i}{2}\,\langle f^{\mu\nu}_{\mp}[u_{\mu},u_{\nu}]\rangle_{2}+ \frac{1}{2}\,\langle f^{\mu\nu}_{\pm}f_{- \, \mu\nu}\rangle_{2}\, .
  \end{aligned}
\end{align}
We are therefore free to choose if we write $\tfrac{(\partial_{\mu}h)}{v} \langle f^{\mu\nu}_{\pm}u_{\nu}\rangle_{2}$ or $\langle u_{\mu} J^{\mu}_{V/A} \rangle_{2} $ as operators in the Lagrangian. The latter directly correspond to the `missing' vector currents of the MUC basis.
\begin{align}
  \begin{aligned}
      \label{eq:dic15}
      \langle u_{\mu} J^{\mu}_{V} \rangle_{2}\, &=\,\tilde{\mathcal{O}}_{\psi V1}^{R}+\tilde{\mathcal{O}}_{\psi V1}^{L} ,\\
      \langle u_{\mu} J^{\mu}_{A} \rangle_{2}\, &=\,\tilde{\mathcal{O}}_{\psi V1}^{R}-\tilde{\mathcal{O}}_{\psi V1}^{L} .
  \end{aligned}
\end{align}
The two bases are therefore equivalent when the bosonic and the fermion-bilinear operators are considered together.

\subsection{Four-fermion operators}

A generic four-fermion operator can be written as
\begin{equation}
  \label{eq:dic16}
  \left(\bar\psi_{\alpha,a}^{i,p}\Gamma_{\alpha\beta}\psi_{\beta,b}^{j,q}\right)\left(\bar\psi_{\gamma,c}^{k,r}\Gamma_{\gamma\delta}\psi_{\delta,d}^{l,s}\right),
\end{equation}
where, $\alpha$--$\delta$ are spinor, $a$--$d$ $SU(3)$, $i$--$l$  $SU(2)$, and $p$--$s$ generation indices. Usually, we suppress the spinor indices, as they are always contracted within each bracket. The generation indices, if necessary, are kept explicit. In order to form a singlet, there are two ways to contract the $SU(2)$ and two ways to contract the $SU(3)$ indices: $\delta_{ij}\delta_{kl}$ and $\delta_{il}\delta_{jk}$, and $\delta_{ab}\delta_{cd}$ and $\delta_{ad}\delta_{bc}$, giving four possibilities to contract all indices in the operator. Using Fierz identities (see Appendix \ref{app:fierz}), we find that only two of them are independent: either $SU(2)$ and $SU(3)$ indices of the same pair of fermions are contracted, or the indices of two different pairs are contracted. Since we want all the indices contracted within each bracket, we use Eq.~\eqref{eq:sun} to relate the two types of operators. As each fermion current has a chiral dimension of two, only currents without $\tau_{L}$ insertions are relevant for our case. The operators that we use are
\begin{align}
  \begin{aligned}
      \label{eq:19}
      \tilde{\mathcal{O}}_{4LL1}\, &\equiv\, (\bar\psi_{L}\gamma_{\mu}\psi_{L})(\bar\psi_{L}\gamma^{\mu}\psi_{L})\, =\, \mathcal{O}_{LL1}\, ,\\
      \tilde{\mathcal{O}}_{4LL2}\, &\equiv\, (\bar\psi_{L}\gamma_{\mu}T^{A}\psi_{L})(\bar\psi_{L}\gamma^{\mu}T^{A}\psi_{L})\, = \,\mathcal{O}_{LL2}+\tfrac{1}{12}\mathcal{O}_{LL1}\, ,\\
      \tilde{\mathcal{O}}_{4LR1}\, &\equiv\, (\bar\psi_{L}\gamma_{\mu}\psi_{L})(\bar\psi_{R}\gamma^{\mu}\psi_{R})\, =\, \mathcal{O}_{LR1}+\mathcal{O}_{LR3}\, ,\\
      \tilde{\mathcal{O}}_{4LR2}\, &\equiv\, (\bar\psi_{L}\gamma_{\mu}T^{A}\psi_{L})(\bar\psi_{R}\gamma^{\mu}T^{A}\psi_{R})\, =\, \mathcal{O}_{LR2}+\mathcal{O}_{LR4}\, ,\\
      \tilde{\mathcal{O}}_{4RR1}\, &\equiv\, (\bar\psi_{R}\gamma_{\mu}\psi_{R})(\bar\psi_{R}\gamma^{\mu}\psi_{R})\, =\, \mathcal{O}_{RR1}+\mathcal{O}_{RR2}+2\mathcal{O}_{RR3}\, ,\\
      \tilde{\mathcal{O}}_{4RR2}\, &\equiv\, (\bar\psi_{R}\gamma_{\mu}T^{A}\psi_{R})(\bar\psi_{R}\gamma^{\mu}T^{A}\psi_{R})\, =\, \tfrac{1}{3}\mathcal{O}_{RR1}+\tfrac{1}{3}\mathcal{O}_{RR2}+2\mathcal{O}_{RR4}\, ,\\
      \tilde{\mathcal{O}}_{4S1}\, &\equiv\, (\bar\psi_{L}U\psi_{R})(\bar\psi_{L}U\psi_{R})+\text{h.c. }\, =\, \mathcal{O}_{FY1}+\mathcal{O}_{FY3}+2\mathcal{O}_{ST5}+\mathrm{h.c.}\, ,\\
      \tilde{\mathcal{O}}_{4S2}\, &\equiv\, (\bar\psi_{L}U\psi_{R})(\bar\psi_{R}U^{\dagger}\psi_{L})\, =\, \mathcal{O}_{FY5}+\mathcal{O}_{FY5}^{\dagger}\\
      & -\,\tfrac{1}{12}(\mathcal{O}_{LR1}+\mathcal{O}_{LR3})-\tfrac{1}{2}(\mathcal{O}_{LR2}+\mathcal{O}_{LR4})+\tfrac{1}{6}(\mathcal{O}_{LR12}-\mathcal{O}_{LR10})-\mathcal{O}_{LR11}+\mathcal{O}_{LR13}\, ,\\
      \tilde{\mathcal{O}}_{4S3}\, &\equiv\, (\bar\psi_{L}U T^{A}\psi_{R})(\bar\psi_{L}U T^{A}\psi_{R})+\text{h.c. }\, =\, \mathcal{O}_{FY2}+\mathcal{O}_{FY4}+2\mathcal{O}_{ST7}+\mathrm{h.c.}\, ,\\
      \tilde{\mathcal{O}}_{4S4}\, &\equiv\, (\bar\psi_{L}U T^{A}\psi_{R})(\bar\psi_{R}U^{\dagger} T^{A}\psi_{L})\, =\, \mathcal{O}_{FY6}+\mathcal{O}_{FY6}^{\dagger}\\
      & -\,\tfrac{1}{9}(\mathcal{O}_{LR1}+\mathcal{O}_{LR3})+\tfrac{1}{12}(\mathcal{O}_{LR2}+\mathcal{O}_{LR4})-\tfrac{5}{72}(\mathcal{O}_{LR12}-\mathcal{O}_{LR10})+\tfrac{1}{6}(\mathcal{O}_{LR11}-\mathcal{O}_{LR13})\, ,\\
      \tilde{\mathcal{O}}_{4T1}\, &\equiv\, (\bar\psi_{L}U\sigma_{\mu\nu}\psi_{R})(\bar\psi_{L}U\sigma^{\mu\nu}\psi_{R})+\text{h.c. }\\
      & =\, -\tfrac{20}{3}(\mathcal{O}_{FY1}+\mathcal{O}_{FY3})-16(\mathcal{O}_{FY2}+\mathcal{O}_{FY4})-8\mathcal{O}_{ST5}-\tfrac{16}{3}\mathcal{O}_{ST6}-32\mathcal{O}_{ST8}+\mathrm{h.c.}\, ,\\
      \tilde{\mathcal{O}}_{4T2}\, &\equiv\, (\bar\psi_{L}U\sigma_{\mu\nu}T^{A}\psi_{R})(\bar\psi_{L}U\sigma^{\mu\nu}T^{A}\psi_{R})+\text{h.c. }\\
      & =\, -\tfrac{32}{9}(\mathcal{O}_{FY1}+\mathcal{O}_{FY3})-\tfrac{4}{3}(\mathcal{O}_{FY2}+\mathcal{O}_{FY4})-\tfrac{64}{9}\mathcal{O}_{ST6}-8\mathcal{O}_{ST7}+\tfrac{16}{3}\mathcal{O}_{ST8}+\mathrm{h.c.}\, .
  \end{aligned}
\end{align}
The tensor-equivalent of the scalar operators $\tilde{\mathcal{O}}_{4S2}$ and $\tilde{\mathcal{O}}_{4S4}$ vanish by using Fierz identities. In the VLC basis, there are 12 four-fermion operators. We list the dictionary in Table \ref{tab:dic:fourfermi}.

{\renewcommand{\arraystretch}{1.5}
\begin{table}[!ht]
    \centering
    \begin{tabular}[t]{|c|c|}
\hline
VLC & MUC \\
\hline
\hline
$\mathcal{O}^{\psi^{4}}_{1}$ & $-\tfrac{1}{6}\tilde{\mathcal{O}}_{4S1}-\tilde{\mathcal{O}}_{4S3}-\tfrac{1}{24}\tilde{\mathcal{O}}_{4T1}-\tfrac{1}{4}\tilde{\mathcal{O}}_{4T2}-\tfrac{1}{3}\tilde{\mathcal{O}}_{4LR1}-2\tilde{\mathcal{O}}_{4LR2} $ \\
\hline
$\mathcal{\tilde O}^{\psi^{4}}_{1}$ & $-\tfrac{1}{3}\tilde{\mathcal{O}}_{4LL1}-2\tilde{\mathcal{O}}_{4LL2}+\tfrac{1}{3}\tilde{\mathcal{O}}_{4RR1}+2\tilde{\mathcal{O}}_{4RR2} $ \\
\hline
$\mathcal{O}^{\psi^{4}}_{2}$ & $\tfrac{1}{6}\tilde{\mathcal{O}}_{4S1}+\tilde{\mathcal{O}}_{4S3}+\tfrac{1}{24}\tilde{\mathcal{O}}_{4T1}+\tfrac{1}{4}\tilde{\mathcal{O}}_{4T2}-\tfrac{1}{3}\tilde{\mathcal{O}}_{4LR1}-2\tilde{\mathcal{O}}_{4LR2} $ \\
\hline
$\mathcal{\tilde O}^{\psi^{4}}_{2}$ &$\tilde{\mathcal{O}}_{4RR1}-\tilde{\mathcal{O}}_{4LL1}  $ \\
\hline
$\mathcal{O}^{\psi^{4}}_{3}$ & $\tilde{\mathcal{O}}_{4S1}+2\tilde{\mathcal{O}}_{4S2}  $ \\
\hline
$\mathcal{O}^{\psi^{4}}_{4}$ & $- \tilde{\mathcal{O}}_{4S1}+2\tilde{\mathcal{O}}_{4S2} $ \\
\hline
$\mathcal{O}^{\psi^{4}}_{5}$ & $\tfrac{1}{3}\tilde{\mathcal{O}}_{4LL1}+\tfrac{1}{3}\tilde{\mathcal{O}}_{4RR1}+2\tilde{\mathcal{O}}_{4LL2}+2\tilde{\mathcal{O}}_{4RR2}-\frac{4}{3}\tilde{\mathcal{O}}_{4S2}-8\tilde{\mathcal{O}}_{4S4}  $ \\
\hline
$\mathcal{O}^{\psi^{4}}_{6}$ & $\tfrac{1}{3}\tilde{\mathcal{O}}_{4LL1}+\tfrac{1}{3}\tilde{\mathcal{O}}_{4RR1}+2\tilde{\mathcal{O}}_{4LL2}+2\tilde{\mathcal{O}}_{4RR2}+\frac{4}{3}\tilde{\mathcal{O}}_{4S2}+8\tilde{\mathcal{O}}_{4S4}  $ \\
\hline
$\mathcal{O}^{\psi^{4}}_{7}$ & $\tilde{\mathcal{O}}_{4LL1}+\tilde{\mathcal{O}}_{4RR1}+2\tilde{\mathcal{O}}_{4LR1}  $ \\
\hline
$\mathcal{O}^{\psi^{4}}_{8}$ & $\tilde{\mathcal{O}}_{4LL1}+\tilde{\mathcal{O}}_{4RR1}-2\tilde{\mathcal{O}}_{4LR1}  $ \\
\hline
$\mathcal{O}^{\psi^{4}}_{9}$ & $-2\tilde{\mathcal{O}}_{4S1}-12\tilde{\mathcal{O}}_{4S3}+\tfrac{1}{6}\tilde{\mathcal{O}}_{4T1}+\tilde{\mathcal{O}}_{4T2}  $ \\
\hline
$\mathcal{O}^{\psi^{4}}_{10}$ & $\tilde{\mathcal{O}}_{4T1}  $ \\
\hline
    \end{tabular}
    \caption{Dictionary of four fermion operators.}
    \label{tab:dic:fourfermi}
  \end{table}
}

\subsection{Further comments}

In the derivation of this dictionary, we found a redundancy in the MUC basis. Using the determinant identity for a $2\times 2$ matrix $U$,
  \begin{equation}
    \label{eq:muc:red1}
    U^{i1}U^{j2}-U^{i2}U^{j1}\, =\, \epsilon_{ij}\, \det{U},
  \end{equation}
and the fact that $\det{U}=1$, we find the following relations among the operators of Ref.~\cite{Buchalla:2013rka}:
\begin{align}
  \begin{aligned}
    \label{eq:muc:red2}
    \mathcal{O}_{ST5}-\mathcal{O}_{ST6}\, &=\,\mathcal{O}_{ST1}\, ,\\
    \mathcal{O}_{ST7}-\mathcal{O}_{ST8}\, &=\,\mathcal{O}_{ST2}\, ,\\
    \mathcal{O}_{ST9}-\mathcal{O}_{ST10}\, &=\,\mathcal{O}_{ST3}\, ,\\
    \mathcal{O}_{ST11}-\mathcal{O}_{ST12}\, &=\,\mathcal{O}_{ST4}\, ,\\
  \end{aligned}
\end{align}
and similarly for their Hermitian conjugates.

If quarks and leptons, or more than one generation of fermions, are present, the dictionary changes as follows:
\begin{itemize}
\item In the bosonic sector, the operators $\mathcal{O}_{9}$ and $\tilde{\mathcal{O}}_{3}$  should be traded for $ g^{(\prime)\,\, 2}   \langle u_{\mu} J^{\mu}_{V/A} \rangle_{2} $, as the latter can incorporate the extended fermionic sector more easily. The other relations in Table \ref{tab:dic:bosonic} do not change.
\item The fermion-bilinear operators can be extended by adding another copy of fermion bilinears $J_{S,P,V,A,T}$ for the leptons and promoting the coefficients to be matrices in generation space. This should also be done for the operators $\langle u_{\mu} J^{\mu}_{V/A} \rangle_{2} $ discussed above. Only the operator $\mathcal{O}^{\psi^{2}}_{8}$ will not get a leptonic copy, as leptons are not charged under $SU(3)_{C}$. The dictionary in Table \ref{tab:dic:bilinear} stays unchanged. The new, leptonic operators are mapped to the corresponding leptonic operators in the MUC basis, trivially extending Eqs. \eqref{eq:dic10} -- \eqref{eq:dic13}.
\item The four-fermion sector becomes more complicated. First, it is possible to construct many more operators. Second, when using Fierz identities as before, we would also generate mixed quark-lepton currents, which are not building blocks in the MUC basis. We can therefore not Fierz as much as before and Table \ref{tab:dic:fourfermi} will not be valid for the enlarged fermion sector.
\end{itemize}

\section{Useful Algebraic Identities}
\label{app:relations}

\subsection{QCD algebra}
\label{app:QCD-algebra}

The QCD gauge fields $G_\mu^a$ are described by the $3\times 3$ matrix
\be
\hat G_\mu\, =\, g_s\,G^a_\mu\,T^a \, ,
\ee
entering the covariant derivatives in Eq.~(\ref{e_covG}).
Its field-strength tensor is then given by
\be
\hat G_{\mu\nu} \, =\, i\, [D_\mu,D_\nu]\,=\, \partial_\mu \hat{G}_\nu -\partial_\nu \hat{G}_\mu - i\, [\hat{G}_\mu,\hat{G}_\nu]
\,=\, g_s\,G_{\mu\nu}^a\,T^a \, ,
\ee
with
\be
G_{\mu\nu}^a \, = \, \partial_\mu G_\nu^a \, - \, \partial_\nu G_\mu^a \, + g_s\,f^{abc}\,G_\mu^b\,G_\nu^c,
\ee
and $a,b,c=1,\ldots, 8$.
The $3\times 3$ matrices $T^a= \frac{1}{2}\lambda^a$ are the $SU(3)_C$ generators
in the fundamental representation. It is useful to note the Lie algebra relations
$$
\bra T^a\,T^b\ket_3\, =\, T_R\,\delta^{ab}\,,\qquad\quad
[T^a,T^b]\, =\, i\,f^{abc}\,T^c\, , \qquad\quad
\sum_{c,d} f^{acd}\,f^{bcd}\, =\, C_A\,\delta^{ab}\, ,
$$
$$
\sum_a T^a_{ij}\,T^a_{jk}\, =\, C_F\,\delta_{ik} \,,
\qquad\qquad
T^a_{ij} T^a_{kl}\, =\, T_R\,\left(\delta_{jk}\,\delta_{il} - \frac{1}{N_C}\,\delta_{ij}\,\delta_{kl}\right)\, ,
$$
\begin{equation}
  \label{eq:sun}
T_R\, =\,\frac{1}{2}\, , \qquad \quad
C_F\, =\,\frac{4}{3}\, , \qquad  \quad
C_A\, =\, N_C=3\, .
\end{equation}

\subsection{Operator relations}

First, we collect some identities that are useful to relate the operators within the MUC basis and within the VLC basis. Then, we give the relations between both bases.

Using the projectors
\begin{align}
  \begin{aligned}
    \label{eq:dic1}
    P_{+}  = \frac{1}{2} + \frac{\sigma^3}{2}  
    = \begin{pmatrix}1 & 0 \\ 0 & 0 \end{pmatrix}, &\hspace{2cm} P_{-}  = \frac{1}{2} - \frac{\sigma^3}{2}  
    = \begin{pmatrix}0 & 0 \\ 0 & 1 \end{pmatrix},\\
    P_{12}  =  \frac{\sigma^1+i\sigma^2}{2}    
    = \begin{pmatrix}0 & 1 \\ 0 & 0 \end{pmatrix}, &\hspace{2cm}   P_{21}  =
      \frac{\sigma^1-i\sigma^2}{2}   
     = \begin{pmatrix}0 & 0 \\ 1 & 0 \end{pmatrix},\\
  \end{aligned}
\end{align}
we can decompose a generic fermion bilinear as
\begin{equation}
  \label{eq:dic2}
  \bar\psi \Gamma X \psi = \bar\psi \Gamma (P_{+}-P_{-}) \psi \;\langle X   \frac{\sigma^3}{2}    
  \rangle_{2} + \bar\psi \Gamma (P_{+}+P_{-}) \psi \;\langle \tfrac{X}{2} \rangle_{2} + \bar\psi \Gamma P_{12} \psi \;\langle X P_{21}\rangle_{2} + \bar\psi \Gamma P_{21} \psi \;\langle X P_{12}\rangle_{2}.
\end{equation}
Here, $\psi$ is a fermion that transforms like a right-handed  field (either $\psi_{R}$ or $U^{\dagger}\psi_{L}$), $\Gamma$ is an element of the basis of $4\times 4$ matrices $(1,\gamma^{\mu}, i\gamma^{5}, \gamma^{\mu}\gamma^{5},\sigma^{\mu\nu}=\tfrac{i}{2}\, [\gamma^{\mu},\gamma^{\nu}])$, and $X$ is an object that transforms as $X\rightarrow g_{R} X g_{R}^{\dagger}$
(like $U^{\dagger}W_{\mu\nu}U$, $ iD_{\mu}U^{\dagger}U=U^{\dagger}L_{\mu}U$, $U^{\dagger}\tau_{L}U$).
Recalling the definitions of the basic building blocks of the MUC basis,\footnote{
They are related to the VLC basis tensors through $\mT= -g' u^\dagger \tau_L u$, $u_\mu=u^\dagger L_\mu u$.}
\begin{equation}
  \label{eq:dic3}
  L_{\mu} \equiv i U D_{\mu}U^{\dagger}\quad \text{and}\quad \tau_{L} \equiv U  \frac{\sigma^3}{2}  
  U^{\dagger},
\end{equation}
we find the following relations
\begin{align}
  \begin{aligned}
    \label{eq:dic4}
    D_{\mu}\tau_{L}\, &=\, i\, [L_{\mu},\tau_{L}]\, ,\\
    D_{\mu}L_{\nu}-D_{\nu}L_{\mu} &= g W_{\mu\nu} - g' B_{\mu\nu}\tau_{L}+ i\, [L_{\mu},L_{\nu}]\, ,\\
    \langle U^{\dagger} L_{[\mu}L_{\nu]}U P_{12}\rangle_{2}\, & =\, -2\, \langle \tau_{L}L_{[\mu}\rangle_{2} \langle U P_{12}U^{\dagger} L_{\nu]}\rangle_{2}\, ,\\
    \langle U^{\dagger} L_{[\mu}L_{\nu]}U P_{21}\rangle_{2}\, & =\, 2\, \langle \tau_{L}L_{[\mu}\rangle_{2} \langle U P_{21}U^{\dagger} L_{\nu]}\rangle_{2}\, .\\
  \end{aligned}
\end{align}
Useful relations in the VLC basis are
\begin{align}
  \begin{aligned}
    \label{eq:dic7}
    \nabla_{\mu}u_{\nu}-\nabla_{\nu}u_{\mu}\, &=\, u \hat B_{\mu\nu} u^{\dagger} - u^{\dagger} \hat W_{\mu\nu}u\, =\, - f^{-}_{\mu\nu}\, ,\\
    \nabla_{\mu}(u^{\dagger} \hat W^{\mu\nu}u)\,  &=\, u^{\dagger} \nabla_{\mu}\hat W^{\mu\nu}u - \tfrac{i}{2}\, [u_{\mu},u^{\dagger} \hat W^{\mu\nu}u]\, ,\\
    \nabla_{\mu}(u \hat B^{\mu\nu} u^{\dagger})\, &=\, u \nabla_{\mu}\hat B^{\mu\nu} u^{\dagger}+ \tfrac{i}{2}\, [u_{\mu},u \hat B^{\mu\nu} u^{\dagger}]\, .
  \end{aligned}
\end{align}
To translate the operators between the two bases, we use the relations discussed in Ref.~\cite{Pich:2016lew}:
\begin{align}
  \begin{aligned}
    \label{eq:dic8}
    u^{2}\,  =\, U\, , \hspace{0.5cm}&\hspace{0.5cm} (u^{\dagger})^{2}\, =\, U^{\dagger}\, ,\\
    \hat W^{\mu}\,  =\, -g W^{\mu,a}
    \frac{\sigma^a}{2}  \, ,\hspace{1cm}  \hat B^{\mu}\, &=\, - g' B^{\mu}   \frac{\sigma^3}{2}  \, , \hspace{1cm}  \hat X^{\mu}\, =\, - g' B^{\mu}\, ,\\
    \mathcal{T}\, =\, - u g' &
    \frac{\sigma^3}{2}  u^{\dagger}\, =\, -g' u^{\dagger} \tau_{L} u\, ,
\\
    f^{\pm}_{\mu\nu}\, =\, -\bigl( g u^{\dagger}W_{\mu\nu}^{a} & \frac{\sigma^a}{2}  u\pm g' u B_{\mu\nu}
    \frac{\sigma^3}{2}  u^{\dagger}\bigr)\, ,\\
    u_{\mu}\, =\, u_{\mu}^{\dagger}\, =\, i u D_{\mu}&U^{\dagger}u\, =\, -i u^{\dagger} D_{\mu}U u^{\dagger}\, .
  \end{aligned}
\end{align}
The fermion currents in the MUC basis are expressed in terms of fermions with defined chirality.
We list the resulting relations in Table \ref{tab:bilinears}.\\
\renewcommand{\arraystretch}{1.5}
\begin{table}[!ht]
    \centering
    \begin{tabular}[t]{|c|c|c|}
\hline
 \multicolumn{2}{|c|}{$\Gamma$} & $(J_{\Gamma})_{mn} =\bar\xi_{n}\Gamma\xi_{m}$\\
\hline
S & $1$ & $(\bar\psi_{L}u)_{n}(u\psi_{R})_{m}+(\bar\psi_{R}u^{\dagger})_{n}(u^{\dagger}\psi_{L})_{m}$\\
\hline
P & $i\gamma^{5}$ & $i(\bar\psi_{L}u)_{n}(u\psi_{R})_{m}-i(\bar\psi_{R}u^{\dagger})_{n}(u^{\dagger}\psi_{L})_{m}$\\
\hline
V & $\gamma_{\mu}$ & $(\bar\psi_{L}u)_{n}\gamma_{\mu}(u^{\dagger}\psi_{L})_{m}+(\bar\psi_{R}u^{\dagger})_{n}\gamma_{\mu}(u\psi_{R})_{m}$\\
\hline
A & $\gamma_{\mu}\gamma^{5}$ & $-(\bar\psi_{L}u)_{n}\gamma_{\mu}(u^{\dagger}\psi_{L})_{m}+(\bar\psi_{R}u^{\dagger})_{n}\gamma_{\mu}(u\psi_{R})_{m}$\\
\hline
T & $\sigma_{\mu\nu}$ & $(\bar\psi_{L}u)_{n}\sigma_{\mu\nu}(u\psi_{R})_{m}+(\bar\psi_{R}u^{\dagger})_{n}\sigma_{\mu\nu}(u^{\dagger}\psi_{L})_{m}$\\
\hline
    \end{tabular}
    \caption{The fermion currents in the two notations.}
    \label{tab:bilinears}
  \end{table}
Additional relations come from the EoM. We find
\begin{align}
(D_{\mu} W^{\mu\nu})^{a}\, &=\, (\partial_{\mu} W^{\mu\nu}+i\,g\, [ W_{\mu}, W^{\mu\nu}])^{a}\, =\, g \bar\psi_{L}\gamma^{\nu}\frac{\sigma^a}{2}\psi_{L}-\frac{g v^{2}}{4}\, ([1+F_{U}(h)]\,(u u^{\nu} u^{\dagger})^{a}\, ,\\ \label{eq:dic5}
\partial_\mu B^{\mu\nu} \, &=\, g'  \bar{\psi}\gamma^\nu \left(\Frac{\rm B-L}{2}\right)\psi   +   g'\bar{\psi}_R\gamma^\nu \Frac{\sigma^3}{2}\psi_R +    \Frac{g' v^2}{4}  [1+F_U(h)]      (u^\dagger u^\nu   u)^3  \, ,
  \end{align}
where we defined the $SU(2)_{L}$ components of $W_{\mu\nu}$ {\it via} $W_{\mu\nu} = W_{\mu\nu}^{a} \tfrac{\sigma^{a}}{2}$, and similarly for $u^{\dagger} u^{\nu}u $ and $u u^{\nu} u^{\dagger}$. Note that in Eq. \eqref{eq:dic5} the right-handed vector current and the $({\rm B-L})$ current combine into the hypercharge current of the SM.

\subsection{Fierz identities}
\label{app:fierz}

With the definition of the Dirac $\Gamma$ basis of Table \ref{tab:bilinears} above, we find the Fierz identities compiled in Table~\ref{tab:Fierz}. Note that this Fierz table assumes that the four-fermion operators have the structure $(\bar\psi\Gamma^{a}\psi)(\bar\psi\Gamma^{a}\psi)$ instead of the commonly used $(\bar\psi\Gamma_{a}\psi)(\bar\psi\Gamma^{a}\psi)$ \cite{Itzykson:1980rh}.

\renewcommand{\arraystretch}{1.5}
\begin{table}[!ht]
    \centering
    \begin{tabular}[t]{|c|c|c|c|c|c|}
\hline
 & S & P & V & A & T \\
\hline
S & $-\frac{1}{4}$ & $\frac{1}{4} $& $-\frac{1}{4}$&$\frac{1}{4}$&$-\frac{1}{8}$ \\
\hline
P &$\frac{1}{4}$ & $-\frac{1}{4} $& $-\frac{1}{4}$&$\frac{1}{4}$&$\frac{1}{8}$ \\
\hline
V &$-1$ & $-1 $& $\frac{1}{2}$&$\frac{1}{2}$&$0$ \\
\hline
A &$1$ & $1$& $\frac{1}{2}$&$\frac{1}{2}$&$0$ \\
\hline
T & $-3$ & $3 $& $0$&$0$&$\frac{1}{2}$ \\
\hline
    \end{tabular}
    \caption{Fierz identities for $\Gamma=(1, i\gamma^{5},\gamma^{\mu}, \gamma^{\mu}\gamma^{5},\sigma^{\mu\nu}=\tfrac{i}{2}\,[\gamma^{\mu},\gamma^{\nu}])$.}
    \label{tab:Fierz}
  \end{table}

\section{The Power Counting with Enhanced Couplings}
\label{app:coupling}

In Section~\ref{sec:res-pc} we discussed the power counting of the EWET with resonance fields and assumed that their couplings satisfy $c_{i}\ll M_{R}^{2}/ v^{2} $. Let us now analyze the case when $c_{i}\sim M_{R}^{2}/ v^{2}$ and the linear-resonance interaction approximation cannot be used. Operators with two and more resonance fields must be taken into account in such a situation. In Ref.~\cite{Buchalla:2016bse}, it was shown that in models with an additional Higgsed scalar singlet an enhancement to $\mathcal{O}(M_R^{2}/f^{2})\gg \mathcal{O}(1)$, with $f^{2}$ the sum of the two scalar vacuum expectation values squared, is possible. Since the enhancement came from the scalar obtaining a vacuum expectation value and its mixing with the light scalar $h$, such effects can only affect scalar, color-singlet, fields. The case of $S_{1}^{1}$ is the same that was discussed in Ref.~\cite{Buchalla:2016bse}. In order to see what happens with the triplet, $S^1_3$, {let us consider the Georgi-Machacek Model \cite{Georgi:1985nv,Chanowitz:1985ug,Gunion:1989ci}. Based on Ref.~\cite{Georgi:1985nv}, we can distinguish two possible scenarios:
\begin{enumerate}
\item The scalars are elementary.\\
We would be interested in a scenario where one of the scalars is light (the $125~\text{GeV}$ Higgs) and all the others are heavy. Ref.~\cite{Chanowitz:1985ug} reports that such an area of parameter space exists, but the stability of the hierarchy under loop corrections is not addressed. (In Ref.~\cite{Buchalla:2016bse} an approximate SO(5) symmetry achieved the stability). In general, these models are only reasonable phenomenologically if there is a custodial $SU(2)$ symmetry protecting the $T$ parameter. This symmetry would then also prevent a mixing of the heavy triplet with the singlets \cite{Gunion:1989ci}. Thus, only the singlets would mix with each other. However, there is mixing induced at the one-loop level, since the SM breaks the custodial symmetry at one loop. These effects will then be suppressed by a loop factor. We therefore conclude that there are no effects of
$\mathcal{O}(M_R^2/f^2)$ for the $S^1_3$, but only for the $S^1_1$.

\item The scalars are composite objects.\\
This case is only discussed in Ref.~\cite{Georgi:1985nv}. By construction, they consider the triplet to be light (a pNGB) as well. Since in Technicolor (or how they called it, Ultracolor) models the electroweak symmetry breaking is generated at a high-energy scale, the $SU(2)$ multiplet structure could also be broken at the same level \cite{Georgi:1985nv}, introducing the dangerous mixing. However, the masses of the scalars would be generated from the dynamical symmetry breaking, not from Higgsing. We therefore also conclude that effects of
$\mathcal{O}(M_R^2/f^2)$ are not present for the $S^1_3$.
\end{enumerate}

The discussion of possible enhanced resonance couplings reduces then to the $S_{1}^{1}$ case already analyzed in Ref.~\cite{Buchalla:2016bse}. Phenomenologically, these contributions are interesting, as they modify $\mathcal{L}_{\text{EWET}}^{(2)}$ at leading order, without any heavy-mass suppression. All other resonance contributions modify $\mathcal{L}_{\text{EWET}}^{(2)}$ and $\mathcal{L}_{\text{EWET}}^{(4)}$ at next-to-leading order, {\it i.e.} at $\mathcal{O}(v^{2}/M_R^{2})$.

\section{Simplification of the fermionic doublet resonance Lagrangian}
\label{app:fermiondiagonalization}

The most general  $CP$-invariant Lagrangian with couplings of the $\Psi$ resonance to light fields, up to $\mathcal{O}(p^{2})$, has the form
\bear
\label{eq:LPsiFull}
\mL_\Psi&=&\bar{\Psi}(i\slashed{D}-M_{\Psi})\Psi + m_\lambda \,\mathcal{O}^{(1)}_{1}+ \widetilde{m}_\lambda \,\widetilde{\mathcal{O}}^{(1)}_{1} +\sum_{j=0}^4\left( \lambda_i\,\mO_i^{(2)}  +\wl_i\,\widetilde\mO_i^{(2)} \right),
\eear
using the operators in Table~\ref{tab:ferm.res} and generic couplings $m_\lambda,\widetilde{m}_\lambda,\lambda_i,$ and $\wl_i$. Note that all these parameters are functions of the singlet Higgs field $h$.

%
\begin{table}[!t] 
\begin{center}
\renewcommand{\arraystretch}{1.8}
\begin{tabular}{|c|c|c|}
\hline
$i$ & $ \mathcal{O}^{(1)}_i$ &   $ \widetilde{\mathcal{O}}^{(1)}_i$
\\ \hline
1  & $- (\bar{\xi}\Psi + \bar{\Psi}\xi)\quad  {}^{(\dagger\dagger)} $  & $-(\bar{\xi}\gamma_{5}\Psi - \bar{\Psi}\gamma_{5}\xi)\quad  {}^{(\dagger\dagger)} $
\\
\hline
\multicolumn{3}{}{}  \vspace{-4mm} \\
\hline
$i$ & $\mathcal{O}^{(2)}_i$ &   $\widetilde{\mathcal{O}}^{(2)}_i$
\\ \hline
0  &  $ \bar{\xi}\mT\Psi + \bar{\Psi}\mT\xi  $   &
 $ -(\bar{\xi}\gamma_{5}\mT\Psi - \bar{\Psi}\gamma_{5}\mT\xi) $
\\ \hline
1  & $ \bar{\xi}\gamma^{\mu}\gamma_{5} u_\mu \Psi+ \bar{\Psi}\gamma^{\mu}\gamma_{5} u_\mu \xi$  &
$ \bar{\xi}\gamma^{\mu} u_\mu \Psi+ \bar{\Psi}\gamma^{\mu} u_\mu \xi $
\\ \hline
2  & $ i \Frac{(\partial_\mu h)}{v} (\bar{\xi}\gamma^{\mu} \Psi- \bar{\Psi}\gamma^{\mu} \xi) $  &
$ i \Frac{(\partial_\mu h)}{v} (\bar{\xi}\gamma^{\mu}\gamma_{5} \Psi- \bar{\Psi}\gamma^{\mu}\gamma_{5} \xi) $
\\ \hline
3  & $\Frac{i}{2} \left[\left( \bar{\xi} \gamma^\mu d_\mu \Psi
- \overline{d_\mu \Psi}\gamma^\mu \xi\right)
+
\left(
\xi\leftrightarrow\Psi
\right)
\right] \quad  {}^{(\dagger\dagger)}$
&
$\Frac{i}{2} \left[\left( \bar{\xi} \gamma^\mu \gamma_5 d_\mu \Psi
- \overline{d_\mu \Psi}\gamma^\mu \gamma_5 \xi\right)
+
\left(
\xi\leftrightarrow\Psi
\right)
\right]\quad  {}^{(\dagger\dagger)}$
\\ \hline
4  & $\Frac{i}{2} \left[\left( \bar{\xi} \gamma^\mu d_\mu \Psi
- \overline{d_\mu \Psi}\gamma^\mu \xi\right)
-
\left(
\xi\leftrightarrow\Psi
\right)
\right] \quad  {}^{(*)} $
&
$\Frac{i}{2} \left[\left( \bar{\xi} \gamma^\mu \gamma_5 d_\mu \Psi
- \overline{d_\mu \Psi}\gamma^\mu \gamma_5 \xi\right)
-
\left(
\xi\leftrightarrow\Psi
\right)
\right]\quad  {}^{(*)} $
\\ [1ex] \hline
\end{tabular}
\caption{\small $CP$-conserving operators of $\cO(p^{1})$ and $\cO(p^{2})$ with one heavy resonance and one light fermion field. $\mO_i$ ($\widetilde\mO_i$) denote $P$-even (odd) structures. Terms that can be rotated away through the diagonalization of the ``kinetic'' and ``mass'' terms are marked with $(\dagger\dagger)$. Terms marked with $(*)$ are redundant.  }
\label{tab:ferm.res}
\end{center}
\end{table}

We will show in the following that the operators $\mathcal{O}^{(1)}_{1}$, $\mathcal{O}^{(2)}_{3}$, $\mathcal{O}^{(2)}_{4}$ and $\widetilde{\mathcal{O}}^{(1)}_{1}$, $\widetilde{\mathcal{O}}^{(2)}_{3}$, $\widetilde{\mathcal{O}}^{(2)}_{4}$ from this table are redundant in Eq.~\eqref{eq:LPsiFull}. Afterwards, we are left with a Lagrangian with canonically normalized kinetic and mass terms, where the contributions from $\lambda_3$, $\wl_3$, $m_\lambda$  and $\widetilde{m}_\lambda$ have been absorbed by the parameters $\lambda_2$, $\wl_2$, $m_\xi$ and $M_\Psi$.

First, the term $\lambda_4\mO_4^{(2)}+\wl_4\widetilde\mO_4^{(2)}$ is redundant and can be fully transformed into a combination of $\mO_2^{(2)}$ and $\widetilde\mO_2^{(2)}$, using integration by parts. The remaining terms, however, can be removed by means of an adequate rotation of the $\xi$ and $\Psi$ fields, following the standard procedure to diagonalize kinetic and mass terms (see {\it e.g.}~\cite{Escribano:2010wt}).

\subsection{Diagonalization of quadratic fermion couplings}}
\label{sec:diagonal-P-odd}

	We denote the fermion fields before the diagonalization as $\xi_B$ and $\Psi_B$ and we put them together in the vector
	\bear
	V_B = \left(\begin{array}{c} V_{L,B} \\   V_{R,B} \end{array}\right)\, ,\qquad
	V_{L,B} = \left(\begin{array}{c} \xi_{L,B} \\ \Psi_{L,B} \end{array}\right)\, ,\qquad
	V_{R,B} = \left(\begin{array}{c}  \xi_{R,B} \\ \Psi_{R,B}\end{array}\right)\, .
	\eear
	Therefore, the $\xi_B$ and $\Psi_B$ kinetic and mass terms can be written in the form
	\bear
	\mL&=& \Frac{i}{2}\left( \overline{V}_B\gamma^\mu A d_\mu V_B
	- \overline{(d_\mu V_B)} \gamma^\mu A V_B\right)
	- \overline{V}_B  B V_B \, ,
	\eear
	with
	\bear
	A =  \left(\begin{array}{cc} A_{LL} & 0  \\ 0 & A_{RR} \end{array}\right) = A^t\,
	\qquad \qquad
	B = \left(\begin{array}{cc} 0 & B_{LR}
		\\ B_{LR}^t & 0 \end{array}\right) = B^t \,,
	\eear
	given the contribution from the available $CP$-even operators
	\bear
	A_{LL\, (RR)} &=&  \left(\begin{array}{cc}
		1 & \lambda_3(h)\mp \wl_3(h)  \\ \lambda_3(h)\mp \wl_3(h) & 1\end{array}\right)
	= A_{LL\, (RR)}^t   \, ,
	\nn\\
	B_{LR} &=& \left(\begin{array}{cc} m_\xi(h) & m_\lambda(h)+ \widetilde{m}_\lambda(h)
		\\ m_\lambda(h)- \widetilde{m}_\lambda(h)  & M_\Psi(h) \end{array}\right)   \,.
	\eear
	
	In order to diagonalize this Lagrangian we follow the standard procedure (see {\it e.g.}~\cite{Escribano:2010wt}):
	first we diagonalized the kinetic term and then, in a second step, the mass term.
	First, the diagonalization of the kinetic term is achieved by means of the symmetric
	transformation:
	\bear
	V_{L,B}= Z_L^{\frac{1}{2}} V_{L,K}\, ,\qquad
	\qquad (L\leftrightarrow R)\, ,
	\eear
	with
	\bear
	&&
	Z_L(h)^{\frac{1}{2}} A_{LL}(h) Z_L(h)^{\frac{1}{2}}={\rm\bf 1}
	\Longrightarrow   Z_L(h)=A_{LL}(h)^{-1}=Z_L(h)^t\, ,
	\quad Z_L(h)^{\frac{1}{2}}=(Z(h)^{\frac{1}{2}})^t\, ,
	\nn\\
	&&
	(L\leftrightarrow R)\, .
	\eear
	This leaves the Lagrangian
	\bear
	\mL&=& \Frac{i}{2}\left( \overline{V}_K\gamma^\mu d_\mu V_K
	- \overline{(d_\mu V_K)} \gamma^\mu V_K\right)
	- (\overline{V}_{L,K}  M_K V_{R,K} + \overline{V}_{R,K}  M_K^\dagger V_{L,K})
	\nn\\
	&&
	+\, \Frac{i}{2} (\partial_\mu h) \overline{V}_{L,K} \gamma^\mu
	\left[ Z_L^{-\frac{1}{2}},  (Z_L^{\frac{1}{2}})' \right]
	V_{L,K}
	+(L \leftrightarrow R)  \, ,
	\eear
	being $M_K(h) = Z_L^{\frac{1}{2}} B_{LR} Z_R^{\frac{1}{2}}$.
	Notice that $Z_{L,\, R}^{\frac{1}{2}}$ is not just a rotation since we transform
	$A_{LL, \, RR}\to {\rm\bf 1}$,
	{\it i.e.}, the eigenvalues are in general different than one.

	Second, one diagonalizes the matrix $M_K$
	making use of the general diagonalization
	(similar to that for the CKM matrix~\cite{Pich:2012sx})
	\bear
	M_K &=& S_K^\dagger M S_K U_K\, ,
	\eear
	where $S_K$ and $U_K$ are unitary matrices and $M=$diag$(m_\xi^{\rm phys},M_\Psi^{\rm phys})$
	is diagonal and positive definite.
	Thus, the transformations
	\bear
	V_{L,K} = S_K^\dagger V_L \, , \qquad
	V_{R,K} =U_K^\dagger S_K^\dagger V_R\, ,
	\eear
	lead to the canonical fermionic kinetic and mass terms
	in terms of the physical fields $V$:
	\bear
	\mL&=& \Frac{i}{2}\left( \overline{V}\gamma^\mu d_\mu V
	- \overline{(d_\mu V)} \gamma^\mu V\right)
	- (\overline{V}_L  M V_R + \overline{V}_R  M V_L)
	\nn\\
	&&
+\, \Frac{i}{2} (\partial_\mu h) \overline{V}_{L}\gamma^\mu
   \left(S_K\left[ Z_L^{-\frac{1}{2}},  (Z_L^{\frac{1}{2}})' \right] S_K^\dagger + S_K(S_K^\dagger)'-(S_K)'(S_K^{\dagger})\right)
 V_{L}
\nn\\
&&
+\, \Frac{i}{2} (\partial_\mu h) \overline{V}_{R}\gamma^\mu
  \left(S_K U_K\left[ Z_R^{-\frac{1}{2}},  (Z_R^{\frac{1}{2}})' \right]   U_K^\dagger S_K^\dagger\right. \nn\\
&&\hspace{3cm} \left.    + S_K U_K(U_K^\dagger S_K^\dagger)'-(S_K U_K)'U_K^\dagger S_K^\dagger\right)
 V_{R}    \, .
	\label{eq:diagonal-P-odd}
	\eear

	After all this procedure, we are left with a Lagrangian with
	canonically normalized kinetic and mass terms and where the operators $\mathcal{O}^{(1)}_{1}$, $\mathcal{O}^{(2)}_{3}$ and $\widetilde{\mathcal{O}}^{(1)}_{1}$, $\widetilde{\mathcal{O}}^{(2)}_{3}$
	have been traded off by the operators $\mO_2^{(2)}$ and $\widetilde{\mO}_2^{(2)}$.

\bibliographystyle{unsrturl}

\begin{thebibliography}{99}

\bibitem{Pich:2016lew}
  A.~Pich, I.~Rosell, J.~Santos and J.~J.~Sanz-Cillero,
  ``Fingerprints of heavy scales in electroweak effective Lagrangians,''
  JHEP {\bf 1704} (2017) 012
  [arXiv:1609.06659 [hep-ph]].

\bibitem{Pich:2015kwa}
  A.~Pich, I.~Rosell, J.~Santos and J.~J.~Sanz-Cillero,
  ``Low-energy signals of strongly-coupled electroweak symmetry-breaking scenarios,''
  Phys.\ Rev.\ D {\bf 93} (2016) no.5,  055041
  [arXiv:1510.03114 [hep-ph]].

\bibitem{Aad:2012tfa}
  G.~Aad {\it et al.} [ATLAS Collaboration],
  ``Observation of a new particle in the search for the Standard Model Higgs boson with the ATLAS detector at the LHC,''
  Phys.\ Lett.\ B {\bf 716} (2012) 1
  [arXiv:1207.7214 [hep-ex]].

\bibitem{Chatrchyan:2012xdj}
  S.~Chatrchyan {\it et al.} [CMS Collaboration],
  ``Observation of a new boson at a mass of 125 GeV with the CMS experiment at the LHC,''
  Phys.\ Lett.\ B {\bf 716} (2012) 30
  [arXiv:1207.7235 [hep-ex]].

\bibitem{Buchalla:2016bse}
  G.~Buchalla, O.~Cat\`a, A.~Celis and C.~Krause,
  ``Standard Model Extended by a Heavy Singlet: Linear vs. Nonlinear EFT,''
  Nucl.\ Phys.\ B {\bf 917} (2017) 209
  [arXiv:1608.03564 [hep-ph]].

\bibitem{Ecker:1988te}
  G.~Ecker, J.~Gasser, A.~Pich and E.~de Rafael,
  ``The Role of Resonances in Chiral Perturbation Theory,''
  Nucl.\ Phys.\ B {\bf 321} (1989) 311.

\bibitem{Ecker:1989yg}
G.~Ecker, J.~Gasser, H.~Leutwyler, A.~Pich and E.~de Rafael,
  ``Chiral Lagrangians for Massive Spin--1 Fields,''
  Phys.\ Lett.\ B {\bf 223} (1989) 425.

\bibitem{Rosell:2017kps}
  I.~Rosell, C.~Krause, A.~Pich, J.~Santos and J.~J.~Sanz-Cillero,
  ``Tracks of resonances in electroweak effective Lagrangians,''
  PoS EPS {\bf -HEP2017} (2018) 334
  [arXiv:1710.06622 [hep-ph]].

\bibitem{Buchalla:2013rka}
  G.~Buchalla, O.~Cat\`a and C.~Krause,
  ``Complete Electroweak Chiral Lagrangian with a Light Higgs at NLO,''
  Nucl.\ Phys.\ B {\bf 880} (2014) 552
   [Erratum: Nucl.\ Phys.\ B {\bf 913} (2016) 475]
  [arXiv:1307.5017 [hep-ph]].

\bibitem{Dobado:1989ax}
  A.~Dobado and M.~J.~Herrero,
  ``Phenomenological Lagrangian Approach to the Symmetry Breaking Sector of the Standard Model,''
  Phys.\ Lett.\ B {\bf 228} (1989) 495.

\bibitem{Dobado:1989ue}
  A.~Dobado and M.~J.~Herrero,
  ``Testing the Hypothesis of Strongly Interacting Longitudinal Weak Bosons in Electron - Positron Collisions at Tev Energies,''
  Phys.\ Lett.\ B {\bf 233} (1989) 505.

\bibitem{Dobado:1990zh}
  A.~Dobado, D.~Espriu and M.~J.~Herrero,
  ``Chiral Lagrangians as a tool to probe the symmetry breaking sector of the SM at LEP,''
  Phys.\ Lett.\ B {\bf 255} (1991) 405.

\bibitem{Espriu:1991vm}
  D.~Espriu and M.~J.~Herrero,
  ``Chiral Lagrangians and precision tests of the symmetry breaking sector of the Standard Model,''
  Nucl.\ Phys.\ B {\bf 373} (1992) 117.

\bibitem{Herrero:1993nc}
  M.~J.~Herrero and E.~Ruiz Morales,
  ``The Electroweak chiral Lagrangian for the Standard Model with a heavy Higgs,''
  Nucl.\ Phys.\ B {\bf 418} (1994) 431
  [hep-ph/9308276].

\bibitem{Herrero:1994iu}
  M.~J.~Herrero and E.~Ruiz Morales,
  ``Nondecoupling effects of the SM higgs boson to one loop,''
  Nucl.\ Phys.\ B {\bf 437} (1995) 319
  [hep-ph/9411207].

\bibitem{Feruglio:1992wf}
  F.~Feruglio,
  ``The Chiral approach to the electroweak interactions,''
  Int.\ J.\ Mod.\ Phys.\ A {\bf 8} (1993) 4937
  [hep-ph/9301281].

\bibitem{Pich:1998xt}
  A.~Pich, ``Effective field theory,''
Proc. Les Houches Summer School of Theoretical
Physics --Probing the Standard Model of Particle Interactions-- (Les Houches, France, 1997), eds. R. Gupta {\it et al.} (Elsevier Science B.V., Amsterdam, 1999), Vol. II, p.~949
  [hep-ph/9806303].

\bibitem{Bagger:1993zf}
  J.~Bagger, V.~D.~Barger, K.~m.~Cheung, J.~F.~Gunion, T.~Han, G.~A.~Ladinsky, R.~Rosenfeld and C.~P.~Yuan,
  ``The Strongly interacting W W system: Gold plated modes,''
  Phys.\ Rev.\ D {\bf 49} (1994) 1246
  [hep-ph/9306256].

\bibitem{Koulovassilopoulos:1993pw}
  V.~Koulovassilopoulos and R.~S.~Chivukula,
  ``The Phenomenology of a nonstandard Higgs boson in W(L) W(L) scattering,''
  Phys.\ Rev.\ D {\bf 50} (1994) 3218
  [hep-ph/9312317].

\bibitem{Burgess:1999ha}
  C.~P.~Burgess, J.~Matias and M.~Pospelov,
  ``A Higgs or not a Higgs? What to do if you discover a new scalar particle,''
  Int.\ J.\ Mod.\ Phys.\ A {\bf 17} (2002) 1841
  [hep-ph/9912459].

\bibitem{Wang:2006im}
  L.~M.~Wang and Q.~Wang,
  ``Electroweak chiral Lagrangian for neutral Higgs boson,''
  Chin.\ Phys.\ Lett.\  {\bf 25} (2008) 1984
  [hep-ph/0605104].

\bibitem{Grinstein:2007iv}
  B.~Grinstein and M.~Trott,
  ``A Higgs-Higgs bound state due to new physics at a TeV,''
  Phys.\ Rev.\ D {\bf 76} (2007) 073002
  [arXiv:0704.1505 [hep-ph]].

\bibitem{Buchalla:2012qq}
  G.~Buchalla and O.~Cat\`a,
  ``Effective Theory of a Dynamically Broken Electroweak Standard Model at NLO,''
  JHEP {\bf 1207} (2012) 101
  [arXiv:1203.6510 [hep-ph]].

\bibitem{Alonso:2012px}
  R.~Alonso, M.~B.~Gavela, L.~Merlo, S.~Rigolin and J.~Yepes,
  ``The Effective Chiral Lagrangian for a Light Dynamical ``Higgs Particle",''
  Phys.\ Lett.\ B {\bf 722} (2013) 330
   [Erratum: Phys.\ Lett.\ B {\bf 726} (2013) 926]
  [arXiv:1212.3305 [hep-ph]].

\bibitem{Buchalla:2013eza}
  G.~Buchalla, O.~Cat\`a and C.~Krause,
  ``On the Power Counting in Effective Field Theories,''
  Phys.\ Lett.\ B {\bf 731} (2014) 80
  [arXiv:1312.5624 [hep-ph]].

\bibitem{Pich:2018ltt}
  A.~Pich,
  ``Effective Field Theory with Nambu-Goldstone Modes,''
  arXiv:1804.05664 [hep-ph].

\bibitem{Dobado:1990jy}
  A.~Dobado, M.~J.~Herrero and J.~Terron,
  ``The Role of Chiral Lagrangians in Strongly Interacting $W$(l) $W$(l) Signals at $p p$ Supercolliders,''
  Z.\ Phys.\ C {\bf 50} (1991) 205.

\bibitem{deBlas:2018tjm}
  J.~de Blas, O.~Eberhardt and C.~Krause,
  ``Current and Future Constraints on Higgs Couplings in the Nonlinear Effective Theory,''
  JHEP {\bf 1807} (2018) 048
  [arXiv:1803.00939 [hep-ph]].

\bibitem{deBlas:2016ojx}
  J.~de Blas, M.~Ciuchini, E.~Franco, S.~Mishima, M.~Pierini, L.~Reina and L.~Silvestrini,
  ``Electroweak precision observables and Higgs-boson signal strengths in the Standard Model and beyond: present and future,''
  JHEP {\bf 1612} (2016) 135
  [arXiv:1608.01509 [hep-ph]].

\bibitem{Pich:2013fea}
  A.~Pich, I.~Rosell and J.~J.~Sanz-Cillero,
  ``Oblique S and T Constraints on Electroweak Strongly-Coupled Models with a Light Higgs,''
  JHEP {\bf 1401} (2014) 157
  [arXiv:1310.3121 [hep-ph]].

\bibitem{Pich:2012dv}
  A.~Pich, I.~Rosell and J.~J.~Sanz-Cillero,
  ``Viability of strongly-coupled scenarios with a light Higgs-like boson,''
  Phys.\ Rev.\ Lett.\  {\bf 110} (2013) 181801
  [arXiv:1212.6769 [hep-ph]].

\bibitem{Appelquist:1980vg}
  T.~Appelquist and C.~W.~Bernard,
  ``Strongly Interacting Higgs Bosons,''
  Phys.\ Rev.\ D {\bf 22} (1980) 200.

\bibitem{Sikivie:1980hm}
  P.~Sikivie, L.~Susskind, M.~B.~Voloshin and V.~I.~Zakharov,
  ``Isospin Breaking in Technicolor Models,''
  Nucl.\ Phys.\ B {\bf 173} (1980) 189.

\bibitem{Longhitano:1980iz}
  A.~C.~Longhitano,
  ``Heavy Higgs Bosons in the Weinberg-Salam Model,''
  Phys.\ Rev.\ D {\bf 22} (1980) 1166.

\bibitem{Longhitano:1980tm}
  A.~C.~Longhitano,
  ``Low-Energy Impact of a Heavy Higgs Boson Sector,''
  Nucl.\ Phys.\ B {\bf 188} (1981) 118.

\bibitem{Peskin:1990zt}
  M.~E.~Peskin and T.~Takeuchi,
  ``A New constraint on a strongly interacting Higgs sector,''
  Phys.\ Rev.\ Lett.\  {\bf 65} (1990) 964.

\bibitem{Peskin:1991sw}
  M.~E.~Peskin and T.~Takeuchi,
  ``Estimation of oblique electroweak corrections,''
  Phys.\ Rev.\ D {\bf 46} (1992) 381.

\bibitem{Haller:2018nnx}
  J.~Haller, A.~Hoecker, R.~Kogler, K.~M\"onig, T.~Peiffer and J.~Stelzer,
  ``Update of the global electroweak fit and constraints on two-Higgs-doublet models,''
  Eur.\ Phys.\ J.\ C {\bf 78} (2018) no.8,  675
  [arXiv:1803.01853 [hep-ph]].

\bibitem{weinberg}
  S.~Weinberg,
  ``Phenomenological Lagrangians,''
  Physica A {\bf 96} (1979) 327.

\bibitem{Buchalla:2016sop}
  G.~Buchalla, O.~Cat\`a, A.~Celis and C.~Krause,
  ``Comment on "Analysis of General Power Counting Rules in Effective Field Theory",''
  arXiv:1603.03062 [hep-ph].

\bibitem{Guo:2015isa}
  F.~K.~Guo, P.~Ruiz-Femenía and J.~J.~Sanz-Cillero,
  ``One loop renormalization of the electroweak chiral Lagrangian with a light Higgs boson,''
  Phys.\ Rev.\ D {\bf 92} (2015) 074005
  [arXiv:1506.04204 [hep-ph]].

\bibitem{Buchalla:2017jlu}
  G.~Buchalla, O.~Cata, A.~Celis, M.~Knecht and C.~Krause,
  ``Complete One-Loop Renormalization of the Higgs-Electroweak Chiral Lagrangian,''
  Nucl.\ Phys.\ B {\bf 928} (2018) 93
  [arXiv:1710.06412 [hep-ph]].

\bibitem{Alonso:2017tdy}
  R.~Alonso, K.~Kanshin and S.~Saa,
  ``Renormalization group evolution of Higgs effective field theory,''
  Phys.\ Rev.\ D {\bf 97} (2018) no.3,  035010
  [arXiv:1710.06848 [hep-ph]].

\bibitem{Coleman:1969sm}
  S.~R.~Coleman, J.~Wess and B.~Zumino,
  ``Structure of phenomenological Lagrangians. 1.,''
  Phys.\ Rev.\  {\bf 177} (1969) 2239.

\bibitem{Callan:1969sn}
  C.~G.~Callan, Jr., S.~R.~Coleman, J.~Wess and B.~Zumino,
  ``Structure of phenomenological Lagrangians. 2.,''
  Phys.\ Rev.\  {\bf 177} (1969) 2247.

\bibitem{Pich:2012jv}
  A.~Pich, I.~Rosell and J.~J.~Sanz-Cillero,
  ``One-Loop Calculation of the Oblique S Parameter in Higgsless Electroweak Models,''
  JHEP {\bf 1208} (2012) 106
  [arXiv:1206.3454 [hep-ph]].

\bibitem{Buchalla:2014eca}
  G.~Buchalla, O.~Cat\`a and C.~Krause,
  ``A Systematic Approach to the SILH Lagrangian,''
  Nucl.\ Phys.\ B {\bf 894} (2015) 602
  [arXiv:1412.6356 [hep-ph]].

\bibitem{Hirn:2005fr}
  J.~Hirn and J.~Stern,
  ``Lepton-number violation and right-handed neutrinos in Higgs-less effective theories,''
  Phys.\ Rev.\ D {\bf 73} (2006) 056001
  [hep-ph/0504277].

\bibitem{Appelquist:1984rr}
  T.~Appelquist, M.~J.~Bowick, E.~Cohler and A.~I.~Hauser,
  ``The Breaking of Isospin Symmetry in Theories With a Dynamical Higgs Mechanism,''
  Phys.\ Rev.\ D {\bf 31} (1985) 1676.

\bibitem{Bagan:1998vu}
  E.~Bagan, D.~Espriu and J.~Manzano,
  ``The Effective electroweak chiral Lagrangian: The Matter sector,''
  Phys.\ Rev.\ D {\bf 60} (1999) 114035
  [hep-ph/9809237].


\bibitem{Burgess:2007pt}
  C.~P.~Burgess,
  ``Introduction to Effective Field Theory,''
  Ann.\ Rev.\ Nucl.\ Part.\ Sci.\  {\bf 57} (2007) 329
  [hep-th/0701053].

\bibitem{Pati:1974yy}
  J.~C.~Pati and A.~Salam,
  ``Lepton Number as the Fourth Color,''
  Phys.\ Rev.\ D {\bf 10} (1974) 275
   [Erratum: Phys.\ Rev.\ D {\bf 11} (1975) 703].

\bibitem{Grober:2016wmf}
  R.~Grober, M.~Muhlleitner and M.~Spira,
  ``Signs of Composite Higgs Pair Production at Next-to-Leading Order,''
  JHEP {\bf 1606} (2016) 080
  [arXiv:1602.05851 [hep-ph]].

\bibitem{Dobado:1989gr}
  A.~Dobado, M.~J.~Herrero and T.~N.~Truong,
  ``Study of the Strongly Interacting Higgs Sector,''
  Phys.\ Lett.\ B {\bf 235} (1990) 129.

\bibitem{Dobado:1990am}
  A.~Dobado, M.~J.~Herrero and J.~Terron,
  ``$W^{\pm}Z^{0}$ signals from the strongly interacting symmetry breaking sector,''
  Z.\ Phys.\ C {\bf 50} (1991) 465.

\bibitem{Dobado:1999xb}
  A.~Dobado, M.~J.~Herrero, J.~R.~Pelaez and E.~Ruiz Morales,
  ``CERN LHC sensitivity to the resonance spectrum of a minimal strongly interacting electroweak symmetry breaking sector,''
  Phys.\ Rev.\ D {\bf 62} (2000) 055011
  [hep-ph/9912224].

\bibitem{Filipuzzi:2012bv}
  A.~Filipuzzi, J.~Portoles and P.~Ruiz-Femenia,
  ``Zeros of the $W_L Z_L \to W_L Z_L$ Amplitude: Where Vector Resonances Stand,''
  JHEP {\bf 1208} (2012) 080
  [arXiv:1205.4682 [hep-ph]].

\bibitem{Delgado:2014dxa}
  R.~L.~Delgado, A.~Dobado and F.~J.~Llanes-Estrada,
  ``Possible new resonance from $W_L W_L$-$hh$ interchannel coupling,''
  Phys.\ Rev.\ Lett.\  {\bf 114} (2015) no.22,  221803
  [arXiv:1408.1193 [hep-ph]].

\bibitem{Delgado:2015kxa}
  R.~L.~Delgado, A.~Dobado and F.~J.~Llanes-Estrada,
  ``Unitarity, analyticity, dispersion relations, and resonances in strongly interacting $W_LW_L$, $Z_LZ_L$, and hh scattering,''
  Phys.\ Rev.\ D {\bf 91} (2015) no.7,  075017
  [arXiv:1502.04841 [hep-ph]].

\bibitem{Delgado:2017cat}
  R.~L.~Delgado, A.~Dobado, M.~Espada, F.~J.~Llanes-Estrada and I.~L.~Merino,
  ``Collider production of Electroweak resonances from photon-photon states,''
  arXiv:1710.07548 [hep-ph].


\bibitem{Aaboud:2017yvp}
  M.~Aaboud {\it et al.} [ATLAS Collaboration],
  ``Search for new phenomena in dijet events using 37 fb$^{-1}$ of $pp$ collision data collected at $\sqrt{s}=$13 TeV with the ATLAS detector,''
  Phys.\ Rev.\ D {\bf 96} (2017) no.5,  052004
  [arXiv:1703.09127 [hep-ex]].

\bibitem{Sirunyan:2017ygf}
  A.~M.~Sirunyan {\it et al.} [CMS Collaboration],
  ``Search for new physics with dijet angular distributions in proton-proton collisions at $ \sqrt{s}=13 $ TeV,''
  JHEP {\bf 1707} (2017) 013
  [arXiv:1703.09986 [hep-ex]].

\bibitem{ATLAS:2014cra}
  The ATLAS collaboration [ATLAS Collaboration],
  ``Search for contact interactions and large extra dimensions in the dilepton channel using proton-proton collisions at $\sqrt{s}$ = 8 TeV with the ATLAS detector,''
  ATLAS-CONF-2014-030.

\bibitem{CMS:2014aea}
  CMS Collaboration [CMS Collaboration],
  ``Search for Contact Interactions in Dilepton Mass Spectra in pp Collisions at sqrt(s) = 8 TeV,''
  CMS-PAS-EXO-12-020.

\bibitem{Schael:2013ita}
  S.~Schael {\it et al.} [ALEPH and DELPHI and L3 and OPAL and LEP Electroweak Collaborations],
  ``Electroweak Measurements in Electron-Positron Collisions at W-Boson-Pair Energies at LEP,''
  Phys.\ Rept.\  {\bf 532} (2013) 119
  [arXiv:1302.3415 [hep-ex]].

\bibitem{Falkowski:2017pss}
  A.~Falkowski, M.~González-Alonso and K.~Mimouni,
  ``Compilation of low-energy constraints on 4-fermion operators in the SMEFT,''
  JHEP {\bf 1708} (2017) 123
  [arXiv:1706.03783 [hep-ph]].

\bibitem{AguilarSaavedra:2018nen}
  J.~A.~Aguilar-Saavedra {\it et al.},
  ``Interpreting top-quark LHC measurements in the standard-model effective field theory,''
  arXiv:1802.07237 [hep-ph].

\bibitem{Zhang:2017mls}
  C.~Zhang,
  ``Constraining $qqtt$ operators from four-top production: a case for enhanced EFT sensitivity,''
  Chin.\ Phys.\ C {\bf 42} (2018) no.2,  023104
  [arXiv:1708.05928 [hep-ph]].

\bibitem{Buckley:2015lku}
  A.~Buckley, C.~Englert, J.~Ferrando, D.~J.~Miller, L.~Moore, M.~Russell and C.~D.~White,
  ``Constraining top quark effective theory in the LHC Run II era,''
  JHEP {\bf 1604} (2016) 015
  [arXiv:1512.03360 [hep-ph]].

\bibitem{Greljo:2017vvb}
  A.~Greljo and D.~Marzocca,
  ``High-$p_T$ dilepton tails and flavor physics,''
  Eur.\ Phys.\ J.\ C {\bf 77} (2017) no.8,  548
  [arXiv:1704.09015 [hep-ph]].

\bibitem{Isidori:2013ez}
  G.~Isidori,
  ``Flavor physics and CP violation,''
  arXiv:1302.0661 [hep-ph].

\bibitem{Jung:2018lfu}
  M.~Jung and D.~M.~Straub,
  ``Constraining new physics in $b\to c\ell\nu$ transitions,''
  arXiv:1801.01112 [hep-ph].

\bibitem{Aaboud:2016lwx}
  M.~Aaboud {\it et al.} [ATLAS Collaboration],
  ``Search for new resonances decaying to a $W$ or $Z$ boson and a Higgs boson in the $\ell^+ \ell^- b\bar b$, $\ell \nu b\bar b$, and $\nu\bar{\nu} b\bar b$ channels with $pp$ collisions at $\sqrt s = 13$ TeV with the ATLAS detector,''
  Phys.\ Lett.\ B {\bf 765} (2017) 32
  [arXiv:1607.05621 [hep-ex]].

\bibitem{Khachatryan:2014hpa}
  V.~Khachatryan {\it et al.} [CMS Collaboration],
  ``Search for massive resonances in dijet systems containing jets tagged as W or Z boson decays in pp collisions at $ \sqrt{s} $ = 8 TeV,''
  JHEP {\bf 1408} (2014) 173
  [arXiv:1405.1994 [hep-ex]].

\bibitem{Aad:2015owa}
  G.~Aad {\it et al.} [ATLAS Collaboration],
  ``Search for high-mass diboson resonances with boson-tagged jets in proton-proton collisions at $ \sqrt{s}=8 $ TeV with the ATLAS detector,''
  JHEP {\bf 1512} (2015) 055
  [arXiv:1506.00962 [hep-ex]].

\bibitem{Aaboud:2017ahz}
  M.~Aaboud {\it et al.} [ATLAS Collaboration],
  ``Search for heavy resonances decaying to a $W$ or $Z$ boson and a Higgs boson in the $q\bar{q}^{(\prime)}b\bar{b}$ final state in $pp$ collisions at $\sqrt{s} = 13$ TeV with the ATLAS detector,''
  Phys.\ Lett.\ B {\bf 774} (2017) 494
  [arXiv:1707.06958 [hep-ex]].

\bibitem{Aaboud:2017cxo}
  M.~Aaboud {\it et al.} [ATLAS Collaboration],
  ``Search for heavy resonances decaying into a $W$ or $Z$ boson and a Higgs boson in final states with leptons and $b$-jets in 36 fb$^{-1}$ of $\sqrt s = 13$ TeV $pp$ collisions with the ATLAS detector,''
  JHEP {\bf 1803} (2018) 174
  [arXiv:1712.06518 [hep-ex]].

\bibitem{Sirunyan:2017wto}
  A.~M.~Sirunyan {\it et al.} [CMS Collaboration],
  ``Search for heavy resonances that decay into a vector boson and a Higgs boson in hadronic final states at $\sqrt{s}=13$ TeV,''
  Eur.\ Phys.\ J.\ C {\bf 77} (2017) no.9,  636
  [arXiv:1707.01303 [hep-ex]].

\bibitem{Aaboud:2016okv}
  M.~Aaboud {\it et al.} [ATLAS Collaboration],
  ``Searches for heavy diboson resonances in $pp$ collisions at $\sqrt{s}=13$ TeV with the ATLAS detector,''
  JHEP {\bf 1609} (2016) 173
  [arXiv:1606.04833 [hep-ex]].

\bibitem{Sirunyan:2018iff}
  A.~M.~Sirunyan {\it et al.} [CMS Collaboration],
  ``Search for a heavy resonance decaying to a pair of vector bosons in the lepton plus merged jet final state at $\sqrt{s} =$ 13 TeV,''
  JHEP {\bf 1805} (2018) 088
  [arXiv:1802.09407 [hep-ex]].

\bibitem{Sirunyan:2017nrt}
  A.~M.~Sirunyan {\it et al.} [CMS Collaboration],
  ``Combination of searches for heavy resonances decaying to WW, WZ, ZZ, WH, and ZH boson pairs in proton–proton collisions at $\sqrt{s}=8$ and 13 TeV,''
  Phys.\ Lett.\ B {\bf 774} (2017) 533
  [arXiv:1705.09171 [hep-ex]].

\bibitem{Sirunyan:2017acf}
  A.~M.~Sirunyan {\it et al.} [CMS Collaboration],
  ``Search for massive resonances decaying into $WW$, $WZ$, $ZZ$, $qW$, and $qZ$ with dijet final states at $\sqrt{s}=13\text{ }\text{ }\mathrm{TeV}$,''
  Phys.\ Rev.\ D {\bf 97} (2018) no.7,  072006
  [arXiv:1708.05379 [hep-ex]].

\bibitem{Liu:2018pkg}
  D.~Liu and L.~T.~Wang,
  ``Precision Measurement with Diboson at the LHC,''
  arXiv:1804.08688 [hep-ph].

\bibitem{Sirunyan:2018ivv}
  A.~M.~Sirunyan {\it et al.} [CMS Collaboration],
  ``Search for a heavy resonance decaying into a Z boson and a vector boson in the $ \nu \overline{\nu}\mathrm{q}\overline{\mathrm{q}} $ final state,''
  JHEP {\bf 1807} (2018) 075
  [arXiv:1803.03838 [hep-ex]].

\bibitem{Sirunyan:2018hsl}
  A.~M.~Sirunyan {\it et al.} [CMS Collaboration],
  ``Search for a new heavy resonance decaying into a Z boson and a Z or W boson in 2$\ell$2q final states at $\sqrt{s}=$ 13 TeV,''
  JHEP {\bf 1809} (2018) 101
  [arXiv:1803.10093 [hep-ex]].

\bibitem{ATLAS:2018tpf}
  The ATLAS collaboration [ATLAS Collaboration],
  ``Search for diboson resonances in hadronic final states in 79.8 fb$^{-1}$ of $pp$ collisions at $\sqrt{s} = 13$ TeV with the ATLAS detector,''
  ATLAS-CONF-2018-016.

\bibitem{Dorigo:2018cbl}
  T.~Dorigo,
  ``Hadron Collider Searches for Diboson Resonances,''
  Prog.\ Part.\ Nucl.\ Phys.\  {\bf 100} (2018) 211
  [arXiv:1802.00354 [hep-ex]].

\bibitem{Pappadopulo:2014qza}
  D.~Pappadopulo, A.~Thamm, R.~Torre and A.~Wulzer,
  ``Heavy Vector Triplets: Bridging Theory and Data,''
  JHEP {\bf 1409} (2014) 060
  [arXiv:1402.4431 [hep-ph]].

\bibitem{Delgado:2017cls}
  R.~L.~Delgado, A.~Dobado, D.~Espriu, C.~Garcia-Garcia, M.~J.~Herrero, X.~Marcano and J.~J.~Sanz-Cillero,
  ``Production of vector resonances at the LHC via WZ-scattering: a unitarized EChL analysis,''
  JHEP {\bf 1711} (2017) 098
  [arXiv:1707.04580 [hep-ph]].

\bibitem{Guo:2007ff}
  Z.~H.~Guo, J.~J.~Sanz Cillero and H.~Q.~Zheng,
  ``Partial waves and large N(C) resonance sum rules,''
  JHEP {\bf 0706} (2007) 030
  [hep-ph/0701232].

\bibitem{Ledwig:2014cla}
  T.~Ledwig, J.~Nieves, A.~Pich, E.~Ruiz Arriola and J.~Ruiz de Elvira,
  ``Large-$N_c$ naturalness in coupled-channel meson-meson scattering,''
  Phys.\ Rev.\ D {\bf 90} (2014) no.11,  114020
  [arXiv:1407.3750 [hep-ph]].

\bibitem{Kaplan:1983fs}
  D.~B.~Kaplan and H.~Georgi,
  ``SU(2) x U(1) Breaking by Vacuum Misalignment,''
  Phys.\ Lett.\  {\bf 136B} (1984) 183.

\bibitem{Kaplan:1983sm}
  D.~B.~Kaplan, H.~Georgi and S.~Dimopoulos,
  ``Composite Higgs Scalars,''
  Phys.\ Lett.\  {\bf 136B} (1984) 187.

\bibitem{Dugan:1984hq}
  M.~J.~Dugan, H.~Georgi and D.~B.~Kaplan,
  ``Anatomy of a Composite Higgs Model,''
  Nucl.\ Phys.\ B {\bf 254} (1985) 299.

\bibitem{Georgi:1984af}
  H.~Georgi and D.~B.~Kaplan,
  ``Composite Higgs and Custodial SU(2),''
  Phys.\ Lett.\  {\bf 145B} (1984) 216.

\bibitem{Georgi:1984ef}
  H.~Georgi, D.~B.~Kaplan and P.~Galison,
  ``Calculation of the Composite Higgs Mass,''
  Phys.\ Lett.\  {\bf 143B} (1984) 152.

\bibitem{Agashe:2004rs}
  K.~Agashe, R.~Contino and A.~Pomarol,
  ``The Minimal composite Higgs model,''
  Nucl.\ Phys.\ B {\bf 719} (2005) 165
  [hep-ph/0412089].

\bibitem{Contino:2006qr}
  R.~Contino, L.~Da Rold and A.~Pomarol,
  ``Light custodians in natural composite Higgs models,''
  Phys.\ Rev.\ D {\bf 75} (2007) 055014
  [hep-ph/0612048].

\bibitem{Marzocca:2012zn}
  D.~Marzocca, M.~Serone and J.~Shu,
  ``General Composite Higgs Models,''
  JHEP {\bf 1208} (2012) 013
  [arXiv:1205.0770 [hep-ph]].

\bibitem{Barnard:2013zea}
  J.~Barnard, T.~Gherghetta and T.~S.~Ray,
  ``UV descriptions of composite Higgs models without elementary scalars,''
  JHEP {\bf 1402} (2014) 002
  [arXiv:1311.6562 [hep-ph]].

\bibitem{Ferretti:2013kya}
  G.~Ferretti and D.~Karateev,
  ``Fermionic UV completions of Composite Higgs models,''
  JHEP {\bf 1403} (2014) 077
  [arXiv:1312.5330 [hep-ph]].

\bibitem{Ferretti:2014qta}
  G.~Ferretti,
  ``UV Completions of Partial Compositeness: The Case for a SU(4) Gauge Group,''
  JHEP {\bf 1406} (2014) 142
  [arXiv:1404.7137 [hep-ph]].

\bibitem{Vecchi:2015fma}
  L.~Vecchi,
  ``A dangerous irrelevant UV-completion of the composite Higgs,''
  JHEP {\bf 1702} (2017) 094
  [arXiv:1506.00623 [hep-ph]].

\bibitem{Belyaev:2016ftv}
  A.~Belyaev, G.~Cacciapaglia, H.~Cai, G.~Ferretti, T.~Flacke, A.~Parolini and H.~Serodio,
  ``Di-boson signatures as Standard Candles for Partial Compositeness,''
  JHEP {\bf 1701} (2017) 094
   [Erratum: JHEP {\bf 1712} (2017) 088]
  [arXiv:1610.06591 [hep-ph]].

\bibitem{Cacciapaglia:2014uja}
  G.~Cacciapaglia and F.~Sannino,
  ``Fundamental Composite (Goldstone) Higgs Dynamics,''
  JHEP {\bf 1404} (2014) 111
  [arXiv:1402.0233 [hep-ph]].

\bibitem{Liu:2016idz}
  D.~Liu, A.~Pomarol, R.~Rattazzi and F.~Riva,
  ``Patterns of Strong Coupling for LHC Searches,''
  JHEP {\bf 1611} (2016) 141
  [arXiv:1603.03064 [hep-ph]].

\bibitem{Yepes:2017pjr}
  J.~Yepes and A.~Zerwekh,
  ``Top partner-resonance interplay in a composite Higgs framework,''
  Int.\ J.\ Mod.\ Phys.\ A {\bf 33} (2018) no.11,  1841008
  [arXiv:1711.10523 [hep-ph]].

\bibitem{Yepes:2018dlw}
  J.~Yepes and A.~Zerwekh,
  ``Modelling top partner-vector resonance phenomenology,''
  arXiv:1806.06694 [hep-ph].

\bibitem{Contino:2011np}
  R.~Contino, D.~Marzocca, D.~Pappadopulo and R.~Rattazzi,
  ``On the effect of resonances in composite Higgs phenomenology,''
  JHEP {\bf 1110}, 081 (2011)
  [arXiv:1109.1570 [hep-ph]].

\bibitem{Vignaroli:2011um}
  N.~Vignaroli,
  ``Phenomenology of heavy fermion and vector resonances in composite Higgs models,''
  arXiv:1112.0218 [hep-ph].

\bibitem{Buschmann:2017ucg}
  M.~Buschmann and F.~Yu,
  ``Collider constraints and new tests of color octet vectors,''
  JHEP {\bf 1709} (2017) 101
  [arXiv:1706.07057 [hep-ph]].

\bibitem{DeGrand:2015zxa}
  T.~DeGrand,
  ``Lattice tests of beyond Standard Model dynamics,''
  Rev.\ Mod.\ Phys.\  {\bf 88} (2016) 015001
  [arXiv:1510.05018 [hep-ph]].

\bibitem{Appelquist:2016viq}
  T.~Appelquist {\it et al.},
  ``Strongly interacting dynamics and the search for new physics at the LHC,''
  Phys.\ Rev.\ D {\bf 93} (2016) no.11,  114514
  [arXiv:1601.04027 [hep-lat]].

\bibitem{Bennett:2017kga}
  E.~Bennett, D.~K.~Hong, J.~W.~Lee, C.-J.~D.~Lin, B.~Lucini, M.~Piai and D.~Vadacchino,
  ``Sp(4) gauge theory on the lattice: towards SU(4)/Sp(4) composite Higgs (and beyond),''
  JHEP {\bf 1803} (2018) 185
  [arXiv:1712.04220 [hep-lat]].

\bibitem{Holdom:2017wpj}
  B.~Holdom and R.~Koniuk,
  ``A bound state model for a light scalar,''
  JHEP {\bf 1712} (2017) 102
  [arXiv:1704.05893 [hep-ph]].

\bibitem{Ayyar:2017qdf}
  V.~Ayyar, T.~DeGrand, M.~Golterman, D.~C.~Hackett, W.~I.~Jay, E.~T.~Neil, Y.~Shamir and B.~Svetitsky,
  ``Spectroscopy of SU(4) composite Higgs theory with two distinct fermion representations,''
  Phys.\ Rev.\ D {\bf 97} (2018) no.7,  074505
  [arXiv:1710.00806 [hep-lat]].

\bibitem{Contino:2010rs}
  R.~Contino,
  ``The Higgs as a Composite Nambu-Goldstone Boson,''
  doi:10.1142/9789814327183\_0005
  arXiv:1005.4269 [hep-ph].

\bibitem{Alonso:2016btr}
  R.~Alonso, E.~E.~Jenkins and A.~V.~Manohar,
  Phys.\ Lett.\ B {\bf 756} (2016) 358
  doi:10.1016/j.physletb.2016.03.032
  [arXiv:1602.00706 [hep-ph]].

\bibitem{Alonso:2016oah}
  R.~Alonso, E.~E.~Jenkins and A.~V.~Manohar,
  JHEP {\bf 1608} (2016) 101
  doi:10.1007/JHEP08(2016)101
  [arXiv:1605.03602 [hep-ph]].

\bibitem{Sanz-Cillero:2017jhb}
  J.~J.~Sanz-Cillero,
  PoS EPS {\bf -HEP2017} (2017) 460
  doi:10.22323/1.314.0460
  [arXiv:1710.07611 [hep-ph]].

\bibitem{Kaplan:1991dc}
  D.~B.~Kaplan,
  ``Flavor at SSC energies: A New mechanism for dynamically generated fermion masses,''
  Nucl.\ Phys.\ B {\bf 365} (1991) 259.

\bibitem{Ferretti:2016upr}
  G.~Ferretti,
  ``Gauge theories of Partial Compositeness: Scenarios for Run-II of the LHC,''
  JHEP {\bf 1606} (2016) 107
  [arXiv:1604.06467 [hep-ph]].

\bibitem{Cacciapaglia:2015eqa}
  G.~Cacciapaglia, H.~Cai, A.~Deandrea, T.~Flacke, S.~J.~Lee and A.~Parolini,
  ``Composite scalars at the LHC: the Higgs, the Sextet and the Octet,''
  JHEP {\bf 1511} (2015) 201
  [arXiv:1507.02283 [hep-ph]].

\bibitem{Jiang:2016czg}
  Y.~Jiang and M.~Trott,
  ``On the non-minimal character of the SMEFT,''
  Phys.\ Lett.\ B {\bf 770} (2017) 108
  [arXiv:1612.02040 [hep-ph]].

\bibitem{delAguila:2000rc}
  F.~del Aguila, M.~Perez-Victoria and J.~Santiago,
  ``Observable contributions of new exotic quarks to quark mixing,''
  JHEP {\bf 0009} (2000) 011
  [hep-ph/0007316].

\bibitem{delAguila:2008pw}
  F.~del Aguila, J.~de Blas and M.~Perez-Victoria,
  ``Effects of new leptons in Electroweak Precision Data,''
  Phys.\ Rev.\ D {\bf 78} (2008) 013010
  [arXiv:0803.4008 [hep-ph]].

\bibitem{delAguila:2010mx}
  F.~del Aguila, J.~de Blas and M.~Perez-Victoria,
  ``Electroweak Limits on General New Vector Bosons,''
  JHEP {\bf 1009} (2010) 033
  [arXiv:1005.3998 [hep-ph]].

\bibitem{deBlas:2014mba}
  J.~de Blas, M.~Chala, M.~Perez-Victoria and J.~Santiago,
  ``Observable Effects of General New Scalar Particles,''
  JHEP {\bf 1504} (2015) 078
  [arXiv:1412.8480 [hep-ph]].

\bibitem{deBlas:2017xtg}
  J.~de Blas, J.~C.~Criado, M.~Perez-Victoria and J.~Santiago,
  ``Effective description of general extensions of the Standard Model: the complete tree-level dictionary,''
  JHEP {\bf 1803} (2018) 109
  [arXiv:1711.10391 [hep-ph]].

\bibitem{Dawson:2017vgm}
  S.~Dawson and C.~W.~Murphy,
  ``Standard Model EFT and Extended Scalar Sectors,''
  Phys.\ Rev.\ D {\bf 96} (2017) no.1,  015041
  [arXiv:1704.07851 [hep-ph]].

\bibitem{Manohar:2006ga}
  A.~V.~Manohar and M.~B.~Wise,
  ``Flavor changing neutral currents, an extended scalar sector, and the Higgs production rate at the CERN LHC,''
  Phys.\ Rev.\ D {\bf 74} (2006) 035009
  [hep-ph/0606172].

\bibitem{Hayreter:2017wra}
  A.~Hayreter and G.~Valencia,
  ``LHC constraints on color octet scalars,''
  Phys.\ Rev.\ D {\bf 96} (2017) no.3,  035004
  [arXiv:1703.04164 [hep-ph]].

\bibitem{Pich:2009sp}
  A.~Pich and P.~Tuz\'on,
  ``Yukawa Alignment in the Two-Higgs-Doublet Model,''
  Phys.\ Rev.\ D {\bf 80} (2009) 091702
  [arXiv:0908.1554 [hep-ph]].

\bibitem{Sannino:2015sel}
  F.~Sannino,
  ``$\alpha_s$ at LHC: Challenging asymptotic freedom,''
  arXiv:1511.09022 [hep-ph].

\bibitem{CMS:2014mna}
  V.~Khachatryan {\it et al.} [CMS Collaboration],
  ``Measurement of the inclusive 3-jet production differential cross section in proton–proton collisions at 7 TeV and determination of the strong coupling constant in the TeV range,''
  Eur.\ Phys.\ J.\ C {\bf 75} (2015) no.5,  186
  [arXiv:1412.1633 [hep-ex]].

\bibitem{Khachatryan:2016mlc}
  V.~Khachatryan {\it et al.} [CMS Collaboration],
  ``Measurement and QCD analysis of double-differential inclusive jet cross sections in pp collisions at $ \sqrt{s}=8 $ TeV and cross section ratios to 2.76 and 7 TeV,''
  JHEP {\bf 1703} (2017) 156
  [arXiv:1609.05331 [hep-ex]].

\bibitem{Krause:2016uhw}
  C.~G.~Krause,
  ``Higgs Effective Field Theories - Systematics and Applications,''
  arXiv:1610.08537 [hep-ph].

\bibitem{Itzykson:1980rh}
  C.~Itzykson and J.~B.~Zuber,
  ``Quantum Field Theory,''
    New York, USA: Mcgraw-hill (1980) 705 P.(International Series In Pure and Applied Physics).

\bibitem{Georgi:1985nv}
  H.~Georgi and M.~Machacek,
  ``Doubly Charged Higgs Bosons,''
  Nucl.\ Phys.\ B {\bf 262} (1985) 463.

\bibitem{Chanowitz:1985ug}
  M.~S.~Chanowitz and M.~Golden,
  ``Higgs Boson Triplets With $M_{W} = M_{Z}\cos \theta_{\omega} $,''
  Phys.\ Lett.\  {\bf 165B} (1985) 105.

\bibitem{Gunion:1989ci}
  J.~F.~Gunion, R.~Vega and J.~Wudka,
  ``Higgs triplets in the standard model,''
  Phys.\ Rev.\ D {\bf 42} (1990) 1673.

\bibitem{Escribano:2010wt}
  R.~Escribano, P.~Masjuan and J.~J.~Sanz-Cillero,
  ``Chiral dynamics predictions for eta' $\rightarrow$ eta pi pi,''
  JHEP {\bf 1105} (2011) 094
  [arXiv:1011.5884 [hep-ph]].

\bibitem{Pich:2012sx}
  A.~Pich,
  ``The Standard Model of Electroweak Interactions,''
  arXiv:1201.0537 [hep-ph].




\end{thebibliography}

\end{document}